\pdfoutput=1
\documentclass[a4paper,american,floatfix,pdftex,superscriptaddress,%
aps,prl,%
reprint,twocolumn,
]{revtex4-2}
\PassOptionsToPackage{unicode}{hyperref} 
\usepackage{amsfonts,amsmath,amssymb}
\usepackage{graphicx}
\usepackage[T1]{fontenc}
\usepackage[utf8]{inputenc}
\usepackage{hypernat}
\usepackage[ams,todos]{CMImacros}

\newcommand*\Np{\ensuremath{\text{N}^+}\xspace}  
\newcommand*\Npp{\ensuremath{\text{N}^{++}}\xspace} 
\newcommand*\Ntp{\ensuremath{\text{N}_2^{+}}\xspace} 
\newcommand*\Ntpp{\ensuremath{\text{N}_2^{++}}\xspace} 
\newcommand*\Ntppp{\ensuremath{\text{N}_2^{3+}}\xspace} 
\newcommand*\Nt{\ensuremath{\text{N}_2}\xspace} 

\graphicspath{{./fig/}} 
\newcommand{\cfeldesy}{\affiliation{Center for Free-Electron Laser Science, Deutsches
      Elektronen-Synchrotron DESY, Notkestrasse 85, 22607 Hamburg, Germany}}%
\newcommand{\uhhcui}{\affiliation{Center for Ultrafast Imaging, Universität Hamburg, Luruper
      Chaussee 149, 22761 Hamburg, Germany}}%
\newcommand{\uhhphys}{\affiliation{Department of Physics, Universität Hamburg, Luruper Chaussee 149,
      22761 Hamburg, Germany}}%
\newcommand{\uhhchem}{\affiliation{Department of Chemistry, Universität Hamburg,
      Martin-Luther-King-Platz 6, 20146 Hamburg, Germany}}%
\newcommand{\desy}{\affiliation{Deutsches Elektronen-Synchrotron DESY, Notkestraße 85, 22607
      Hamburg, Germany}}%
\newcommand{\xfel}{\affiliation{European XFEL, Holzkoppel 4, 22869 Schenefeld, Germany}}%
\newcommand{\stemail}{\email[Email:~]{sebastian.trippel@cfel.de}}%
\newcommand{\cmiweb}{\homepage[\\Website:~]{https://www.controlled-molecule-imaging.org}}%

\begin{document}
\title{Single-shot MHz velocity-map-imaging using two Timepix3 cameras}
\author{Hubertus Bromberger}\cfeldesy%
\author{Christopher Passow}\desy%
\author{David Pennicard}\desy %
\author{Rebecca Boll}\xfel%
\author{Jonathan Correa}\desy%
\author{Lanhai He}\cfeldesy%
\author{Melby Johny}\cfeldesy\uhhcui\uhhphys%
\author{Christina Papadopoulou}\desy%
\author{Atia Tul-Noor}\desy%
\author{Joss Wiese}\cfeldesy\uhhcui\uhhchem%
\author{Sebastian Trippel}\stemail\cmiweb\cfeldesy\uhhcui%
\author{Benjamin Erk}\desy%
\author{Jochen Küpper}\cfeldesy\uhhcui\uhhphys\uhhchem%

\begin{abstract}\noindent%
   We demonstrate the application of event-driven Timepix3-based detectors in combination with a
   double-sided velocity-map-imaging spectrometer to record the full 3D momentum of charged
   particles at the free-electron-laser facility FLASH. We measured the XUV induced fragmentation of
   \Nt using 250~kHz FLASH bursts with sub-pixel spatial resolution and up to 1.7~ns temporal
   resolution for photoelectrons. To further demonstrate the capabilities of this camera at even
   higher repetition rates we measured single-shot images of He($1s$) photoelectrons for bursts with
   a repetition rate of 1~MHz. Overall, with the Timepix3 camera we overcome limitations of
   standard-camera technology for advanced-imaging experiments with requirements on high event-rates
   and high spatio-temporal resolution.
\end{abstract}
\date{\today}%
\maketitle%

\section{Introduction}
\label{sec:introduction}
The underlying theme of the Grand Challenges being posed to physics, chemistry, biology, and
materials science is to understand, predict, and ultimately control the properties of
matter~\cite{Young:JPB51:032003, Heinz:OpportunitiesXFELUlrafast:2017}. To help us in this quest, in
recent years ever more advanced light sources have been developed. After the success of the first
superconducting free-electron-laser (FEL), the Free-electron LASer in Hamburg
(FLASH)~\cite{Ackermann:NatPhoton1:336}, the next generation of FELs, such as the European X-ray
Free-Electron Laser facility (XFEL)~\cite{Abela:DESY:2006, Tschentscher:ApplSci} and the upcoming Linac
Coherent Light Source II (LCLS-II)~\cite{SLAC:New-Science-LCLS-II:2016}, operates with repetition
rates of up to 4.5~MHz which are far higher than those of normal-conducting
FELs, potentially allowing for more data to be collected at a much faster
pace.  However, exploiting the advantages provided by these next-generation light sources requires
detectors to collect data at the same rate. This need created a drive for new detectors to match the
potential of the new FELs, with a few developments being reflected in
AGIPD~\cite{Henrich:NIMA633:S11}, ePix~\cite{Blaj:AIPConfProc1741:040012},
tPix~\cite{Markovic:IEEE}, and DSSC~\cite{Porro:NS68:1334}.

Whilst these new facilities provide opportunities for a vast range of experiments, those
investigating the chemical dynamics of molecules and clusters in the gas phase often utilize ion and
electron imaging~\cite{Whitaker:Imaging:2003}. Ion imaging~\cite{Chandler:JCP87:1445} traditionally
uses 2D position-sensitive detectors in tandem with velocity-map-imaging (VMI)
techniques~\cite{Eppink:RSI68:3477} to map the transverse momenta of charged particles. This method
allowed for a deeper insight into the dynamics of molecules~\cite{Chandler:JCP147:013601,
Suits:ACS770:1, Reid:MolPhys3:131}. Common VMI applications use a wide range of light sources, such
as gas-discharge lamps~\cite{Takahashi:RSI71:1337}, pico- and femtosecond laser
systems~\cite{Larsen:JCP109:8857, Roeterdink:PCCP4:601, Trippel:MP111:1738}, high-harmonics
sources~\cite{Aseyev:PRL91:223902, Rading:AS8:998, Kling:JInst9:P05005} as well as studies performed
on large-scale facilities such as synchrotron-radiation sources~\cite{Garcia:RSI76:053302,
Rolles:NIMB261:170} and FELs~\cite{Johnsson:JModOpt55:2693, Burt:PRA96:043415, Erk:JSR25:1529,
Osipov:RSI89:035112} and even enabled the investigations of
chemical-reaction-dynamics~\cite{Whitaker:Imaging:2003, Mikosch:Science319:183}.

Many 2D detectors for VMI purposes are realized through the use of microchannel plates (MCPs) in
combination with a phosphor screen and a camera~\cite{Chichinin:IRPC28:607, Debrah:RSI91:023316,
Cheng:PRA102:052813}. Alternative methods use delay-line-detectors (DLDs)~\cite{Oelsner:RSI72:3968}
instead of the camera-phosphor combination~\cite{Chichinin:RSI73:1856}, which potentially allow even
for 3D ion momentum detection~\cite{Chichinin:RSI73:1856}. However, these detectors come at the
disadvantage of a poorer multi-hit capability compared to the phosphor-screen approach, which limits
the applications for high flux experiments as typically performed at FELs. This has pushed
experiments that use the more traditional 2D detectors to find novel ways to obtain full 3D momentum
distributions. For example, methods like Abel inversion~\cite{Montgomery:JQSRT39:367,
Vrakking:RSI72:4084, Garcia:RSI75:4989} can be utilized, but come at the cost of requiring a
cylindrical symmetry of the observed physical problem. An alternative experimental approach to
inversion techniques for capturing the 3D momentum is slice imaging~\cite{Gebhardt:RSI72:3848}. It
requires to sequentially gate or slice on the time-of-flight distribution for the charged particles under
investigation for a fraction of its flight time over as many slices as needed~\cite{Amini:RSI86:103113}.

In parallel to the emergence of VMI, coincidence momentum-imaging methods, such as cold target
recoil-ion momentum spectroscopy (COLTRIMS)~\cite{Doerner:PhysRep330:95, Ullrich:RPP66:1463,
Ullrich:JPB30:2917} and related techniques~\cite{Davies:JCP111:1, Lafosse:PRL84:5987,
Dowek:NIMA477:323}, also developed into powerful methods for imaging gas-phase photoionization and
photofragmentation dynamics. With the detection of both electrons and cations emitted from the same
atom, molecule, or cluster, the technique allows access to the full kinematics of a single reaction
and, therefore, systematic studies of the underlying physical mechanisms.

In recent years, several ``hybrid'' spectrometers were developed that use elements of both
traditional VMI and COLTRIMS spectrometers, thereby overcoming some limitations of each of the
techniques~\cite{Ablikim:RSI90:055103}. Examples of such double-sided hybrid spectrometers for
coincident electron and ion detection include combinations of a VMI- with a traditional
time-of-flight spectrometer or an electrostatic cylindrical analyzer~\cite{Garcia:RSI75:4989,
Garcia:RSI76:053302, Rolles:NIMB261:170, Pesic:JESRP155:155, Bodi:RSI80:034101,
Keeffe:RSI82:033109}. Furthermore, double-sided VMI spectrometers with two position-sensitive
detectors for coincident momentum-resolved imaging of electrons and ions were
developed~\cite{Strueder:NIMA614:483, Rolles:JPB47:124035, Takahashi:RSI71:1337,
Vredenborg:RSI79:0034}. Many of these spectrometers were specifically designed for photoionization
studies using synchrotron radiation~\cite{Garcia:RSI84:053112, Bomme:RSI84:103104,
Sztaray:JCP147:013944, Hosaka:JJAP45:1841, Tang:RSI80:1131101, Bodi:RSI83:083105}.

In the last decade, several methods were established to record the photons emitted from a phosphor
screen outside the vacuum chamber with nanosecond time resolution. Two important devices providing
such capabilities for visible-light detection are the Pixel Imaging Mass Spectrometry (PImMS)
camera~\cite{Clark:JPCA116:10897, John:JInst7:C08001, Slater:PRA89:011401} and Timepix-based
cameras~\cite{Poikela:JInst9:C05013, Zhao:RSI88:113104, FisherLevine:JInst11:C03016,
Nomerotski:NIMA937:26, Roberts:JInst14:P06001}.  Both PImMS and Timepix cameras employ a
silicon-based sensor technology and are event triggered, \ie, only pixels which collected charges
above a certain threshold are read out and contribute to the data stream. For PImMS~II, a maximum
rate of 60~Hz with a temporal resolution of 12.5~ns can be achieved~\cite{Amini:RSI86:103113}. The
performance of the current Timepix generation cameras, Timepix3, is given by a maximum event rate of
80~Mpixel/s with a temporal resolution of 1.6~ns~\cite{Nomerotski:NIMA937:26}. With few nanosecond
time resolution, it has become possible to measure the full 3D momenta of ions also in experiments
without cylindrical symmetry~\cite{Vallance:PCCP16:383, Lee:CommChem3:72, Debrah:RSI91:023316,
Amini:RSI86:103113}.

Furthermore, it allows for the recording of data at high repetition rates, \eg, provided by
synchrotrons, Optical Parametric Chirped-pulse Amplificatiers (OPCPAs)~\cite{Mecseki:OptLett44:1257},
FELs providing trains of individual pulses at high-repetition-rates in bursts, such as
FLASH~\cite{Ackermann:NatPhoton1:336} or the European XFEL~\cite{Abela:DESY:2006}, and the upcoming
LCLS-II FEL~\cite{SLAC:New-Science-LCLS-II:2016} with a plain 1~MHz repetition rate. Therefore,
time-resolving pixelated detectors permit the combination of position-sensitive phosphor-screen
based detectors with high repetition rate light sources.

Here, we present results from an experimental setup utilizing a double-sided VMI spectrometer with
two Timepix3 cameras to capture the signals from both ions and electrons simultaneously at FLASH. We
captured the 3D momenta of ions and electrons produced from nitrogen and helium by XUV ionization at
repetition rates up to 1~MHz bursts for electrons and 250~kHz bursts for ions.

\section{Method}
\label{sec:method}
\begin{figure}
   \includegraphics[width=\linewidth]{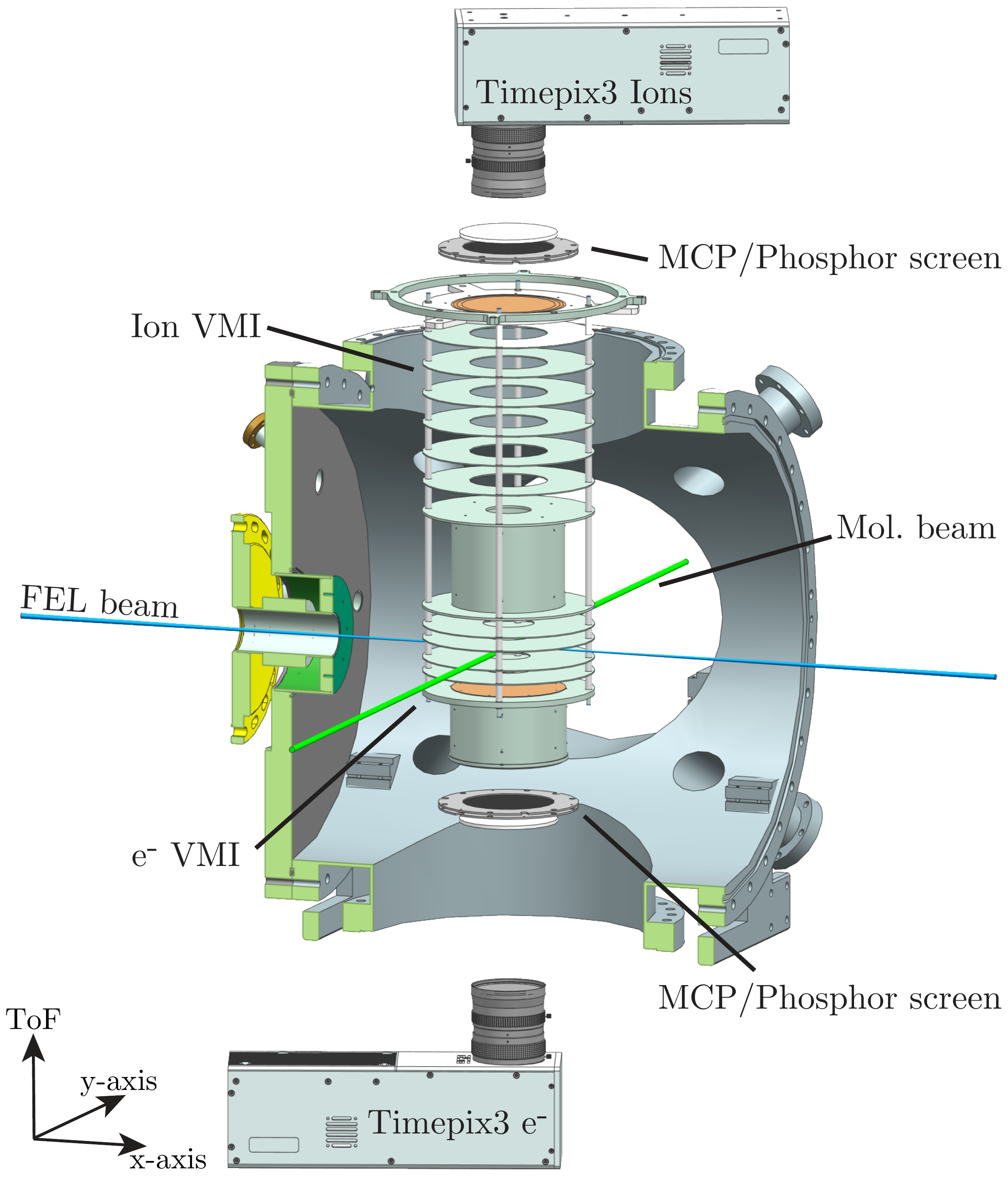}%
   \caption{Schematic of the experimental setup: The FEL (blue) and molecular beam (green)
      intersected in the center of the double-sided VMI. Ions and electrons are detected using
      MCP-phosphor-screen combinations. Photons emitted by the phosphor screens are recorded by
      Timepix3 cameras.}
   \label{fig:experimental_setup}
\end{figure}

The experiments were conducted using the CAMP endstation~\cite{Erk:JSR25:1529} at beamline BL1 at
FLASH~\cite{Ackermann:NatPhoton1:336, Tiedtke:NJP11:023029}. The setup comprises a double-sided VMI
spectrometer capable of simultaneously measuring ions and electrons~\cite{Erk:JSR25:1529}. Details
on the experimental setup were published elsewhere~\cite{Erk:JSR25:1529} and only a brief summary is
provided here. \autoref{fig:experimental_setup} shows a sketch of the detection setup.

Nitrogen or helium gases were expanded through a 30~\um diameter orifice from a stagnation pressure
of $\ordsim1$~bar into vacuum forming a continuous supersonic jet. After 5~mm the jet was collimated
by two skimmers with orifice diameters of 150~\um and 350~\um with a distance of 18~mm of one
another. The interaction region was $\ordsim78$~cm downstream of the 2nd skimmer. The sample and
XUV-photon beams crossed each other in the interaction center of a double-sided VMI. The wavelength
of the FEL was 8.37~nm (148.1~eV) with an averaged pulse energy of $0.9\pm0.2$~µJ and $0.7\pm0.1$~µJ
for the helium and \Nt measurements, respectively, focused to a nominal diameter of 10~µm. For the
\Nt measurements the FEL was configured to provide 8~pulses at a repetition rate of 250~kHz within
the pulse train, limited by the available duration of the radiofrequency window of $\ordsim30$~µs.
Datasets were recorded for 72007 FEL pulse trains with on average 6.5 detected ions per FEL pulse.
For the electrons from He, 30~pulses at a repetition rate of 1~MHz within the train were used with
the same limitation on the radiofrequency window as for the ions. Overall, electron data of 6287
pulse trains was recorded. Here, on average, $\ordsim9$ electrons were detected for a single FEL
pulse. The charged particles were detected by position-sensitive detectors consisting of two MCPs in
chevron configuration and a fast P47 phosphor screen (Photonis, APD~2~PS~75/32/25/8~I~60:1~NR~P47).
Both detectors had a diameter of 80~mm. The photons emitted by the phosphors were recorded by two
time-stamping cameras consisting of an optical-imaging element and an assembly of a light-sensitive
silicon sensor on the Timepix3 chip (Amsterdam Scientific Instruments, TPX3CAM). Both Timepix3
cameras were synchronized with a trigger from the FEL. To match the data from the cameras with the
diagnostics provided by the facility and the beamline, the ``Train ID'' from the FEL corresponding
to each specific XUV shot was recorded simultaneously and stored alongside the Timepix
data~\cite{Fernandes:gitlab}. This allowed for later correlation, sorting, and normalization based
on individual triggers and FEL parameters read from the data acquisition of the
FEL~\cite{Savelyev:NJP19:043009, Ott:PRL123:163201}. Timepix3 itself is an event driven detector
where the data from each pixel consists of a tuple with 4~elements, namely the x and y position of
the pixel on the camera sensor, the time-of-arrival (ToA), which gets translated into the
time-of-flight (ToF) of the particle in the VMI by taking into account the simultaneously recorded
trigger event, and the time-over-threshold (ToT), a measure of the pixel
intensity~\cite{Poikela:JInst9:C05013}.

\begin{figure}
   \includegraphics[width=\columnwidth]{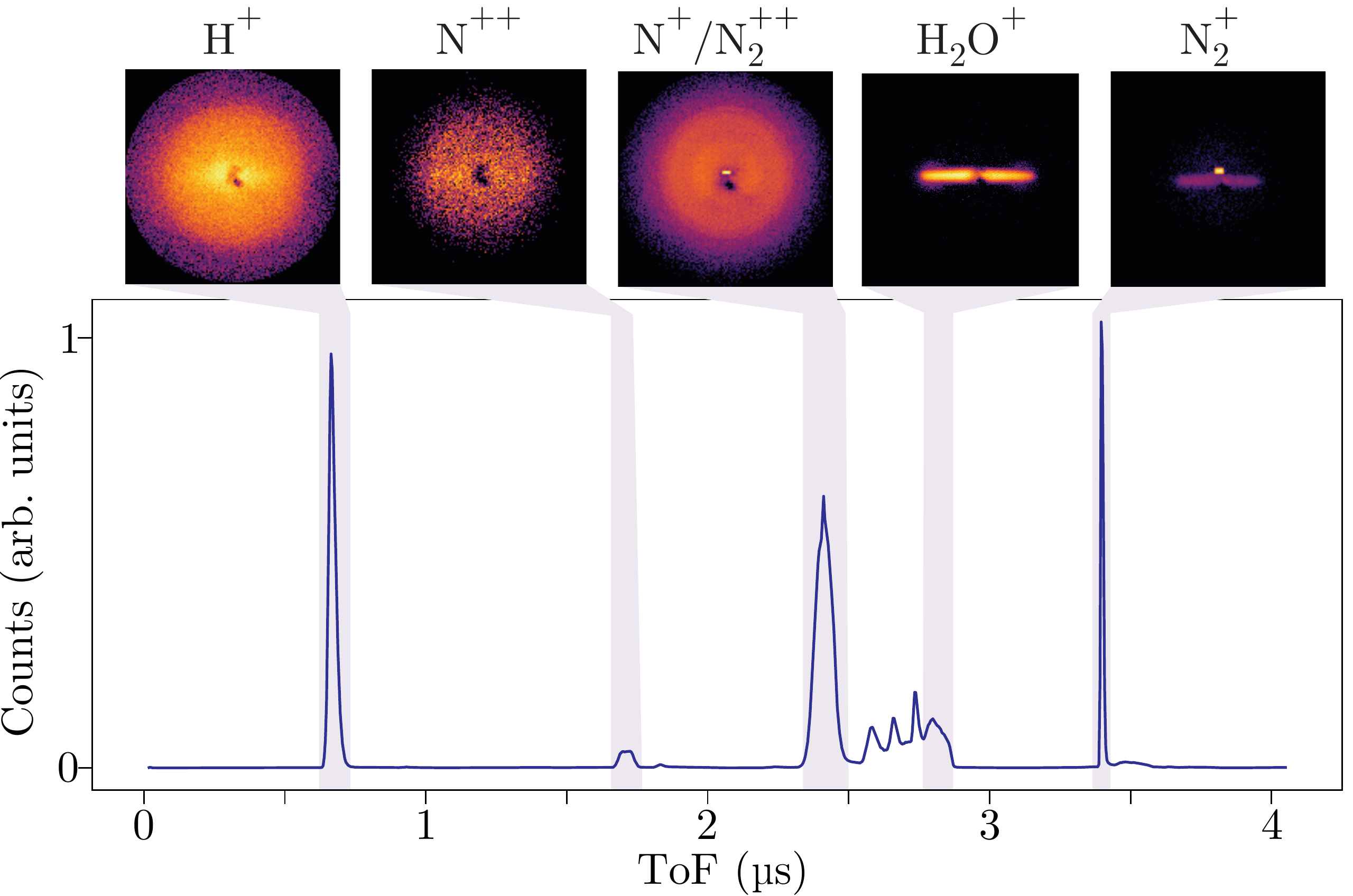}%
   \caption{Ion time-of-flight spectrum from \Nt and background water. VMI images of the highlighted
   ToF ranges are depicted in the insets above. The area of reduced sensitivity visible in the
   center part of the images is due detector damage. Horizontal lines in
   \Ntp and \HHO are from residual gas ionized along the FEL propagation.}
   \label{fig:N2ToFSpectrum}
\end{figure}

Event-based detectors can record VMI images not only for a single arrival time, but for the complete
ToF spectrum~\cite{FisherLevine:JSR25:336, Johny:protection:inprep}. \autoref{fig:N2ToFSpectrum}
shows the ToF spectrum recorded for the XUV ionization of the \Nt beam.
The insets with 2D images show the corresponding VMIs from the highlighted ToF-ranges, which were
all recorded simultaneously. The control of the Timepix3 camera and online analysis of data stream
was performed using an updated version of the open-source library
PymePix~\cite{AlRefaie:JInst14:P10003}. Data analysis made use of numerous Python packages, most
importantly NumPy~\cite{Harris:Nature585:7825}, Scikit-learn~\cite{Pedregosa:JMLR12:2825},
SciPy.ndimage~\cite{Pauli:NatMeth17:261}, Dask, Jupyter, Matplotlib~\cite{Hunter:CISE9:90},
Pandas~\cite{McKinney::56}, HoloViews, and PyAbel~\cite{Hickstein:RSI90:065115}.

Hit positions were determined by clustering and centroiding of triggered-pixel events. The position
was defined as the center of mass of the individual blobs of triggered pixels. For determination of
the ToF, the temporal information of the individual pixels was time-walk corrected by a pixel-based
correlation curve between ToT and ToA~\cite{Tsigaridas:NIMA930:185, Turecek:JInst12:C12065,
Pitters:JInst14:P05022}. The correlation curve was obtained from electron data for both detectors
using the assumption that all electrons arrive at the same time within the resolution of the
Timepix3 camera. The ToT-weighted expectation value of ToF was calculated from the time-walk
corrected data, taking into account all pixels in a blob. This value was time-walk corrected
making use of a second mean ToT-ToA correlation curve.

The spatio-temporal $(x,y,\text{ToF})$ data was used to determine the initial 3D velocity vector of
all ions in the interaction region. Assuming ideal velocity mapping, each $(x,y)$ position of an ion
on the 2D detector was converted to its initial parallel velocity components $v_x$ and
$v_y$~\cite{Chichinin:IRPC28:607}:
\begin{equation}
   v_x = \frac{x-x_0}{C_s \mathrm{ToF}} \quad \text{and} \quad v_y = \frac{y-y_0}{C_s \mathrm{ToF}}.
   \label{eq:vx_vy}
\end{equation}
The conversion depends on the velocity origin in spatial coordinates $(x_0, y_0)$, which itself
depends on the mass-over-charge ratio of the specific ion and the transverse velocity of the
molecular beam. Furthermore, it depends on the specific ToF as well as a spatial magnification
factor $C_s=1.18$, which was determined from simulations~\cite{Simion:8.1}. The velocity component
$v_z$, perpendicular to the detector surface, was evaluated from the specific temporal deviation
$\Delta t$ from the mean of the individual ToF-peak $t_0$. This is in first order given
by~\cite{Li:RSI76:063106}:
\begin{equation}
   v_z = -e \abs{\vec{E_z}} \Delta t/m \, .
   \label{eq:vz}
\end{equation}
with the electric field strength in the interaction region perpendicular to the detector surface
$E_z$, the Coulomb constant $e$, and the mass $m$ of the specific ion. $E_z=269.6$~V/cm was
determined from simulations. Combining \eqref{eq:vx_vy} and \eqref{eq:vz} yielded all three
components of the initial-ion-velocity vector in the interaction center.

\section{Results and Discussion}
\label{sec:results}

\subsection{Spatial resolution of the detection system}
To classify the scientific results obtained, we first discuss the spatio-temporal resolution of the
detection system after centroiding. To provide an upper limit for the spatial resolution, the
smallest structures observed in our data were taken into account.
\autoref[a]{fig:spatial_resolution} shows a zoom into the bright \Ntpp peak shown in the
corresponding inset of \autoref{fig:N2ToFSpectrum}, which reveals a rhomboid structure. This is
visible throughout the whole detector area, and results from the mesh placed in front of the
electron detector, which has a pitch of 224~\um and a wire thickness of 30~\um. The image shown in
\autoref[a]{fig:spatial_resolution} was rotated by \degree{7} to align one row of gaps with the $y$
axis.
\begin{figure}
   \includegraphics[width=\linewidth]{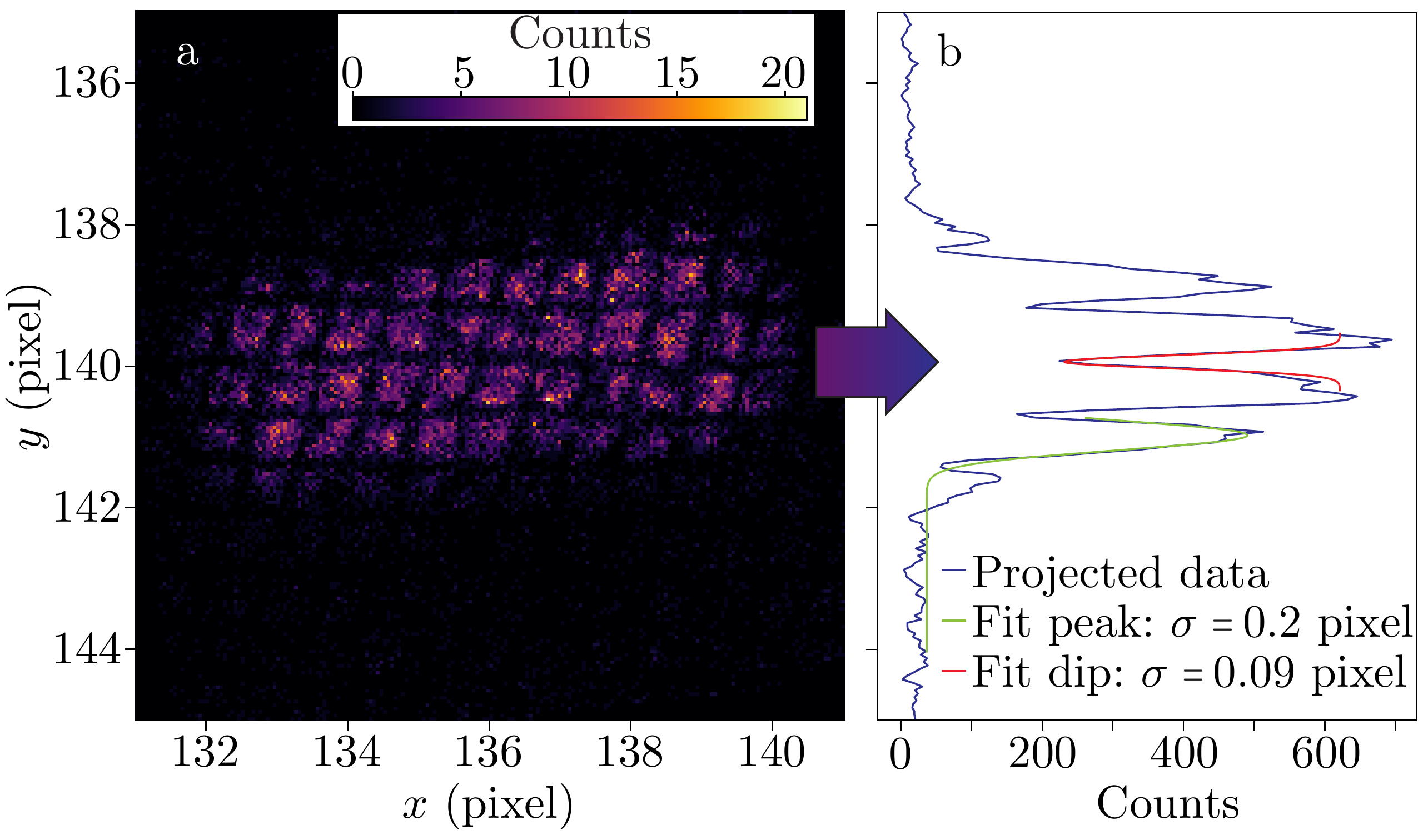}%
   \caption{Illustration of the spatial resolution. 
     (a) Zoom into the \Ntpp peak from \autoref{fig:N2ToFSpectrum}. 
     (b) Projection along the $x$ axis (blue), Gaussian fits of one of the peaks
     (green) and one of the dips (red). Standard deviations for the fits are
     indicated with the specific label.
     }
  \label{fig:spatial_resolution}
\end{figure}
The size of the overall structure along the $x$ axis is attributed to the size of the molecular beam
with a diameter of 6~mm and non-perfect VMI conditions. The finite spread along the $y$ axis is
ascribed to the recoil from the leaving electrons with kinetic energies up to 130~eV.

\autoref[b]{fig:spatial_resolution} shows the projection of the rhomboid structure along the
horizontal axis as a blue line. The minima and maxima are clearly visible. Applying a Gaussian fit
to one of the maxima results in a width corresponding to a standard deviation of $\sigma=0.2$~pixel.
On the other hand, fitting one of the dips between two maxima with an inverse Gaussian yielded a
standard deviation of $\sigma=0.09$~pixel.

This gain in resolution due to centroiding, by about one order of magnitude in a single spatial
dimension, is similar to the ones observed before~\cite{Kella:NIMA329:440, Wester:NIMA413:379,
Chang:RSI69:1665}. In our case, on average 8.8~pixel were illuminated for a single ion impinging
onto the detector. Obtaining such an increase in resolution by calculating the intensity-weighted
center of mass of the blob distribution clearly indicates that the ToT is a good measure of the
energy deposited in a specific single pixel. Overall, an effective spatial resolution of the
detector equivalent to 8~Mpixel without centroiding was achieved, compared to the 65~kpixel sensor array
of the Timepix3 chip.

\subsection{Temporal resolution of the detection system}
Ion and electron data were used to determine the temporal resolution of the detection system. On the
timescales of the camera system, electrons from the origin effectively arrive at the same time, and
are thus well suited to determine the temporal resolution~\cite{Zhao:RSI88:113104}.
\autoref[a]{fig:temporal_resolution} shows a VMI where electrons from the XUV-ionization of \Nt were
detected.
\begin{figure}
   \includegraphics[width=\linewidth]{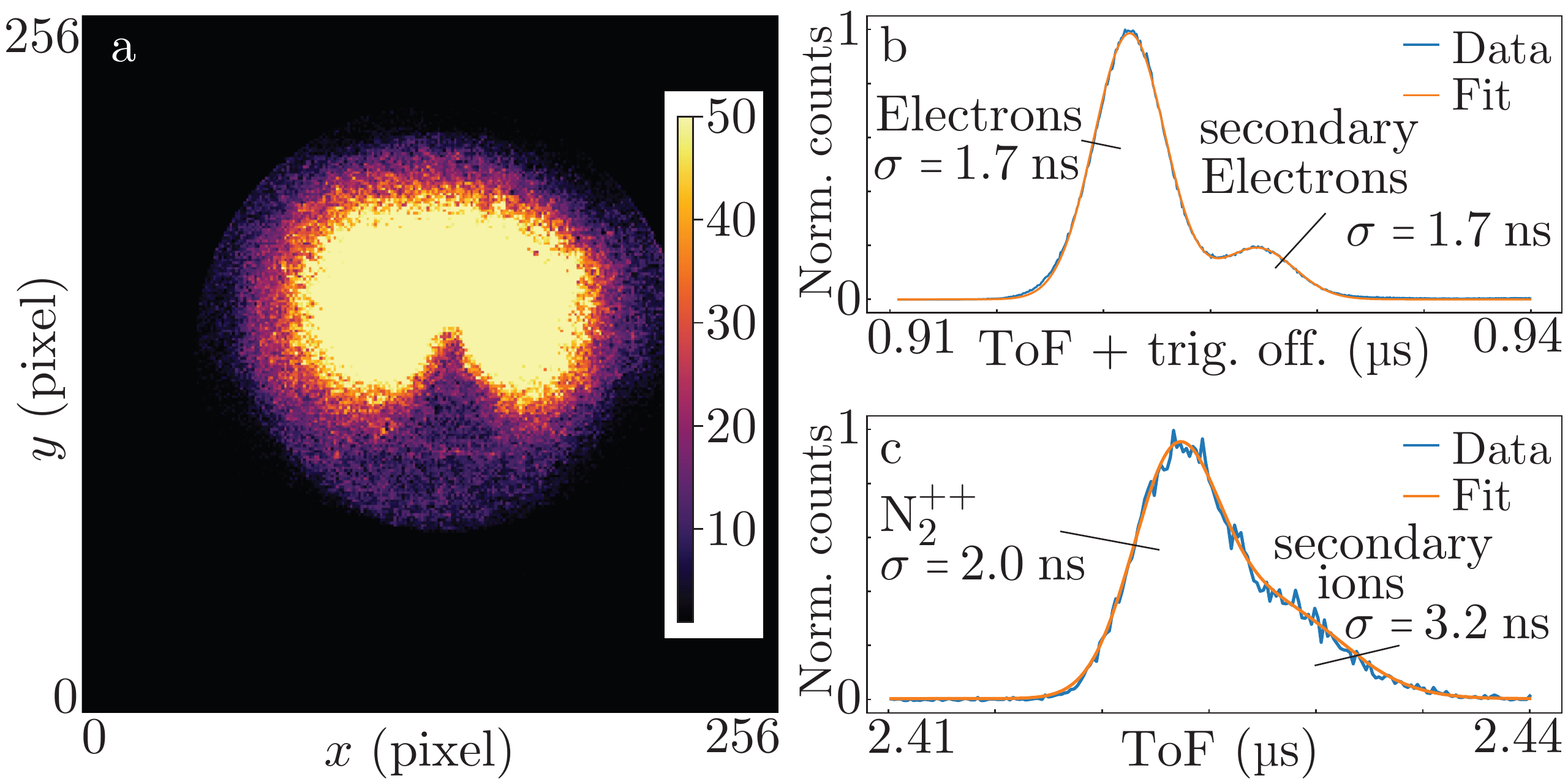}%
   \caption{(a) VMI pixel map containing electrons of \Nt.
            (b) Temporal profile of (a).
            (c) Temporal profile of the \Ntpp ion data shown in \autoref{fig:Nitrogen_coll}.}
   \label{fig:temporal_resolution}
\end{figure}
It shows signal on a considerable part of the detector and, therefore, the temporal resolution
obtained is representative for the whole detector. The corresponding ToF histogram is plotted in
\autoref[b]{fig:temporal_resolution}. All entries were time-walk corrected as discussed above. For
each particle hit in the detector, multiple pixels recorded the time; the weighted average was
calculated, using the ToT values as weights since pixels with larger signals will tend to give
more accurate measurements. Two peaks were observed, corresponding to the signal from the actual
electrons with background (large amplitude) and secondary electrons with only background (small
amplitude), respectively. The best fit of a sum of two Gaussians to this data is achieved with a
common $\sigma=1.7$~ns for both peaks.

In \autoref[c]{fig:temporal_resolution} the temporal profile for ions of the \Ntpp peak is plotted.
For the small region where \Ntpp was detected, we also expect no significant spreading of the
measured ToA due to the initial finite velocities of the ions. Here, the best fit of two Gaussians
yielded a standard deviation of $\sigma=2$~ns for the 1st peak and $\sigma=3.2$~ns for the second
peak. The first and narrower peak is attributed to the actual \Ntpp ions whereas the second peak are
contributions from unspecified residual background which cannot be separated from the ions of the
target but have slightly different ToF due to the non-ideal VMI conditions. The 18~\% larger width
of the ion peak as compared to the electrons, is attributed to detector artifacts: The mean blob
size in this region of the detector was 5.7~pixel whereas the mean blob size for electrons was
7.1~pixel. This is ascribed to the fact that the \Ntpp peak is located in a less sensitive area of
the MCP, and it resulted in a lower number of pixels that can be taken into account for the
time-walk correction. The area of reduced sensitivity is clearly visible in the projected image,
\autoref[a]{fig:Nitrogen_coll}, around and below the central bright \Ntpp peak. On top, the less
intense pixels also suffered from a larger temporal jitter due to time-walk.

Calculating the weighted average of the time-walk corrected ToA did not result in a superior
resolution than the 1.56~ns clock period. Previously, a 720~ps resolution obtained after a
pixel-by-pixel calibration of the time response in the case of direct irradiation with
minimum-ionizing particles was reported in a dedicated Timepix3 setup~\cite{Pitters:JInst14:P05022}.
However, due to the lack of flexibility in terms of detector tuning and differences in the sensors
of our Timepix detector with respect the one previously reported, a pixel-by-pixel calibration did
not result in an improved resolution.

Overall, we attribute the observed limit in temporal resolution to the relatively long phosphor
screen rise time of 6.7~ns~\cite{Winter:RSI85:0034} in combination with the dynamics of the
charge-collection and signal response of the Timepix3 sensor-chip assembly itself. However, the
observed temporal resolution is still better than the 2~ns previously reported for electrons imaging
using Timepix3~\cite{Zhao:RSI88:113104} as well as the 12.5~ns reported for
PImMS~\cite{Amini:RSI86:103113}. 
On top, faster phosphors would improve the achievable temporal resolution. Rise times with a factor of 2
faster than the used P47 have been realized~\cite{Winter:RSI85:0034}. In addition, these scintillating
materials come with a higher brightness and a much faster decay time, which will improve on the ToA
and ToT determination.

\subsection{Comparison of spatial and temporal resolution}
To directly compare the spatial resolution with the temporal resolution, the ratio of \eqref{eq:vz}
and \eqref{eq:vx_vy} can be used: $R=(C\Delta t)/(\Delta r \sqrt{m/q}$) with the temporal and
spatial resolution denoted as $\Delta t$ and $\Delta p$, respectively. The constant
$C=2.33\times10^{-4}$~$\sqrt{\text{Joule}}\frac{\text{pixel}}{\text{meter}}$ depends only on the
geometry of the spectrometer and the specific voltages applied to the electrodes. For \mq{14} the
temporal and spatial resolutions were determined to $\Delta t=2$~ns and $\Delta r=0.09$~pixel,
respectively, which results in $R=34$. Therefore, the spatial resolution is a factor of 34 better
than the temporal resolution for \mq{14}. However, due to its inverse-square-root mass-over-charge
dependency, this ratio improves in favor of the temporal velocity resolution toward larger ToA, \ie,
mass-to-charge ratios. For \mq{16\,000} a similar resolution in temporal and spatial velocity
coordinates would be obtained for the voltage settings used here.

\subsection{Velocity-map imaging of ions from \Nt}
\autoref[a]{fig:Nitrogen_coll} shows the $(v_x,v_y)$ projected velocity map for ions from XUV
ionization with a mass-over-charge ratio $\mq{14}$.
\begin{figure}
   \includegraphics[width=\linewidth]{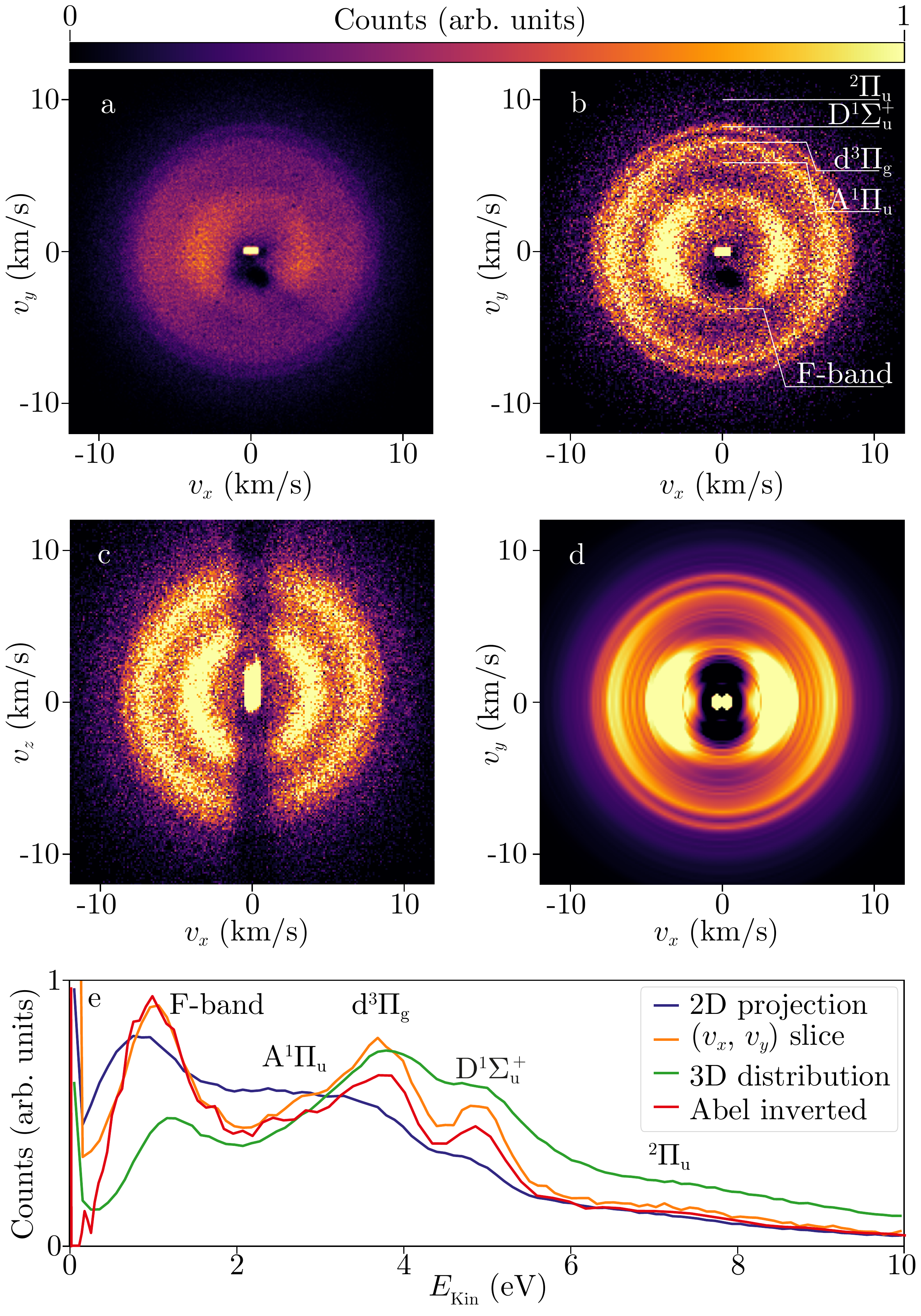}%
   \caption{(a) $v_z$-projected VMI image of \mq{14}.
      (b) $\Delta v_z=2$~km/s wide $(v_x, v_y)$ slice distribution through (a).  The labels specify the respective dissociative
      states in \Ntpp and \Ntppp.
      (c) $\Delta v_y=2$~km/s wide $(v_x, v_z)$ slice distribution through (a).
      (d) Abel-inverted image of (a).
      (e) Kinetic energy distributions for the projected (blue), $(v_x, v_y)$ slice (orange), 3D
      (green), and the Abel-inverted case (red).
      Refer to text for details.
      }
   \label{fig:Nitrogen_coll}
\end{figure}
The projection was carried out between ${v_z}=\pm12$~km/s. The central bright peak is attributed to
\Ntpp, whereas the outer structures arise from the fragmentation of \Ntpp and \Ntppp into \Np +
rest. These features were described in detail before~\cite{Jiang:PRL102:123002, Wu:PCCP13:18398,
   Eckstein:JPCL6:419, Rading:AS8:998} and the corresponing peak velocities and energies are labeled
by the description $\text{F}$ band and the final-state term symbols $\text{A}^1\Pi_u$,
$\text{d}^3\Pi_\text{g}$, $\text{D}^1\Sigma_\text{u}^+$, and $^2\Pi_\text{u}$ in
\autoref[b~and~e]{fig:Nitrogen_coll}.

\autoref[b]{fig:Nitrogen_coll} shows a slice through the velocity distribution of
\autoref[a]{fig:Nitrogen_coll} through the center of the $v_z$ distribution with a full width of
$\Delta v_z=2$~km/s. The central bright peak is again attributed to \Ntpp. The 4 circular structures
arising from \Np are visible in the outer regions peaking at radial velocities given by
$v_r=3.9,~7.1,~8.2,~\text{and}~10$~km/s. The two inner rings show an angular asymmetry, whereas the
outer rings are isotropic.

\autoref[c]{fig:Nitrogen_coll} shows a slice through the velocity distribution of
\autoref[a]{fig:Nitrogen_coll} through the center of the $v_y$ distribution with a width of
$\Delta v_y=2$~km/s. The assignment of the central peak and the rings are as in
\autoref[b]{fig:Nitrogen_coll}. In contrast to \autoref[b]{fig:Nitrogen_coll}, a strong depletion of
signal along $v_x=0$ is observed. This striking difference is ascribed to the less sensitive area of
the detector visible in the center of the projected image toward negative $v_y$ velocities.

Collecting ToA-resolved data allowed to select specific arrival time windows from the complete data
to obtain mass-over-charge-specific velocity maps. This is a significant advantage over traditional
slicing techniques, which generally use gated detectors and thus only capture ions in a specific
arrival time range~\cite{Gebhardt:RSI72:3848}. Whilst this would produce the same images as seen
here, it would come with the caveat that if one wishes to look at another arrival time/slice, a
repetition of the experiment with different gate settings would have to be performed. Moreover, in
that case, correlations between fragments differing in the mass-over-charge ratio are
lost~\cite{Slater:PRA89:011401}.

Furthermore, the recorded full 3D momentum distributions allow for arbitrarily oriented slices or
subsets of the data. For instance, the data in \autoref[c]{fig:Nitrogen_coll} corresponds to a slice
rotated around the $v_x$ axis by \degree{90} relative to the one in \autoref[b]{fig:Nitrogen_coll}.
Both of these are contained in the single recorded dataset as well as the data for any other slice.
Due to the axial symmetry of our experiment with respect to the $v_x$ axis one expects
the same image. We attribute the observed differences in sharpness between
\autoref[b]{fig:Nitrogen_coll} and \autoref[c]{fig:Nitrogen_coll} to the different resolutions
obtained for the spatial and temporal coordinates, as described above. Compared to traditional
gating, which is also very challenging on the sub 10~ns level, we want to stress that
these data for all the different ToAs of the different masses were all recorded simultaneously, in a
time that would otherwise have been needed for a single slice imaging measurement. This demonstrates
the power of recording the 3D datasets for advanced analysis opportunities.

\autoref[d]{fig:Nitrogen_coll} shows the Abel-inverted image of \autoref[a]{fig:Nitrogen_coll}
obtained using the rBasex method~\cite{Hickstein:RSI90:065115}. Again, the inner \Ntpp structure and
the four main circular features are observed.
It shows smooth angular structures in
comparison to the slice images, which are expected for the polar-coordinate-based inversion method
used. However, the statistical noise and systematic errors, such as the less sensitive area, lead
to strong artificial radial features in the Abel-inverted image and thus fine ring-like structures
that are not real. These cannot always be distinguished from the real signal easily, as compared to
the sliced images, and as such this could be seen as a disadvantage of such inversion methods. On
top, Abel-inversion requires for the experiment a cylindrical symmetry of the observed physical
problem.

\autoref[e]{fig:Nitrogen_coll} shows the energy distributions obtained from the projected image
(blue), the $(v_x, v_y)$ slice (orange), the 3D velocity distribution (green), and the Abel-inverted
image (red). The 4 circular structures are clearly visible for all distributions. The increased
resolution obtained by slicing is also observed in the kinetic energy distributions in
\autoref[e]{fig:Nitrogen_coll}. The three central circular structures are visible for the
projected-image distribution, \autoref[a]{fig:Nitrogen_coll}. However, here, the peaks and thus
energies for the specific channels are observed at too low energies due to the projection along
$v_z$.

Furthermore, all ions within the \mq{14} range were used to determine a 3D energy distribution from
the three-dimensional velocity vector that, in principle, would provide the full information with
the best statistics. In practice, the obtained resolution of the 3D distribution is worse compared
to the one obtained by slice imaging, which is again attributed to the temporal resolution being
worse than the spatial one for \mq{14}. Referencing to the Abel-inversion, the peaks of the 3D
distribution are observed at slightly to high energies when compared to slice imaging. We ascribe
this to their overlap in combination with an increasing volume element from the Jacobian as a
function of increasing energy.

The energy distribution obtained from the Abel-inverted image is similar to the one received from
the slice distribution. For the relatively narrow slices used, this is expected for the four \Np
rings. Here, a shift of the peak energy as well as a peak broadening below the percent level with
respect to the peaks obtained from Abel inversion due to the finite slice width is expected.
However, the central part of the image is not well sliced. Especially for radial ion velocities
below 1~km/s the momenta were effectively projected on the $(v_x,v_y)$ plane again. The distribution
from slice imaging overestimates in terms of counts the real distribution in this area.
Therefore, in this part of the image, the energy distribution obtained from Abel inversion is
clearly favorable.

\subsection{Covariance mapping of ions from \Nt}
\begin{figure}
   \includegraphics[width=\linewidth]{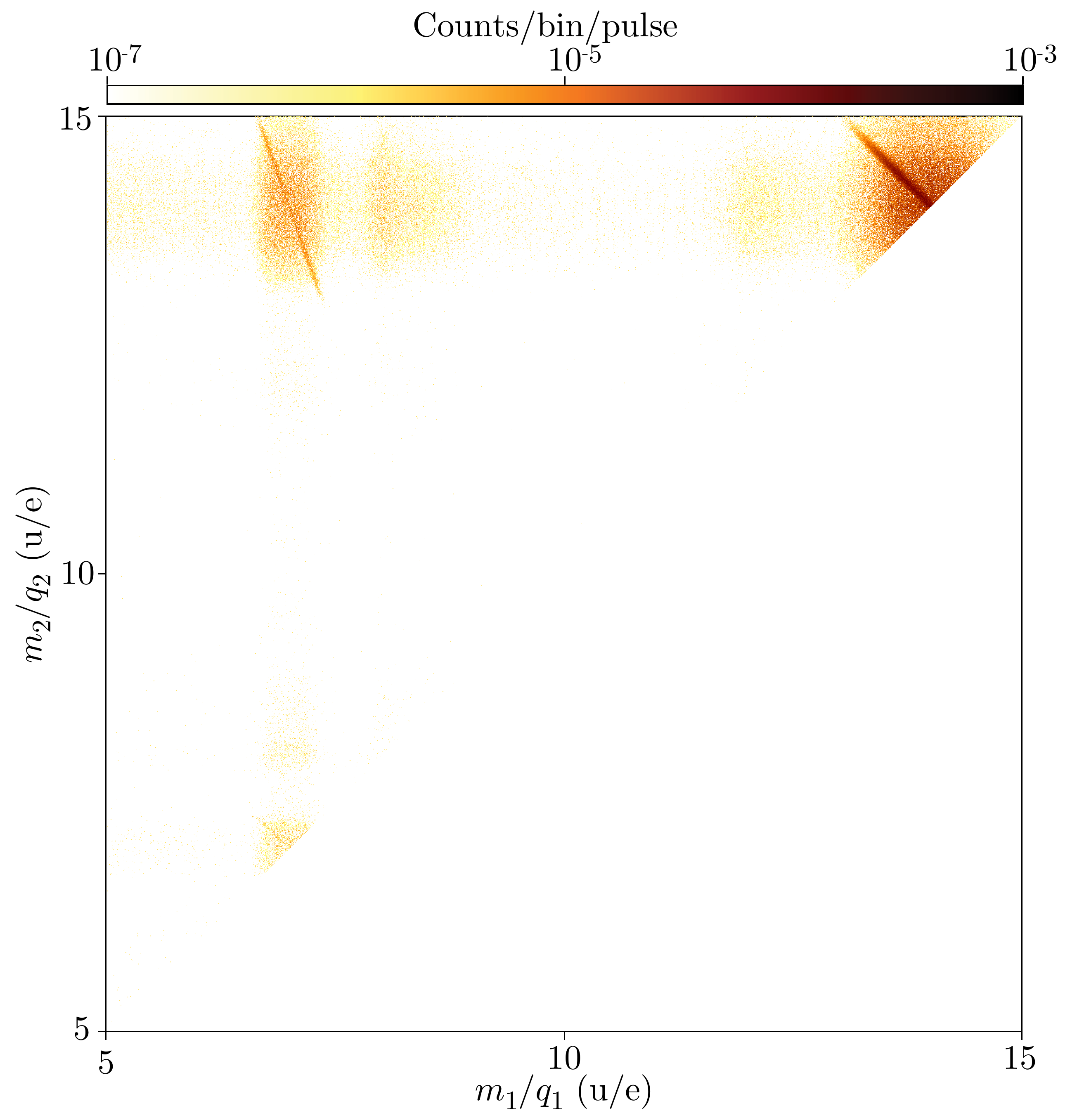}
   \caption{Time-of-flight mass-spectra covariance map. Correlations can be observed at \mqi{14}{1}
   and \mqi{14}{2}, \mqi{7}{1} and \mqi{14}{2} as well as \mqi{7}{1} and \mqi{7}{2}. Refer to text for
   details.
   }
  \label{fig:PIPICO}
\end{figure}

To further investigate the fragmentation dynamics of \Nt after XUV ionization, we employed a
covariance-mapping technique~\cite{Frasinski:Science246:1989, Pickering:JCP144:161105,
Frasinski:JPB49:152004}. \autoref{fig:PIPICO} shows the time-of-flight mass-spectra covariance map
obtained from the temporal component of the Timepix3 data.
Diagonal structures observed between ion pairs in the covariance map are attributed to fragmentation
channels involving at least two charged particles; all other structures arose due to false
coincidences. The line at \mqi{14}{1} and \mqi{14}{2} corresponds to the channel
$\Nt+h\nu\rightarrow\Np+\Np$, whereas the steeper line at \mqi{7}{1} and \mqi{14}{2} is ascribed to
$\Nt+h\nu\rightarrow\Npp+\Np$. The very week line at \mqi{7}{1} and \mqi{7}{2} corresponds to the
channel $\Nt+h\nu\rightarrow\Npp+\Npp$ as documented before~\cite{Dooley:PRA68:023406,
   Kornilov:JPB46:164028, Eckstein:JPCL6:419, Lehmann:PRA94:013426, Wu:PCCP13:18398}. These previous
studies observed further correlation channels, which were not visible in our measurement at lower
FEL intensity, resulting in ionization only to low-charge states $z\le3$ in our case. Nevertheless,
the presence of the three covariance lines shows that our method has the sensitivity to apply these
techniques to the recorded dataset. Further analysis demonstrated good agreement with the
experimentally determined Timpix3 detection efficiency given by 98~\% obtained from the blob-size
distribution.

\subsection{Imaging photoelectrons from He}
The VMI map for electrons from XUV ionized helium is shown in \autoref[a]{fig:Electrons}.
\begin{figure}
   \includegraphics[width=\linewidth]{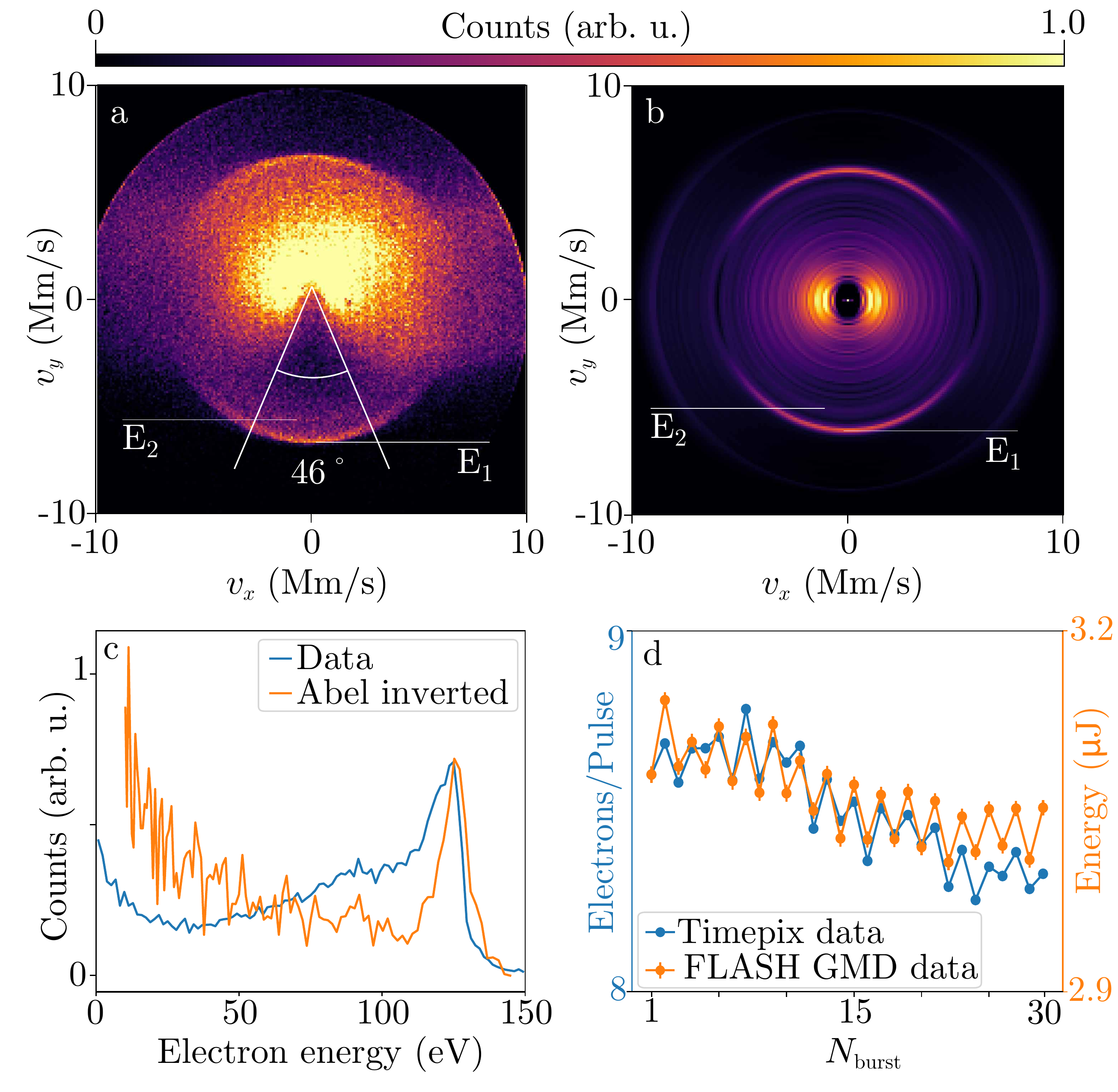}
   \caption{
    (a) Projected electron VMI for ionized He. The outer non-centered ring is the detector edge and
    the bright contribution in the center results from background/stray light electrons. Two lines
    labeled $\text{E}_1$ and $\text{E}_2$ indicate the two photolines.
    (b) Abel-inverted image of (a).
    (c) Kinetic energy distributions from (a) (blue) using a cut of
    \degree{46} as indicated in (a) and from the Abel inversion (orange).
    (d) Mean number of detected electrons per shot for each XUV pulse within a burst
    in blue, and the corresponding averaged FEL intensity per pulse measured with the GMD in
    orange.}
  \label{fig:Electrons}
\end{figure}
Helium was chosen for its simplicity and the clearly detected photoemission line, which for nitrogen
photoelectron imaging was not visible due to the excessive background. The anisotropic He($1s$)
photoemission line labeled $\text{E}_1$ corresponding to the first ionization energy with 24.6~eV of
He at a velocity of 6.6~Mm/s can clearly be observed. Electrons emitted with the energy from the
second ionization energy of 54.4~eV are visible at the line with the label $\text{E}_2$. The inner
bright structure in the image is attributed to undefined anisotropic background signal of undefined
shape, which makes it impossible to subtract it from the actual signal to improve the resolution of
the photoemission line. The outer non-centered ring visible in the VMI is the detector edge. To
determine the correct width of the photoemission line, we also present the Abel-inverted image in
\autoref[b]{fig:Electrons} which clearly shows the expected $\beta\approx2$ angular distribution of
the photoemission line.

\autoref[c]{fig:Electrons} shows the energy distributions for the projected data (blue), cut within
a full opening angle of \degree{46}, as indicated in \autoref[a]{fig:Electrons},
in order to mitigate the effects of the large background contributions in other regions of the
detector. The energy distribution obtained from the Abel inverted data is also shown (orange). Both
curves were normalized to the same peak intensity of the photoemission line. The spectrum was
calibrated using the known photo line energy given by 123.5~eV and a discussion of the accuracy of
the mapping regarding the absolute-velocities determination was thus not possible.

The energy width of the photoemission line in terms of the standard deviation obtained from the
Abel-inverted image is given by $\sigma=4.8$~eV. The width is attributed to three effects: Firstly,
the 1~\% root-mean-square (RMS) bandwidth and 1~\% RMS jitter of the FEL gave rise to an RMS of
about 2~eV at the photon energy of 148.1~eV. Secondly, the limited instrument resolution of the VMI
given by $\text{d}E/E=1.5\text{–}2~\%$. Thirdly, the strong background signals observed for
electrons which broadens the line additionally. Overall, in such typical VMI setups, the spatial
resolution of the detector is not the limiting factor for precise measurements of photoemission
lines.

To further demonstrate the capabilities of the Timepix3 at even higher repetition rates,
photoelectrons from He were recorded for FEL pulse rates of 1~MHz. Due to limitations in the
available radiofrequency window of $\ordsim30$~µs, only 30~pulses were added into the pulse train.
For the short time period of $\Delta{t}=1~\us$ between XUV pulses only electrons were recorded as
the ToF of ions is generally much longer. Disentangling such overlapping ToFs from multiple pulse
times is beyond the scope of the current work.

The results for these measurements are depicted in \autoref[d]{fig:Electrons}, where the average
number of electrons per shot as a function of the specific XUV pulse $N_\text{burst}$ within a pulse
train recorded with the Timepix is plotted in blue. The FEL pulse energy measured with the gas
monitor detector (GMD) of the beamline is shown in orange~\cite{Sorokin:JSR26:1092}. To match both
data sets, the data of the GMD were scaled to the number of electrons measured with the Timepix in
the first burst pulse. An overall decrease of XUV energy
and mean electron number over the course of the pulse train due to the FEL performance is
observed. Starting at around the twentieth pulse in the train, the Timepix electron count starts to
deviate from the GMD energy due to the properties of the detection system. The principal result from
these measurements is that our Timepix3 setup is fully capable of recording data at such high
repetition rates in burst-mode operation.

The 30~pulses within a burst at a burst-to-burst repetition rate of 10~Hz resulted in a data rate of
approximately 20\,000~pixels per second per detector. This is well below the maximum bandwidth of
the camera given by 80~Mpixel/s. However, the full bandwidth of the camera is likely to be reached
at upcoming research facilities like LCLS-II~\cite{SLAC:New-Science-LCLS-II:2016}. In principle
10~ions or electrons with an average blob size of 8~pixels at the full repetition rate of 1~MHz,
corresponding to 10~MHits/s, could be detected. This is still one order of magnitude lower than the
state of the art FPGA DLD technology~\cite{Garzetti:IEEE:1}. The temporal resolution in our
experiment of 1.7~ns is between the 10~ps and 12.5~ns obtained for DLDs and PImMS, respectively. The
Timepix3 spatial resolution of 25~\um is comparable to the ones realized with PImMS and DLDs.
Important advantages of our approach with respect to DLDs are the easier incorporation of the camera
into existing VMI setups, which does not even require to break vacuum, as well as the better
multi-hit capability. Furthermore, for the next generation of Timepix4 with an improved temporal
resolution of $\ordsim200$~ps and a 30~times higher bandwidth, the expected performance gets
considerably closer to that of DLDs~\cite{Ballabriga:RadMeas136:106271, Llopart:JINST:inprep}.

\subsection{Detector dead-time and saturation}
\label{sec:dead-time}
For the last pulse in the trains a deviation in the expected signal intensity of about 2~\% with
respect to the GMD data was observed, see \autoref[d]{fig:Electrons}. This slightly decreased
electron detection efficiency for late pulses in the pulse train was attributed to saturation
effects and will be further investigated in future work.

To minimize saturation, an even distribution of the signal across the detector is advantageous. For
the data presented in \autoref{fig:Nitrogen_coll} and \autoref{fig:Electrons} on average 9~electrons
and 7~ions per XUV pulse were recorded with an average blob size of 7~pixels for electrons and
9~pixels for ions. The dead time for a single pixel is given as a sum of the Timepix3 read-out time
of 475~ns~\cite{Poikela:JInst9:C05013} and the ToT for that pixel. The average ToT for the \Nt data
presented was approximately 500~ns. This results in an averaged pixel dead time of around 1~µs.
However, this does not provide a good estimate for the dead time of the pixels of a whole blob:
Generally, a broad distribution of ToTs is present for the pixels in a single blob, which could lead
to missing pixels in a potential second blob arriving within the dead time of the highest intensity
pixel. Therefore, a much better estimate for the blob dead time results from taking the ToT from
only the pixels with the highest intensity in each individual blob. For our experiment, this was
$\mathrm{ToT}=950$~ns, which results in an average dead time at the blob level of approximately
1.5~\us.

\section{Conclusions}
\label{sec:conclusions}
We demonstrated the implementation of two Timepix3 cameras at a high-repetition-rate x-ray facility
and the ability to record the 3D momentum distribution of all ions from XUV ionization of \Nt and
2D-projected photoelectron distributions from He with the PymePix software. The data was recorded
using a double-sided-VMI setup in the CAMP endstation at the free-electron laser FLASH operating
with pulse-train structures with 250~kHz and 1~MHz pulse repetition rates.

The use of the Timepix3 for acquiring electron images from 30 XUV pulses evenly spaced over 30~\us
showed saturation effects in the order of 2~\% toward the end of the burst, allowing for single ion
and electron detection for all pulses in the train. Through clustering and centroiding of the
signals, we achieved a spatial resolution of $\ordsim0.1$~pixel and thus the equivalent of an
8~Mpixel camera. The arrival times of electrons were determined with a temporal resolution of
$\sigma=1.7$~ns limited by the photoemission dynamics of the phosphor screen and a complex interplay
of camera internals. Here, faster phosphor screens are required for further improvements, which
becomes an even more pressing matter with the developments of Timepix4 on the horizon.

To further exhibit the potential of the Timepix3 detector, a covariance analysis for the ToF-MS was
performed providing fragmentation correlations that are usually lost in VMI spectrometers combined
with regular cameras.

Overall, our results demonstrate the potential of the Timepix hardware platform combined with
appropriate data acquisition and analysis software not only for laboratory experiments but also for
highly demanding experiments at large-scale facilities operating at very high repetition rates such
as FLASH, the European XFEL, or LCLS-II.

\begin{acknowledgments}
   \section*{Acknowledgments}%
   We thank Daniel Ramm for preparing the experimental-setup, Sören Grunewald for helping with IT
   infrastructure, Matthew Robinson for proofreading the manuscript, and Amsterdam Scientific
   Instruments B.~V.\ for collaborative information sharing and discussions.

   \section*{Funding}
   We acknowledge financial support by Deutsches Elektronen-Synchrotron DESY, a member of the
   Helmholtz Association (HGF), also for the provision of experimental facilities and for the use of
   the Maxwell computational resources operated at DESY. Parts of this research were carried out at
   FLASH.

   We acknowledge the Max Planck Society for funding the development and the initial operation of
   the CAMP end-station within the Max Planck Advanced Study Group at CFEL and for providing this
   equipment for CAMP@FLASH. The installation of CAMP@FLASH was partially funded by the BMBF grants
   05K10KT2, 05K13KT2, 05K16KT3, and 05K10KTB from FSP-302.

   The research was further supported by the European Union's Horizon 2020 research and innovation
   program under the Marie Skłodowska-Curie Grant Agreement 641789 ``Molecular Electron Dynamics
   investigated by Intense Fields and Attosecond Pulses'' (MEDEA) and the Cluster of Excellence
   ``Advanced Imaging of Matter'' (AIM, EXC~2056, ID~390715994) of the Deutsche
   Forschungsgemeinschaft (DFG).
\end{acknowledgments}

\bibliography{string,cmi}

\begin{thebibliography}{105}%
\makeatletter
\providecommand \@ifxundefined [1]{%
 \@ifx{#1\undefined}
}%
\providecommand \@ifnum [1]{%
 \ifnum #1\expandafter \@firstoftwo
 \else \expandafter \@secondoftwo
 \fi
}%
\providecommand \@ifx [1]{%
 \ifx #1\expandafter \@firstoftwo
 \else \expandafter \@secondoftwo
 \fi
}%
\providecommand \natexlab [1]{#1}%
\providecommand \enquote  [1]{``#1''}%
\providecommand \bibnamefont  [1]{#1}%
\providecommand \bibfnamefont [1]{#1}%
\providecommand \citenamefont [1]{#1}%
\providecommand \href@noop [0]{\@secondoftwo}%
\providecommand \href [0]{\begingroup \@sanitize@url \@href}%
\providecommand \@href[1]{\@@startlink{#1}\@@href}%
\providecommand \@@href[1]{\endgroup#1\@@endlink}%
\providecommand \@sanitize@url [0]{\catcode `\\12\catcode `\$12\catcode
  `\&12\catcode `\#12\catcode `\^12\catcode `\_12\catcode `\%12\relax}%
\providecommand \@@startlink[1]{}%
\providecommand \@@endlink[0]{}%
\providecommand \url  [0]{\begingroup\@sanitize@url \@url }%
\providecommand \@url [1]{\endgroup\@href {#1}{\urlprefix }}%
\providecommand \urlprefix  [0]{URL }%
\providecommand \Eprint [0]{\href }%
\providecommand \doibase [0]{https://doi.org/}%
\providecommand \selectlanguage [0]{\@gobble}%
\providecommand \bibinfo  [0]{\@secondoftwo}%
\providecommand \bibfield  [0]{\@secondoftwo}%
\providecommand \translation [1]{[#1]}%
\providecommand \BibitemOpen [0]{}%
\providecommand \bibitemStop [0]{}%
\providecommand \bibitemNoStop [0]{.\EOS\space}%
\providecommand \EOS [0]{\spacefactor3000\relax}%
\providecommand \BibitemShut  [1]{\csname bibitem#1\endcsname}%
\let\auto@bib@innerbib\@empty
\bibitem [{\citenamefont {Young}\ \emph {et~al.}(2018)\citenamefont {Young},
  \citenamefont {Ueda}, \citenamefont {Gühr}, \citenamefont {Bucksbaum},
  \citenamefont {Simon}, \citenamefont {Mukamel}, \citenamefont {Rohringer},
  \citenamefont {Prince}, \citenamefont {Masciovecchio}, \citenamefont {Meyer},
  \citenamefont {Rudenko}, \citenamefont {Rolles}, \citenamefont {Bostedt},
  \citenamefont {Fuchs}, \citenamefont {Reis}, \citenamefont {Santra},
  \citenamefont {Kapteyn}, \citenamefont {Murnane}, \citenamefont {Ibrahim},
  \citenamefont {Légaré}, \citenamefont {Vrakking}, \citenamefont {Isinger},
  \citenamefont {Kroon}, \citenamefont {Gisselbrecht}, \citenamefont
  {L’Huillier}, \citenamefont {Wörner},\ and\ \citenamefont
  {Leone}}]{Young:JPB51:032003}%
  \BibitemOpen
  \bibfield  {author} {\bibinfo {author} {\bibfnamefont {L.}~\bibnamefont
  {Young}}, \bibinfo {author} {\bibfnamefont {K.}~\bibnamefont {Ueda}},
  \bibinfo {author} {\bibfnamefont {M.}~\bibnamefont {Gühr}}, \bibinfo
  {author} {\bibfnamefont {P.~H.}\ \bibnamefont {Bucksbaum}}, \bibinfo {author}
  {\bibfnamefont {M.}~\bibnamefont {Simon}}, \bibinfo {author} {\bibfnamefont
  {S.}~\bibnamefont {Mukamel}}, \bibinfo {author} {\bibfnamefont
  {N.}~\bibnamefont {Rohringer}}, \bibinfo {author} {\bibfnamefont {K.~C.}\
  \bibnamefont {Prince}}, \bibinfo {author} {\bibfnamefont {C.}~\bibnamefont
  {Masciovecchio}}, \bibinfo {author} {\bibfnamefont {M.}~\bibnamefont
  {Meyer}}, \bibinfo {author} {\bibfnamefont {A.}~\bibnamefont {Rudenko}},
  \bibinfo {author} {\bibfnamefont {D.}~\bibnamefont {Rolles}}, \bibinfo
  {author} {\bibfnamefont {C.}~\bibnamefont {Bostedt}}, \bibinfo {author}
  {\bibfnamefont {M.}~\bibnamefont {Fuchs}}, \bibinfo {author} {\bibfnamefont
  {D.~A.}\ \bibnamefont {Reis}}, \bibinfo {author} {\bibfnamefont
  {R.}~\bibnamefont {Santra}}, \bibinfo {author} {\bibfnamefont
  {H.}~\bibnamefont {Kapteyn}}, \bibinfo {author} {\bibfnamefont
  {M.}~\bibnamefont {Murnane}}, \bibinfo {author} {\bibfnamefont
  {H.}~\bibnamefont {Ibrahim}}, \bibinfo {author} {\bibfnamefont
  {F.}~\bibnamefont {Légaré}}, \bibinfo {author} {\bibfnamefont {M.~J.~J.}\
  \bibnamefont {Vrakking}}, \bibinfo {author} {\bibfnamefont {M.}~\bibnamefont
  {Isinger}}, \bibinfo {author} {\bibfnamefont {D.}~\bibnamefont {Kroon}},
  \bibinfo {author} {\bibfnamefont {M.}~\bibnamefont {Gisselbrecht}}, \bibinfo
  {author} {\bibfnamefont {A.}~\bibnamefont {L’Huillier}}, \bibinfo {author}
  {\bibfnamefont {H.~J.}\ \bibnamefont {Wörner}},\ and\ \bibinfo {author}
  {\bibfnamefont {S.~R.}\ \bibnamefont {Leone}},\ }\bibfield  {title} {\bibinfo
  {title} {Roadmap of ultrafast x-ray atomic and molecular physics},\ }\href
  {https://doi.org/10.1088/1361-6455/aa9735} {\bibfield  {journal} {\bibinfo
  {journal} {J. Phys. B}\ }\textbf {\bibinfo {volume} {51}},\ \bibinfo {pages}
  {032003} (\bibinfo {year} {2018})}\BibitemShut {NoStop}%
\bibitem [{\citenamefont {Heinz}\ \emph {et~al.}(2017)\citenamefont {Heinz},
  \citenamefont {Shpyrko}, \citenamefont {Basov}, \citenamefont {Berrah},
  \citenamefont {Bucksbaum}, \citenamefont {Devereaux}, \citenamefont {Fritz},
  \citenamefont {Gaffney}, \citenamefont {Gessner}, \citenamefont {Gopalan},
  \citenamefont {Hasan}, \citenamefont {Lanzara}, \citenamefont {Martinez},
  \citenamefont {Millis}, \citenamefont {Mukamel}, \citenamefont {Murnane},
  \citenamefont {Nelson}, \citenamefont {Prasankumar}, \citenamefont {Reis},
  \citenamefont {Schafer}, \citenamefont {Scholes}, \citenamefont {Shen},
  \citenamefont {Stolow}, \citenamefont {Wen}, \citenamefont {Wolf},
  \citenamefont {Xiao}, \citenamefont {Young}, \citenamefont {Garrett},
  \citenamefont {Horton}, \citenamefont {Kerch}, \citenamefont {Krause},
  \citenamefont {Settersten}, \citenamefont {Wilson}, \citenamefont {Runkles},
  \citenamefont {Anderson}, \citenamefont {Chui},\ and\ \citenamefont
  {Rutherford}}]{Heinz:OpportunitiesXFELUlrafast:2017}%
  \BibitemOpen
  \bibfield  {author} {\bibinfo {author} {\bibfnamefont {T.}~\bibnamefont
  {Heinz}}, \bibinfo {author} {\bibfnamefont {O.}~\bibnamefont {Shpyrko}},
  \bibinfo {author} {\bibfnamefont {D.}~\bibnamefont {Basov}}, \bibinfo
  {author} {\bibfnamefont {N.}~\bibnamefont {Berrah}}, \bibinfo {author}
  {\bibfnamefont {P.}~\bibnamefont {Bucksbaum}}, \bibinfo {author}
  {\bibfnamefont {T.}~\bibnamefont {Devereaux}}, \bibinfo {author}
  {\bibfnamefont {D.}~\bibnamefont {Fritz}}, \bibinfo {author} {\bibfnamefont
  {K.}~\bibnamefont {Gaffney}}, \bibinfo {author} {\bibfnamefont
  {O.}~\bibnamefont {Gessner}}, \bibinfo {author} {\bibfnamefont
  {V.}~\bibnamefont {Gopalan}}, \bibinfo {author} {\bibfnamefont
  {Z.}~\bibnamefont {Hasan}}, \bibinfo {author} {\bibfnamefont
  {A.}~\bibnamefont {Lanzara}}, \bibinfo {author} {\bibfnamefont
  {T.}~\bibnamefont {Martinez}}, \bibinfo {author} {\bibfnamefont
  {A.}~\bibnamefont {Millis}}, \bibinfo {author} {\bibfnamefont
  {S.}~\bibnamefont {Mukamel}}, \bibinfo {author} {\bibfnamefont
  {M.}~\bibnamefont {Murnane}}, \bibinfo {author} {\bibfnamefont
  {K.}~\bibnamefont {Nelson}}, \bibinfo {author} {\bibfnamefont
  {R.}~\bibnamefont {Prasankumar}}, \bibinfo {author} {\bibfnamefont
  {D.}~\bibnamefont {Reis}}, \bibinfo {author} {\bibfnamefont {K.}~\bibnamefont
  {Schafer}}, \bibinfo {author} {\bibfnamefont {G.}~\bibnamefont {Scholes}},
  \bibinfo {author} {\bibfnamefont {Z.-X.}\ \bibnamefont {Shen}}, \bibinfo
  {author} {\bibfnamefont {A.}~\bibnamefont {Stolow}}, \bibinfo {author}
  {\bibfnamefont {H.}~\bibnamefont {Wen}}, \bibinfo {author} {\bibfnamefont
  {M.}~\bibnamefont {Wolf}}, \bibinfo {author} {\bibfnamefont {D.}~\bibnamefont
  {Xiao}}, \bibinfo {author} {\bibfnamefont {L.}~\bibnamefont {Young}},
  \bibinfo {author} {\bibfnamefont {B.}~\bibnamefont {Garrett}}, \bibinfo
  {author} {\bibfnamefont {L.}~\bibnamefont {Horton}}, \bibinfo {author}
  {\bibfnamefont {H.}~\bibnamefont {Kerch}}, \bibinfo {author} {\bibfnamefont
  {J.}~\bibnamefont {Krause}}, \bibinfo {author} {\bibfnamefont
  {T.}~\bibnamefont {Settersten}}, \bibinfo {author} {\bibfnamefont
  {L.}~\bibnamefont {Wilson}}, \bibinfo {author} {\bibfnamefont
  {K.}~\bibnamefont {Runkles}}, \bibinfo {author} {\bibfnamefont
  {T.}~\bibnamefont {Anderson}}, \bibinfo {author} {\bibfnamefont
  {G.}~\bibnamefont {Chui}},\ and\ \bibinfo {author} {\bibfnamefont
  {E.}~\bibnamefont {Rutherford}},\ }\href {https://doi.org/10.2172/1616251}
  {\emph {\bibinfo {title} {Basic Energy Sciences Roundtable: Opportunities for
  Basic Research at the Frontiers of XFEL Ultrafast Science}}},\ \bibinfo
  {type} {Tech. Rep.}\ (\bibinfo {year} {2017})\BibitemShut {NoStop}%
\bibitem [{\citenamefont {Ackermann}\ \emph {et~al.}(2007)\citenamefont
  {Ackermann}, \citenamefont {Asova}, \citenamefont {Ayvazyan}, \citenamefont
  {Azima}, \citenamefont {Baboi}, \citenamefont {B{\"a}hr}, \citenamefont
  {Balandin}, \citenamefont {Beutner}, \citenamefont {Brandt}, \citenamefont
  {Bolzmann}, \citenamefont {Brinkmann}, \citenamefont {Brovko}, \citenamefont
  {Castellano}, \citenamefont {Castro}, \citenamefont {Catani}, \citenamefont
  {Chiadroni}, \citenamefont {Choroba}, \citenamefont {Cianchi}, \citenamefont
  {Costello}, \citenamefont {Cubaynes}, \citenamefont {Dardis}, \citenamefont
  {Decking}, \citenamefont {Delsim-Hashemi}, \citenamefont {Delserieys},
  \citenamefont {Di~Pirro}, \citenamefont {Dohlus}, \citenamefont
  {D{\"u}sterer}, \citenamefont {Eckhardt}, \citenamefont {Edwards},
  \citenamefont {Faatz}, \citenamefont {Feldhaus}, \citenamefont
  {Fl{\"o}ttmann}, \citenamefont {Frisch}, \citenamefont {Fr{\"o}hlich},
  \citenamefont {Garvey}, \citenamefont {Gensch}, \citenamefont {Gerth},
  \citenamefont {G{\"o}rler}, \citenamefont {Golubeva}, \citenamefont
  {Grabosch}, \citenamefont {Grecki}, \citenamefont {Grimm}, \citenamefont
  {Hacker}, \citenamefont {Hahn}, \citenamefont {Han}, \citenamefont
  {Honkavaara}, \citenamefont {Hott}, \citenamefont {H{\"u}ning}, \citenamefont
  {Ivanisenko}, \citenamefont {Jaeschke}, \citenamefont {Jalmuzna},
  \citenamefont {Jezynski}, \citenamefont {Kammering}, \citenamefont {Katalev},
  \citenamefont {Kavanagh}, \citenamefont {Kennedy}, \citenamefont
  {Khodyachykh}, \citenamefont {Klose}, \citenamefont {Kocharyan},
  \citenamefont {K{\"o}rfer}, \citenamefont {Kollewe}, \citenamefont {Koprek},
  \citenamefont {Korepanov}, \citenamefont {Kostin}, \citenamefont
  {Krassilnikov}, \citenamefont {Kube}, \citenamefont {Kuhlmann}, \citenamefont
  {Lewis}, \citenamefont {Lilje}, \citenamefont {Limberg}, \citenamefont
  {Lipka}, \citenamefont {L{\"o}hl}, \citenamefont {Luna}, \citenamefont
  {Luong}, \citenamefont {Martins}, \citenamefont {Meyer}, \citenamefont
  {Michelato}, \citenamefont {Miltchev}, \citenamefont {M{\"o}ller},
  \citenamefont {Monaco}, \citenamefont {M{\"u}ller}, \citenamefont
  {Napieralski}, \citenamefont {Napoly}, \citenamefont {Nicolosi},
  \citenamefont {N{\"o}lle}, \citenamefont {Nu{\~n}ez}, \citenamefont {Oppelt},
  \citenamefont {Pagani}, \citenamefont {Paparella}, \citenamefont {Pchalek},
  \citenamefont {Pedregosa-Gutierrez}, \citenamefont {Petersen}, \citenamefont
  {Petrosyan}, \citenamefont {Petrosyan}, \citenamefont {Petrosyan},
  \citenamefont {Pfl{\"u}ger}, \citenamefont {Pl{\"o}njes}, \citenamefont
  {Poletto}, \citenamefont {Pozniak}, \citenamefont {Prat}, \citenamefont
  {Proch}, \citenamefont {Pucyk}, \citenamefont {Radcliffe}, \citenamefont
  {Redlin}, \citenamefont {Rehlich}, \citenamefont {Richter}, \citenamefont
  {Roehrs}, \citenamefont {Roensch}, \citenamefont {Romaniuk}, \citenamefont
  {Ross}, \citenamefont {Rossbach}, \citenamefont {Rybnikov}, \citenamefont
  {Sachwitz}, \citenamefont {Saldin}, \citenamefont {Sandner}, \citenamefont
  {Schlarb}, \citenamefont {Schmidt}, \citenamefont {Schmitz}, \citenamefont
  {Schm{\"u}ser}, \citenamefont {Schneider}, \citenamefont {Schneidmiller},
  \citenamefont {Schnepp}, \citenamefont {Schreiber}, \citenamefont {Seidel},
  \citenamefont {Sertore}, \citenamefont {Shabunov}, \citenamefont {Simon},
  \citenamefont {Simrock}, \citenamefont {Sombrowski}, \citenamefont {Sorokin},
  \citenamefont {Spanknebel}, \citenamefont {Spesyvtsev}, \citenamefont
  {Staykov}, \citenamefont {Steffen}, \citenamefont {Stephan}, \citenamefont
  {Stulle}, \citenamefont {Thom}, \citenamefont {Tiedtke}, \citenamefont
  {Tischer}, \citenamefont {Toleikis}, \citenamefont {Treusch}, \citenamefont
  {Trines}, \citenamefont {Tsakov}, \citenamefont {Vogel}, \citenamefont
  {Weiland}, \citenamefont {Weise}, \citenamefont {Wellh{\"o}fer},
  \citenamefont {Wendt}, \citenamefont {Will}, \citenamefont {Winter},
  \citenamefont {Wittenburg}, \citenamefont {Wurth}, \citenamefont {Yeates},
  \citenamefont {Yurkov}, \citenamefont {Zagorodnov},\ and\ \citenamefont
  {Zapfe}}]{Ackermann:NatPhoton1:336}%
  \BibitemOpen
  \bibfield  {author} {\bibinfo {author} {\bibfnamefont {W.}~\bibnamefont
  {Ackermann}}, \bibinfo {author} {\bibfnamefont {G.}~\bibnamefont {Asova}},
  \bibinfo {author} {\bibfnamefont {V.}~\bibnamefont {Ayvazyan}}, \bibinfo
  {author} {\bibfnamefont {A.}~\bibnamefont {Azima}}, \bibinfo {author}
  {\bibfnamefont {N.}~\bibnamefont {Baboi}}, \bibinfo {author} {\bibfnamefont
  {J.}~\bibnamefont {B{\"a}hr}}, \bibinfo {author} {\bibfnamefont
  {V.}~\bibnamefont {Balandin}}, \bibinfo {author} {\bibfnamefont
  {B.}~\bibnamefont {Beutner}}, \bibinfo {author} {\bibfnamefont
  {A.}~\bibnamefont {Brandt}}, \bibinfo {author} {\bibfnamefont
  {A.}~\bibnamefont {Bolzmann}}, \bibinfo {author} {\bibfnamefont
  {R.}~\bibnamefont {Brinkmann}}, \bibinfo {author} {\bibfnamefont {O.~I.}\
  \bibnamefont {Brovko}}, \bibinfo {author} {\bibfnamefont {M.}~\bibnamefont
  {Castellano}}, \bibinfo {author} {\bibfnamefont {P.}~\bibnamefont {Castro}},
  \bibinfo {author} {\bibfnamefont {L.}~\bibnamefont {Catani}}, \bibinfo
  {author} {\bibfnamefont {E.}~\bibnamefont {Chiadroni}}, \bibinfo {author}
  {\bibfnamefont {S.}~\bibnamefont {Choroba}}, \bibinfo {author} {\bibfnamefont
  {A.}~\bibnamefont {Cianchi}}, \bibinfo {author} {\bibfnamefont {J.~T.}\
  \bibnamefont {Costello}}, \bibinfo {author} {\bibfnamefont {D.}~\bibnamefont
  {Cubaynes}}, \bibinfo {author} {\bibfnamefont {J.}~\bibnamefont {Dardis}},
  \bibinfo {author} {\bibfnamefont {W.}~\bibnamefont {Decking}}, \bibinfo
  {author} {\bibfnamefont {H.}~\bibnamefont {Delsim-Hashemi}}, \bibinfo
  {author} {\bibfnamefont {A.}~\bibnamefont {Delserieys}}, \bibinfo {author}
  {\bibfnamefont {G.}~\bibnamefont {Di~Pirro}}, \bibinfo {author}
  {\bibfnamefont {M.}~\bibnamefont {Dohlus}}, \bibinfo {author} {\bibfnamefont
  {S.}~\bibnamefont {D{\"u}sterer}}, \bibinfo {author} {\bibfnamefont
  {A.}~\bibnamefont {Eckhardt}}, \bibinfo {author} {\bibfnamefont {H.~T.}\
  \bibnamefont {Edwards}}, \bibinfo {author} {\bibfnamefont {B.}~\bibnamefont
  {Faatz}}, \bibinfo {author} {\bibfnamefont {J.}~\bibnamefont {Feldhaus}},
  \bibinfo {author} {\bibfnamefont {K.}~\bibnamefont {Fl{\"o}ttmann}}, \bibinfo
  {author} {\bibfnamefont {J.}~\bibnamefont {Frisch}}, \bibinfo {author}
  {\bibfnamefont {L.}~\bibnamefont {Fr{\"o}hlich}}, \bibinfo {author}
  {\bibfnamefont {T.}~\bibnamefont {Garvey}}, \bibinfo {author} {\bibfnamefont
  {U.}~\bibnamefont {Gensch}}, \bibinfo {author} {\bibfnamefont
  {C.}~\bibnamefont {Gerth}}, \bibinfo {author} {\bibfnamefont
  {M.}~\bibnamefont {G{\"o}rler}}, \bibinfo {author} {\bibfnamefont
  {N.}~\bibnamefont {Golubeva}}, \bibinfo {author} {\bibfnamefont {H.~J.}\
  \bibnamefont {Grabosch}}, \bibinfo {author} {\bibfnamefont {M.}~\bibnamefont
  {Grecki}}, \bibinfo {author} {\bibfnamefont {O.}~\bibnamefont {Grimm}},
  \bibinfo {author} {\bibfnamefont {K.}~\bibnamefont {Hacker}}, \bibinfo
  {author} {\bibfnamefont {U.}~\bibnamefont {Hahn}}, \bibinfo {author}
  {\bibfnamefont {J.~H.}\ \bibnamefont {Han}}, \bibinfo {author} {\bibfnamefont
  {K.}~\bibnamefont {Honkavaara}}, \bibinfo {author} {\bibfnamefont
  {T.}~\bibnamefont {Hott}}, \bibinfo {author} {\bibfnamefont {M.}~\bibnamefont
  {H{\"u}ning}}, \bibinfo {author} {\bibfnamefont {Y.}~\bibnamefont
  {Ivanisenko}}, \bibinfo {author} {\bibfnamefont {E.}~\bibnamefont
  {Jaeschke}}, \bibinfo {author} {\bibfnamefont {W.}~\bibnamefont {Jalmuzna}},
  \bibinfo {author} {\bibfnamefont {T.}~\bibnamefont {Jezynski}}, \bibinfo
  {author} {\bibfnamefont {R.}~\bibnamefont {Kammering}}, \bibinfo {author}
  {\bibfnamefont {V.}~\bibnamefont {Katalev}}, \bibinfo {author} {\bibfnamefont
  {K.}~\bibnamefont {Kavanagh}}, \bibinfo {author} {\bibfnamefont {E.~T.}\
  \bibnamefont {Kennedy}}, \bibinfo {author} {\bibfnamefont {S.}~\bibnamefont
  {Khodyachykh}}, \bibinfo {author} {\bibfnamefont {K.}~\bibnamefont {Klose}},
  \bibinfo {author} {\bibfnamefont {V.}~\bibnamefont {Kocharyan}}, \bibinfo
  {author} {\bibfnamefont {M.}~\bibnamefont {K{\"o}rfer}}, \bibinfo {author}
  {\bibfnamefont {M.}~\bibnamefont {Kollewe}}, \bibinfo {author} {\bibfnamefont
  {W.}~\bibnamefont {Koprek}}, \bibinfo {author} {\bibfnamefont
  {S.}~\bibnamefont {Korepanov}}, \bibinfo {author} {\bibfnamefont
  {D.}~\bibnamefont {Kostin}}, \bibinfo {author} {\bibfnamefont
  {M.}~\bibnamefont {Krassilnikov}}, \bibinfo {author} {\bibfnamefont
  {G.}~\bibnamefont {Kube}}, \bibinfo {author} {\bibfnamefont {M.}~\bibnamefont
  {Kuhlmann}}, \bibinfo {author} {\bibfnamefont {C.~L.~S.}\ \bibnamefont
  {Lewis}}, \bibinfo {author} {\bibfnamefont {L.}~\bibnamefont {Lilje}},
  \bibinfo {author} {\bibfnamefont {T.}~\bibnamefont {Limberg}}, \bibinfo
  {author} {\bibfnamefont {D.}~\bibnamefont {Lipka}}, \bibinfo {author}
  {\bibfnamefont {F.}~\bibnamefont {L{\"o}hl}}, \bibinfo {author}
  {\bibfnamefont {H.}~\bibnamefont {Luna}}, \bibinfo {author} {\bibfnamefont
  {M.}~\bibnamefont {Luong}}, \bibinfo {author} {\bibfnamefont
  {M.}~\bibnamefont {Martins}}, \bibinfo {author} {\bibfnamefont
  {M.}~\bibnamefont {Meyer}}, \bibinfo {author} {\bibfnamefont
  {P.}~\bibnamefont {Michelato}}, \bibinfo {author} {\bibfnamefont
  {V.}~\bibnamefont {Miltchev}}, \bibinfo {author} {\bibfnamefont {W.~D.}\
  \bibnamefont {M{\"o}ller}}, \bibinfo {author} {\bibfnamefont
  {L.}~\bibnamefont {Monaco}}, \bibinfo {author} {\bibfnamefont {W.~F.~O.}\
  \bibnamefont {M{\"u}ller}}, \bibinfo {author} {\bibfnamefont
  {O.}~\bibnamefont {Napieralski}}, \bibinfo {author} {\bibfnamefont
  {O.}~\bibnamefont {Napoly}}, \bibinfo {author} {\bibfnamefont
  {P.}~\bibnamefont {Nicolosi}}, \bibinfo {author} {\bibfnamefont
  {D.}~\bibnamefont {N{\"o}lle}}, \bibinfo {author} {\bibfnamefont
  {T.}~\bibnamefont {Nu{\~n}ez}}, \bibinfo {author} {\bibfnamefont
  {A.}~\bibnamefont {Oppelt}}, \bibinfo {author} {\bibfnamefont
  {C.}~\bibnamefont {Pagani}}, \bibinfo {author} {\bibfnamefont
  {R.}~\bibnamefont {Paparella}}, \bibinfo {author} {\bibfnamefont
  {N.}~\bibnamefont {Pchalek}}, \bibinfo {author} {\bibfnamefont
  {J.}~\bibnamefont {Pedregosa-Gutierrez}}, \bibinfo {author} {\bibfnamefont
  {B.}~\bibnamefont {Petersen}}, \bibinfo {author} {\bibfnamefont
  {B.}~\bibnamefont {Petrosyan}}, \bibinfo {author} {\bibfnamefont
  {G.}~\bibnamefont {Petrosyan}}, \bibinfo {author} {\bibfnamefont
  {L.}~\bibnamefont {Petrosyan}}, \bibinfo {author} {\bibfnamefont
  {J.}~\bibnamefont {Pfl{\"u}ger}}, \bibinfo {author} {\bibfnamefont
  {E.}~\bibnamefont {Pl{\"o}njes}}, \bibinfo {author} {\bibfnamefont
  {L.}~\bibnamefont {Poletto}}, \bibinfo {author} {\bibfnamefont
  {K.}~\bibnamefont {Pozniak}}, \bibinfo {author} {\bibfnamefont
  {E.}~\bibnamefont {Prat}}, \bibinfo {author} {\bibfnamefont {D.}~\bibnamefont
  {Proch}}, \bibinfo {author} {\bibfnamefont {P.}~\bibnamefont {Pucyk}},
  \bibinfo {author} {\bibfnamefont {P.}~\bibnamefont {Radcliffe}}, \bibinfo
  {author} {\bibfnamefont {H.}~\bibnamefont {Redlin}}, \bibinfo {author}
  {\bibfnamefont {K.}~\bibnamefont {Rehlich}}, \bibinfo {author} {\bibfnamefont
  {M.}~\bibnamefont {Richter}}, \bibinfo {author} {\bibfnamefont
  {M.}~\bibnamefont {Roehrs}}, \bibinfo {author} {\bibfnamefont
  {J.}~\bibnamefont {Roensch}}, \bibinfo {author} {\bibfnamefont
  {R.}~\bibnamefont {Romaniuk}}, \bibinfo {author} {\bibfnamefont
  {M.}~\bibnamefont {Ross}}, \bibinfo {author} {\bibfnamefont {J.}~\bibnamefont
  {Rossbach}}, \bibinfo {author} {\bibfnamefont {V.}~\bibnamefont {Rybnikov}},
  \bibinfo {author} {\bibfnamefont {M.}~\bibnamefont {Sachwitz}}, \bibinfo
  {author} {\bibfnamefont {E.~L.}\ \bibnamefont {Saldin}}, \bibinfo {author}
  {\bibfnamefont {W.}~\bibnamefont {Sandner}}, \bibinfo {author} {\bibfnamefont
  {H.}~\bibnamefont {Schlarb}}, \bibinfo {author} {\bibfnamefont
  {B.}~\bibnamefont {Schmidt}}, \bibinfo {author} {\bibfnamefont
  {M.}~\bibnamefont {Schmitz}}, \bibinfo {author} {\bibfnamefont
  {P.}~\bibnamefont {Schm{\"u}ser}}, \bibinfo {author} {\bibfnamefont {J.~R.}\
  \bibnamefont {Schneider}}, \bibinfo {author} {\bibfnamefont {E.~A.}\
  \bibnamefont {Schneidmiller}}, \bibinfo {author} {\bibfnamefont
  {S.}~\bibnamefont {Schnepp}}, \bibinfo {author} {\bibfnamefont
  {S.}~\bibnamefont {Schreiber}}, \bibinfo {author} {\bibfnamefont
  {M.}~\bibnamefont {Seidel}}, \bibinfo {author} {\bibfnamefont
  {D.}~\bibnamefont {Sertore}}, \bibinfo {author} {\bibfnamefont {A.~V.}\
  \bibnamefont {Shabunov}}, \bibinfo {author} {\bibfnamefont {C.}~\bibnamefont
  {Simon}}, \bibinfo {author} {\bibfnamefont {S.}~\bibnamefont {Simrock}},
  \bibinfo {author} {\bibfnamefont {E.}~\bibnamefont {Sombrowski}}, \bibinfo
  {author} {\bibfnamefont {A.~A.}\ \bibnamefont {Sorokin}}, \bibinfo {author}
  {\bibfnamefont {P.}~\bibnamefont {Spanknebel}}, \bibinfo {author}
  {\bibfnamefont {R.}~\bibnamefont {Spesyvtsev}}, \bibinfo {author}
  {\bibfnamefont {L.}~\bibnamefont {Staykov}}, \bibinfo {author} {\bibfnamefont
  {B.}~\bibnamefont {Steffen}}, \bibinfo {author} {\bibfnamefont
  {F.}~\bibnamefont {Stephan}}, \bibinfo {author} {\bibfnamefont
  {F.}~\bibnamefont {Stulle}}, \bibinfo {author} {\bibfnamefont
  {H.}~\bibnamefont {Thom}}, \bibinfo {author} {\bibfnamefont {K.}~\bibnamefont
  {Tiedtke}}, \bibinfo {author} {\bibfnamefont {M.}~\bibnamefont {Tischer}},
  \bibinfo {author} {\bibfnamefont {S.}~\bibnamefont {Toleikis}}, \bibinfo
  {author} {\bibfnamefont {R.}~\bibnamefont {Treusch}}, \bibinfo {author}
  {\bibfnamefont {D.}~\bibnamefont {Trines}}, \bibinfo {author} {\bibfnamefont
  {I.}~\bibnamefont {Tsakov}}, \bibinfo {author} {\bibfnamefont
  {E.}~\bibnamefont {Vogel}}, \bibinfo {author} {\bibfnamefont
  {T.}~\bibnamefont {Weiland}}, \bibinfo {author} {\bibfnamefont
  {H.}~\bibnamefont {Weise}}, \bibinfo {author} {\bibfnamefont
  {M.}~\bibnamefont {Wellh{\"o}fer}}, \bibinfo {author} {\bibfnamefont
  {M.}~\bibnamefont {Wendt}}, \bibinfo {author} {\bibfnamefont
  {I.}~\bibnamefont {Will}}, \bibinfo {author} {\bibfnamefont {A.}~\bibnamefont
  {Winter}}, \bibinfo {author} {\bibfnamefont {K.}~\bibnamefont {Wittenburg}},
  \bibinfo {author} {\bibfnamefont {W.}~\bibnamefont {Wurth}}, \bibinfo
  {author} {\bibfnamefont {P.}~\bibnamefont {Yeates}}, \bibinfo {author}
  {\bibfnamefont {M.~V.}\ \bibnamefont {Yurkov}}, \bibinfo {author}
  {\bibfnamefont {I.}~\bibnamefont {Zagorodnov}},\ and\ \bibinfo {author}
  {\bibfnamefont {K.}~\bibnamefont {Zapfe}},\ }\bibfield  {title} {\bibinfo
  {title} {Operation of a free-electron laser from the extreme ultraviolet to
  the water window},\ }\href {https://doi.org/10.1038/nphoton.2007.76}
  {\bibfield  {journal} {\bibinfo  {journal} {Nat. Photon.}\ }\textbf {\bibinfo
  {volume} {1}},\ \bibinfo {pages} {336} (\bibinfo {year} {2007})}\BibitemShut
  {NoStop}%
\bibitem [{\citenamefont {Abela}\ \emph {et~al.}(2006)\citenamefont {Abela},
  \citenamefont {Witte}, \citenamefont {Schwarz}, \citenamefont {Redlin},
  \citenamefont {Eckoldt}, \citenamefont {Hartrott}, \citenamefont
  {Floettmann}, \citenamefont {Fajardo}, \citenamefont {Katalev}, \citenamefont
  {Schilcher}, \citenamefont {Tiedtke}, \citenamefont {Baehr}, \citenamefont
  {Singer}, \citenamefont {Sombrowski}, \citenamefont {Casalbuoni},
  \citenamefont {Hajdu}, \citenamefont {Schneider}, \citenamefont {Dipirro},
  \citenamefont {Kim}, \citenamefont {Brinkmann}, \citenamefont {Racky},
  \citenamefont {Chen}, \citenamefont {Nienhaus}, \citenamefont {Chapman},
  \citenamefont {Lierl}, \citenamefont {Gadwinkel}, \citenamefont {Gerth},
  \citenamefont {Robinson}, \citenamefont {Choroba}, \citenamefont {Maslov},
  \citenamefont {Wenndorff}, \citenamefont {Pagani}, \citenamefont {Tallents},
  \citenamefont {Grabosch}, \citenamefont {Jaeschke}, \citenamefont
  {Paparella}, \citenamefont {Seller}, \citenamefont {Piot}, \citenamefont
  {Tsakov}, \citenamefont {Kraemer}, \citenamefont {Winter}, \citenamefont
  {Sachwitz}, \citenamefont {Cianchi}, \citenamefont {Weddig}, \citenamefont
  {Sandner}, \citenamefont {Schroer}, \citenamefont {Nunez}, \citenamefont
  {Castro}, \citenamefont {Napoly}, \citenamefont {Kong}, \citenamefont
  {Simrock}, \citenamefont {Gensch}, \citenamefont {Thom}, \citenamefont
  {Bratos}, \citenamefont {Altucci}, \citenamefont {Dobson}, \citenamefont
  {Masciovecchio}, \citenamefont {Quack}, \citenamefont {Kienberger},
  \citenamefont {Wendt}, \citenamefont {Chergui}, \citenamefont {Grimm},
  \citenamefont {Plech}, \citenamefont {Obier}, \citenamefont {Schmidt},
  \citenamefont {Sytchev}, \citenamefont {Rossbach}, \citenamefont {Kozlov},
  \citenamefont {Kuhlmann}, \citenamefont {Senf}, \citenamefont {Wolf},
  \citenamefont {Vogel}, \citenamefont {Prenting}, \citenamefont {Stephan},
  \citenamefont {Syresin}, \citenamefont {Sertore}, \citenamefont {Decking},
  \citenamefont {Baboi}, \citenamefont {Vogel}, \citenamefont {Juha},
  \citenamefont {Brandt}, \citenamefont {Collet}, \citenamefont {Filipov},
  \citenamefont {Vartaniants}, \citenamefont {Marangos}, \citenamefont
  {Golubeva}, \citenamefont {Smith}, \citenamefont {Techert}, \citenamefont
  {Fateev}, \citenamefont {Wulff}, \citenamefont {Haenisch}, \citenamefont
  {Huening}, \citenamefont {Honkim{\"a}ki}, \citenamefont {Treusch},
  \citenamefont {Ihee}, \citenamefont {Prat}, \citenamefont {Petrosyan},
  \citenamefont {Prat}, \citenamefont {Tolan}, \citenamefont {Honkavaara},
  \citenamefont {Tsakanov}, \citenamefont {Ramert}, \citenamefont {Ilday},
  \citenamefont {Schulte-Schrepping}, \citenamefont {Moeller}, \citenamefont
  {Schuch}, \citenamefont {Shabunov}, \citenamefont {Sellmann}, \citenamefont
  {Wark}, \citenamefont {Sch{\"a}fer}, \citenamefont {Stulle}, \citenamefont
  {Nicolosi}, \citenamefont {Castellano}, \citenamefont {Trines}, \citenamefont
  {Fominykh}, \citenamefont {Ross}, \citenamefont {Kim}, \citenamefont
  {McCormick}, \citenamefont {Tschentscher}, \citenamefont {Toral},
  \citenamefont {Sytchev}, \citenamefont {Reschke}, \citenamefont {Anfinrud},
  \citenamefont {Will}, \citenamefont {Toleikis}, \citenamefont {Schlarb},
  \citenamefont {Br{\"u}ck}, \citenamefont {Monaco}, \citenamefont {Ploenjes},
  \citenamefont {Johnson}, \citenamefont {Schmueser}, \citenamefont
  {Aghababyan}, \citenamefont {Krause}, \citenamefont {Han}, \citenamefont
  {Tesch}, \citenamefont {Klose}, \citenamefont {Feldhaus}, \citenamefont
  {Nagl}, \citenamefont {Staack}, \citenamefont {Ischebeck}, \citenamefont
  {Kocharyan}, \citenamefont {Vuilleumier}, \citenamefont {French},
  \citenamefont {Wittenburg}, \citenamefont {Gr{\"u}bel}, \citenamefont
  {Riemann}, \citenamefont {Froehlich}, \citenamefont {Wochner}, \citenamefont
  {Dehler}, \citenamefont {Garc{\'\i}a-Tabar{\'e}s}, \citenamefont {Seidel},
  \citenamefont {Celik}, \citenamefont {Lilje}, \citenamefont {David},
  \citenamefont {Edwards}, \citenamefont {Gel'mukhanov}, \citenamefont
  {Poletto}, \citenamefont {Bostedt}, \citenamefont {Wabnitz}, \citenamefont
  {Bressler}, \citenamefont {Delsim-Hashemi}, \citenamefont {Kapitza},
  \citenamefont {Kostin}, \citenamefont {Bohnet}, \citenamefont {Lange},
  \citenamefont {Richter}, \citenamefont {Matheisen}, \citenamefont {Kook},
  \citenamefont {Zambolin}, \citenamefont {Griogoryan}, \citenamefont {Hahn},
  \citenamefont {Wohlenberg}, \citenamefont {Miltchev}, \citenamefont
  {Schmitz}, \citenamefont {Petrosyan}, \citenamefont {Schreiber},
  \citenamefont {Havlicek}, \citenamefont {Roensch}, \citenamefont {Saldin},
  \citenamefont {Gareta}, \citenamefont {Matzen}, \citenamefont {Schrader},
  \citenamefont {Catani}, \citenamefont {Kammering}, \citenamefont {Jablonka},
  \citenamefont {Lipka}, \citenamefont {Ullrich}, \citenamefont {Blome},
  \citenamefont {Petersen}, \citenamefont {Goerler}, \citenamefont
  {Krasilnikov}, \citenamefont {Loehl}, \citenamefont {Jensen}, \citenamefont
  {Proch}, \citenamefont {Laich}, \citenamefont {Balandin}, \citenamefont
  {Tischer}, \citenamefont {Charalambidis}, \citenamefont {Garvey},
  \citenamefont {Luong}, \citenamefont {Rybnikov}, \citenamefont {Minty},
  \citenamefont {Sinn}, \citenamefont {Cavalleri}, \citenamefont {Variola},
  \citenamefont {M{\"o}ller}, \citenamefont {Weiland}, \citenamefont
  {P{\"o}plau}, \citenamefont {Becker}, \citenamefont {Yurkov}, \citenamefont
  {Zeitoun}, \citenamefont {Bandelmann}, \citenamefont {Pflueger},
  \citenamefont {Wanzenberg}, \citenamefont {Larsson}, \citenamefont {Steffen},
  \citenamefont {Lindenberg}, \citenamefont {Riley}, \citenamefont {Rehlich},
  \citenamefont {Remde}, \citenamefont {Michelato}, \citenamefont {Hacker},
  \citenamefont {Krzywinski}, \citenamefont {Pugachov}, \citenamefont {Lee},
  \citenamefont {Graeff}, \citenamefont {Schneidmiller}, \citenamefont
  {Noelle}, \citenamefont {Bolzmann}, \citenamefont {Faatz}, \citenamefont
  {Carneiro}, \citenamefont {Kollewe}, \citenamefont {Meyners}, \citenamefont
  {Gutt}, \citenamefont {Weise}, \citenamefont {Duesterer}, \citenamefont
  {Petrov}, \citenamefont {Kaerntner}, \citenamefont {Altarelli}, \citenamefont
  {May}, \citenamefont {Brovko}, \citenamefont {Schlott}, \citenamefont
  {Ayvazyan}, \citenamefont {Neubauer}, \citenamefont {Frisch}, \citenamefont
  {Kube}, \citenamefont {van Rienen}, \citenamefont {Zagorodnov}, \citenamefont
  {Sorokin}, \citenamefont {Clausen}, \citenamefont {Bozhko}, \citenamefont
  {Stephenson}, \citenamefont {Graafsma}, \citenamefont {Maquet}, \citenamefont
  {Zolotov}, \citenamefont {Schloesser}, \citenamefont {Sekutowicz},
  \citenamefont {Matyushevskiy}, \citenamefont {Koerfer}, \citenamefont
  {Howells}, \citenamefont {Dohlus}, \citenamefont {McNulty}, \citenamefont
  {Oppelt}, \citenamefont {Rosmej}, \citenamefont {Khodyachykh}, \citenamefont
  {Wichmann}, \citenamefont {Amatuni}, \citenamefont {Wojtkiewicz},
  \citenamefont {Meulen}, \citenamefont {Eckhardt}, \citenamefont {Limberg},
  \citenamefont {Mildner}, \citenamefont {Krebs}, \citenamefont {Reininger},
  \citenamefont {Koehler}, \citenamefont {Hensler}, \citenamefont {Staykov},
  \citenamefont {Schotte}, \citenamefont {Jensch}, \citenamefont {Ziemann},
  \citenamefont {Springate}, \citenamefont {Keil}, \citenamefont {Audebert},
  \citenamefont {Pedersen}, \citenamefont {Danared}, \citenamefont {Magne},
  \citenamefont {Leuschner}, \citenamefont {Beutner}, \citenamefont {Follath},
  \citenamefont {Zapfe}, \citenamefont {Hott}, \citenamefont {Ludwig},\ and\
  \citenamefont {Richter}}]{Abela:DESY:2006}%
  \BibitemOpen
  \bibfield  {author} {\bibinfo {author} {\bibfnamefont {R.}~\bibnamefont
  {Abela}}, \bibinfo {author} {\bibfnamefont {K.}~\bibnamefont {Witte}},
  \bibinfo {author} {\bibfnamefont {A.}~\bibnamefont {Schwarz}}, \bibinfo
  {author} {\bibfnamefont {H.}~\bibnamefont {Redlin}}, \bibinfo {author}
  {\bibfnamefont {H.~J.}\ \bibnamefont {Eckoldt}}, \bibinfo {author}
  {\bibfnamefont {M.}~\bibnamefont {Hartrott}}, \bibinfo {author}
  {\bibfnamefont {K.}~\bibnamefont {Floettmann}}, \bibinfo {author}
  {\bibfnamefont {M.}~\bibnamefont {Fajardo}}, \bibinfo {author} {\bibfnamefont
  {V.}~\bibnamefont {Katalev}}, \bibinfo {author} {\bibfnamefont
  {T.}~\bibnamefont {Schilcher}}, \bibinfo {author} {\bibfnamefont
  {K.}~\bibnamefont {Tiedtke}}, \bibinfo {author} {\bibfnamefont
  {J.}~\bibnamefont {Baehr}}, \bibinfo {author} {\bibfnamefont
  {W.}~\bibnamefont {Singer}}, \bibinfo {author} {\bibfnamefont
  {E.}~\bibnamefont {Sombrowski}}, \bibinfo {author} {\bibfnamefont
  {S.}~\bibnamefont {Casalbuoni}}, \bibinfo {author} {\bibfnamefont
  {J.}~\bibnamefont {Hajdu}}, \bibinfo {author} {\bibfnamefont
  {J.}~\bibnamefont {Schneider}}, \bibinfo {author} {\bibfnamefont
  {G.}~\bibnamefont {Dipirro}}, \bibinfo {author} {\bibfnamefont {J.~W.}\
  \bibnamefont {Kim}}, \bibinfo {author} {\bibfnamefont {R.}~\bibnamefont
  {Brinkmann}}, \bibinfo {author} {\bibfnamefont {B.}~\bibnamefont {Racky}},
  \bibinfo {author} {\bibfnamefont {J.}~\bibnamefont {Chen}}, \bibinfo {author}
  {\bibfnamefont {A.}~\bibnamefont {Nienhaus}}, \bibinfo {author}
  {\bibfnamefont {H.}~\bibnamefont {Chapman}}, \bibinfo {author} {\bibfnamefont
  {H.}~\bibnamefont {Lierl}}, \bibinfo {author} {\bibfnamefont
  {E.}~\bibnamefont {Gadwinkel}}, \bibinfo {author} {\bibfnamefont
  {C.}~\bibnamefont {Gerth}}, \bibinfo {author} {\bibfnamefont
  {I.}~\bibnamefont {Robinson}}, \bibinfo {author} {\bibfnamefont
  {S.}~\bibnamefont {Choroba}}, \bibinfo {author} {\bibfnamefont
  {M.}~\bibnamefont {Maslov}}, \bibinfo {author} {\bibfnamefont
  {R.}~\bibnamefont {Wenndorff}}, \bibinfo {author} {\bibfnamefont
  {C.}~\bibnamefont {Pagani}}, \bibinfo {author} {\bibfnamefont
  {G.}~\bibnamefont {Tallents}}, \bibinfo {author} {\bibfnamefont {H.~J.}\
  \bibnamefont {Grabosch}}, \bibinfo {author} {\bibfnamefont {E.}~\bibnamefont
  {Jaeschke}}, \bibinfo {author} {\bibfnamefont {R.}~\bibnamefont {Paparella}},
  \bibinfo {author} {\bibfnamefont {P.}~\bibnamefont {Seller}}, \bibinfo
  {author} {\bibfnamefont {P.}~\bibnamefont {Piot}}, \bibinfo {author}
  {\bibfnamefont {I.}~\bibnamefont {Tsakov}}, \bibinfo {author} {\bibfnamefont
  {D.}~\bibnamefont {Kraemer}}, \bibinfo {author} {\bibfnamefont
  {A.}~\bibnamefont {Winter}}, \bibinfo {author} {\bibfnamefont
  {M.}~\bibnamefont {Sachwitz}}, \bibinfo {author} {\bibfnamefont
  {A.}~\bibnamefont {Cianchi}}, \bibinfo {author} {\bibfnamefont
  {H.}~\bibnamefont {Weddig}}, \bibinfo {author} {\bibfnamefont
  {W.}~\bibnamefont {Sandner}}, \bibinfo {author} {\bibfnamefont
  {C.}~\bibnamefont {Schroer}}, \bibinfo {author} {\bibfnamefont
  {T.}~\bibnamefont {Nunez}}, \bibinfo {author} {\bibfnamefont
  {P.}~\bibnamefont {Castro}}, \bibinfo {author} {\bibfnamefont
  {O.}~\bibnamefont {Napoly}}, \bibinfo {author} {\bibfnamefont
  {Q.}~\bibnamefont {Kong}}, \bibinfo {author} {\bibfnamefont {S.}~\bibnamefont
  {Simrock}}, \bibinfo {author} {\bibfnamefont {U.}~\bibnamefont {Gensch}},
  \bibinfo {author} {\bibfnamefont {H.}~\bibnamefont {Thom}}, \bibinfo {author}
  {\bibfnamefont {S.}~\bibnamefont {Bratos}}, \bibinfo {author} {\bibfnamefont
  {C.}~\bibnamefont {Altucci}}, \bibinfo {author} {\bibfnamefont
  {B.}~\bibnamefont {Dobson}}, \bibinfo {author} {\bibfnamefont
  {C.}~\bibnamefont {Masciovecchio}}, \bibinfo {author} {\bibfnamefont
  {H.}~\bibnamefont {Quack}}, \bibinfo {author} {\bibfnamefont
  {R.}~\bibnamefont {Kienberger}}, \bibinfo {author} {\bibfnamefont
  {M.}~\bibnamefont {Wendt}}, \bibinfo {author} {\bibfnamefont
  {M.}~\bibnamefont {Chergui}}, \bibinfo {author} {\bibfnamefont
  {O.}~\bibnamefont {Grimm}}, \bibinfo {author} {\bibfnamefont
  {A.}~\bibnamefont {Plech}}, \bibinfo {author} {\bibfnamefont
  {F.}~\bibnamefont {Obier}}, \bibinfo {author} {\bibfnamefont
  {B.}~\bibnamefont {Schmidt}}, \bibinfo {author} {\bibfnamefont
  {V.}~\bibnamefont {Sytchev}}, \bibinfo {author} {\bibfnamefont
  {J.}~\bibnamefont {Rossbach}}, \bibinfo {author} {\bibfnamefont
  {O.}~\bibnamefont {Kozlov}}, \bibinfo {author} {\bibfnamefont
  {M.}~\bibnamefont {Kuhlmann}}, \bibinfo {author} {\bibfnamefont
  {F.}~\bibnamefont {Senf}}, \bibinfo {author} {\bibfnamefont {A.}~\bibnamefont
  {Wolf}}, \bibinfo {author} {\bibfnamefont {J.}~\bibnamefont {Vogel}},
  \bibinfo {author} {\bibfnamefont {J.}~\bibnamefont {Prenting}}, \bibinfo
  {author} {\bibfnamefont {F.}~\bibnamefont {Stephan}}, \bibinfo {author}
  {\bibfnamefont {E.}~\bibnamefont {Syresin}}, \bibinfo {author} {\bibfnamefont
  {D.}~\bibnamefont {Sertore}}, \bibinfo {author} {\bibfnamefont
  {W.}~\bibnamefont {Decking}}, \bibinfo {author} {\bibfnamefont
  {N.}~\bibnamefont {Baboi}}, \bibinfo {author} {\bibfnamefont
  {E.}~\bibnamefont {Vogel}}, \bibinfo {author} {\bibfnamefont
  {L.}~\bibnamefont {Juha}}, \bibinfo {author} {\bibfnamefont {A.}~\bibnamefont
  {Brandt}}, \bibinfo {author} {\bibfnamefont {E.}~\bibnamefont {Collet}},
  \bibinfo {author} {\bibfnamefont {Y.}~\bibnamefont {Filipov}}, \bibinfo
  {author} {\bibfnamefont {I.}~\bibnamefont {Vartaniants}}, \bibinfo {author}
  {\bibfnamefont {J.}~\bibnamefont {Marangos}}, \bibinfo {author}
  {\bibfnamefont {N.}~\bibnamefont {Golubeva}}, \bibinfo {author}
  {\bibfnamefont {R.}~\bibnamefont {Smith}}, \bibinfo {author} {\bibfnamefont
  {S.}~\bibnamefont {Techert}}, \bibinfo {author} {\bibfnamefont
  {A.}~\bibnamefont {Fateev}}, \bibinfo {author} {\bibfnamefont
  {M.}~\bibnamefont {Wulff}}, \bibinfo {author} {\bibfnamefont
  {L.}~\bibnamefont {Haenisch}}, \bibinfo {author} {\bibfnamefont
  {M.}~\bibnamefont {Huening}}, \bibinfo {author} {\bibfnamefont
  {V.}~\bibnamefont {Honkim{\"a}ki}}, \bibinfo {author} {\bibfnamefont
  {R.}~\bibnamefont {Treusch}}, \bibinfo {author} {\bibfnamefont
  {H.}~\bibnamefont {Ihee}}, \bibinfo {author} {\bibfnamefont {S.}~\bibnamefont
  {Prat}}, \bibinfo {author} {\bibfnamefont {B.}~\bibnamefont {Petrosyan}},
  \bibinfo {author} {\bibfnamefont {E.}~\bibnamefont {Prat}}, \bibinfo {author}
  {\bibfnamefont {M.}~\bibnamefont {Tolan}}, \bibinfo {author} {\bibfnamefont
  {K.}~\bibnamefont {Honkavaara}}, \bibinfo {author} {\bibfnamefont
  {V.}~\bibnamefont {Tsakanov}}, \bibinfo {author} {\bibfnamefont
  {D.}~\bibnamefont {Ramert}}, \bibinfo {author} {\bibfnamefont {F.~{\"O}.}\
  \bibnamefont {Ilday}}, \bibinfo {author} {\bibfnamefont {H.}~\bibnamefont
  {Schulte-Schrepping}}, \bibinfo {author} {\bibfnamefont {W.~D.}\ \bibnamefont
  {Moeller}}, \bibinfo {author} {\bibfnamefont {R.}~\bibnamefont {Schuch}},
  \bibinfo {author} {\bibfnamefont {A.}~\bibnamefont {Shabunov}}, \bibinfo
  {author} {\bibfnamefont {D.}~\bibnamefont {Sellmann}}, \bibinfo {author}
  {\bibfnamefont {J.~S.}\ \bibnamefont {Wark}}, \bibinfo {author}
  {\bibfnamefont {J.}~\bibnamefont {Sch{\"a}fer}}, \bibinfo {author}
  {\bibfnamefont {F.}~\bibnamefont {Stulle}}, \bibinfo {author} {\bibfnamefont
  {P.}~\bibnamefont {Nicolosi}}, \bibinfo {author} {\bibfnamefont
  {M.}~\bibnamefont {Castellano}}, \bibinfo {author} {\bibfnamefont
  {D.}~\bibnamefont {Trines}}, \bibinfo {author} {\bibfnamefont
  {B.}~\bibnamefont {Fominykh}}, \bibinfo {author} {\bibfnamefont
  {M.}~\bibnamefont {Ross}}, \bibinfo {author} {\bibfnamefont {Y.}~\bibnamefont
  {Kim}}, \bibinfo {author} {\bibfnamefont {D.}~\bibnamefont {McCormick}},
  \bibinfo {author} {\bibfnamefont {T.}~\bibnamefont {Tschentscher}}, \bibinfo
  {author} {\bibfnamefont {F.}~\bibnamefont {Toral}}, \bibinfo {author}
  {\bibfnamefont {K.}~\bibnamefont {Sytchev}}, \bibinfo {author} {\bibfnamefont
  {D.}~\bibnamefont {Reschke}}, \bibinfo {author} {\bibfnamefont
  {P.}~\bibnamefont {Anfinrud}}, \bibinfo {author} {\bibfnamefont
  {I.}~\bibnamefont {Will}}, \bibinfo {author} {\bibfnamefont {S.}~\bibnamefont
  {Toleikis}}, \bibinfo {author} {\bibfnamefont {H.}~\bibnamefont {Schlarb}},
  \bibinfo {author} {\bibfnamefont {H.}~\bibnamefont {Br{\"u}ck}}, \bibinfo
  {author} {\bibfnamefont {L.}~\bibnamefont {Monaco}}, \bibinfo {author}
  {\bibfnamefont {E.}~\bibnamefont {Ploenjes}}, \bibinfo {author}
  {\bibfnamefont {S.}~\bibnamefont {Johnson}}, \bibinfo {author} {\bibfnamefont
  {P.}~\bibnamefont {Schmueser}}, \bibinfo {author} {\bibfnamefont
  {A.}~\bibnamefont {Aghababyan}}, \bibinfo {author} {\bibfnamefont
  {B.}~\bibnamefont {Krause}}, \bibinfo {author} {\bibfnamefont {J.~H.}\
  \bibnamefont {Han}}, \bibinfo {author} {\bibfnamefont {N.}~\bibnamefont
  {Tesch}}, \bibinfo {author} {\bibfnamefont {K.}~\bibnamefont {Klose}},
  \bibinfo {author} {\bibfnamefont {J.}~\bibnamefont {Feldhaus}}, \bibinfo
  {author} {\bibfnamefont {M.}~\bibnamefont {Nagl}}, \bibinfo {author}
  {\bibfnamefont {M.}~\bibnamefont {Staack}}, \bibinfo {author} {\bibfnamefont
  {R.}~\bibnamefont {Ischebeck}}, \bibinfo {author} {\bibfnamefont
  {V.}~\bibnamefont {Kocharyan}}, \bibinfo {author} {\bibfnamefont
  {R.}~\bibnamefont {Vuilleumier}}, \bibinfo {author} {\bibfnamefont
  {M.}~\bibnamefont {French}}, \bibinfo {author} {\bibfnamefont
  {K.}~\bibnamefont {Wittenburg}}, \bibinfo {author} {\bibfnamefont
  {G.}~\bibnamefont {Gr{\"u}bel}}, \bibinfo {author} {\bibfnamefont
  {S.}~\bibnamefont {Riemann}}, \bibinfo {author} {\bibfnamefont
  {L.}~\bibnamefont {Froehlich}}, \bibinfo {author} {\bibfnamefont
  {P.}~\bibnamefont {Wochner}}, \bibinfo {author} {\bibfnamefont
  {M.}~\bibnamefont {Dehler}}, \bibinfo {author} {\bibfnamefont
  {L.}~\bibnamefont {Garc{\'\i}a-Tabar{\'e}s}}, \bibinfo {author}
  {\bibfnamefont {M.}~\bibnamefont {Seidel}}, \bibinfo {author} {\bibfnamefont
  {S.}~\bibnamefont {Celik}}, \bibinfo {author} {\bibfnamefont
  {L.}~\bibnamefont {Lilje}}, \bibinfo {author} {\bibfnamefont
  {C.}~\bibnamefont {David}}, \bibinfo {author} {\bibfnamefont
  {H.}~\bibnamefont {Edwards}}, \bibinfo {author} {\bibfnamefont
  {F.}~\bibnamefont {Gel'mukhanov}}, \bibinfo {author} {\bibfnamefont
  {L.}~\bibnamefont {Poletto}}, \bibinfo {author} {\bibfnamefont
  {C.}~\bibnamefont {Bostedt}}, \bibinfo {author} {\bibfnamefont
  {H.}~\bibnamefont {Wabnitz}}, \bibinfo {author} {\bibfnamefont
  {C.}~\bibnamefont {Bressler}}, \bibinfo {author} {\bibfnamefont
  {H.}~\bibnamefont {Delsim-Hashemi}}, \bibinfo {author} {\bibfnamefont
  {H.}~\bibnamefont {Kapitza}}, \bibinfo {author} {\bibfnamefont
  {D.}~\bibnamefont {Kostin}}, \bibinfo {author} {\bibfnamefont
  {I.}~\bibnamefont {Bohnet}}, \bibinfo {author} {\bibfnamefont
  {R.}~\bibnamefont {Lange}}, \bibinfo {author} {\bibfnamefont
  {D.}~\bibnamefont {Richter}}, \bibinfo {author} {\bibfnamefont
  {A.}~\bibnamefont {Matheisen}}, \bibinfo {author} {\bibfnamefont
  {W.}~\bibnamefont {Kook}}, \bibinfo {author} {\bibfnamefont {P.}~\bibnamefont
  {Zambolin}}, \bibinfo {author} {\bibfnamefont {B.}~\bibnamefont
  {Griogoryan}}, \bibinfo {author} {\bibfnamefont {U.}~\bibnamefont {Hahn}},
  \bibinfo {author} {\bibfnamefont {T.}~\bibnamefont {Wohlenberg}}, \bibinfo
  {author} {\bibfnamefont {V.}~\bibnamefont {Miltchev}}, \bibinfo {author}
  {\bibfnamefont {M.}~\bibnamefont {Schmitz}}, \bibinfo {author} {\bibfnamefont
  {L.}~\bibnamefont {Petrosyan}}, \bibinfo {author} {\bibfnamefont
  {S.}~\bibnamefont {Schreiber}}, \bibinfo {author} {\bibfnamefont
  {J.}~\bibnamefont {Havlicek}}, \bibinfo {author} {\bibfnamefont
  {J.}~\bibnamefont {Roensch}}, \bibinfo {author} {\bibfnamefont
  {E.}~\bibnamefont {Saldin}}, \bibinfo {author} {\bibfnamefont {J.~J.}\
  \bibnamefont {Gareta}}, \bibinfo {author} {\bibfnamefont {O.}~\bibnamefont
  {Matzen}}, \bibinfo {author} {\bibfnamefont {S.}~\bibnamefont {Schrader}},
  \bibinfo {author} {\bibfnamefont {L.}~\bibnamefont {Catani}}, \bibinfo
  {author} {\bibfnamefont {R.}~\bibnamefont {Kammering}}, \bibinfo {author}
  {\bibfnamefont {M.}~\bibnamefont {Jablonka}}, \bibinfo {author}
  {\bibfnamefont {D.}~\bibnamefont {Lipka}}, \bibinfo {author} {\bibfnamefont
  {F.~R.}\ \bibnamefont {Ullrich}}, \bibinfo {author} {\bibfnamefont
  {C.}~\bibnamefont {Blome}}, \bibinfo {author} {\bibfnamefont
  {B.}~\bibnamefont {Petersen}}, \bibinfo {author} {\bibfnamefont
  {M.}~\bibnamefont {Goerler}}, \bibinfo {author} {\bibfnamefont
  {M.}~\bibnamefont {Krasilnikov}}, \bibinfo {author} {\bibfnamefont
  {F.}~\bibnamefont {Loehl}}, \bibinfo {author} {\bibfnamefont {J.~P.}\
  \bibnamefont {Jensen}}, \bibinfo {author} {\bibfnamefont {D.}~\bibnamefont
  {Proch}}, \bibinfo {author} {\bibfnamefont {H.}~\bibnamefont {Laich}},
  \bibinfo {author} {\bibfnamefont {V.}~\bibnamefont {Balandin}}, \bibinfo
  {author} {\bibfnamefont {M.}~\bibnamefont {Tischer}}, \bibinfo {author}
  {\bibfnamefont {D.}~\bibnamefont {Charalambidis}}, \bibinfo {author}
  {\bibfnamefont {T.}~\bibnamefont {Garvey}}, \bibinfo {author} {\bibfnamefont
  {M.}~\bibnamefont {Luong}}, \bibinfo {author} {\bibfnamefont
  {V.}~\bibnamefont {Rybnikov}}, \bibinfo {author} {\bibfnamefont
  {M.}~\bibnamefont {Minty}}, \bibinfo {author} {\bibfnamefont
  {H.}~\bibnamefont {Sinn}}, \bibinfo {author} {\bibfnamefont {A.}~\bibnamefont
  {Cavalleri}}, \bibinfo {author} {\bibfnamefont {A.}~\bibnamefont {Variola}},
  \bibinfo {author} {\bibfnamefont {T.}~\bibnamefont {M{\"o}ller}}, \bibinfo
  {author} {\bibfnamefont {T.}~\bibnamefont {Weiland}}, \bibinfo {author}
  {\bibfnamefont {G.}~\bibnamefont {P{\"o}plau}}, \bibinfo {author}
  {\bibfnamefont {J.}~\bibnamefont {Becker}}, \bibinfo {author} {\bibfnamefont
  {M.}~\bibnamefont {Yurkov}}, \bibinfo {author} {\bibfnamefont
  {P.}~\bibnamefont {Zeitoun}}, \bibinfo {author} {\bibfnamefont
  {R.}~\bibnamefont {Bandelmann}}, \bibinfo {author} {\bibfnamefont
  {J.}~\bibnamefont {Pflueger}}, \bibinfo {author} {\bibfnamefont
  {R.}~\bibnamefont {Wanzenberg}}, \bibinfo {author} {\bibfnamefont
  {M.}~\bibnamefont {Larsson}}, \bibinfo {author} {\bibfnamefont
  {B.}~\bibnamefont {Steffen}}, \bibinfo {author} {\bibfnamefont
  {A.}~\bibnamefont {Lindenberg}}, \bibinfo {author} {\bibfnamefont
  {D.}~\bibnamefont {Riley}}, \bibinfo {author} {\bibfnamefont
  {K.}~\bibnamefont {Rehlich}}, \bibinfo {author} {\bibfnamefont
  {H.}~\bibnamefont {Remde}}, \bibinfo {author} {\bibfnamefont
  {P.}~\bibnamefont {Michelato}}, \bibinfo {author} {\bibfnamefont
  {K.}~\bibnamefont {Hacker}}, \bibinfo {author} {\bibfnamefont
  {J.}~\bibnamefont {Krzywinski}}, \bibinfo {author} {\bibfnamefont
  {D.}~\bibnamefont {Pugachov}}, \bibinfo {author} {\bibfnamefont {R.~W.}\
  \bibnamefont {Lee}}, \bibinfo {author} {\bibfnamefont {W.}~\bibnamefont
  {Graeff}}, \bibinfo {author} {\bibfnamefont {E.}~\bibnamefont
  {Schneidmiller}}, \bibinfo {author} {\bibfnamefont {D.}~\bibnamefont
  {Noelle}}, \bibinfo {author} {\bibfnamefont {A.}~\bibnamefont {Bolzmann}},
  \bibinfo {author} {\bibfnamefont {B.}~\bibnamefont {Faatz}}, \bibinfo
  {author} {\bibfnamefont {J.~P.}\ \bibnamefont {Carneiro}}, \bibinfo {author}
  {\bibfnamefont {M.}~\bibnamefont {Kollewe}}, \bibinfo {author} {\bibfnamefont
  {N.}~\bibnamefont {Meyners}}, \bibinfo {author} {\bibfnamefont
  {C.}~\bibnamefont {Gutt}}, \bibinfo {author} {\bibfnamefont {H.}~\bibnamefont
  {Weise}}, \bibinfo {author} {\bibfnamefont {S.}~\bibnamefont {Duesterer}},
  \bibinfo {author} {\bibfnamefont {A.}~\bibnamefont {Petrov}}, \bibinfo
  {author} {\bibfnamefont {F.}~\bibnamefont {Kaerntner}}, \bibinfo {author}
  {\bibfnamefont {M.}~\bibnamefont {Altarelli}}, \bibinfo {author}
  {\bibfnamefont {H.~J.}\ \bibnamefont {May}}, \bibinfo {author} {\bibfnamefont
  {O.}~\bibnamefont {Brovko}}, \bibinfo {author} {\bibfnamefont
  {V.}~\bibnamefont {Schlott}}, \bibinfo {author} {\bibfnamefont
  {V.}~\bibnamefont {Ayvazyan}}, \bibinfo {author} {\bibfnamefont
  {G.}~\bibnamefont {Neubauer}}, \bibinfo {author} {\bibfnamefont
  {J.}~\bibnamefont {Frisch}}, \bibinfo {author} {\bibfnamefont
  {G.}~\bibnamefont {Kube}}, \bibinfo {author} {\bibfnamefont {U.}~\bibnamefont
  {van Rienen}}, \bibinfo {author} {\bibfnamefont {I.}~\bibnamefont
  {Zagorodnov}}, \bibinfo {author} {\bibfnamefont {A.~A.}\ \bibnamefont
  {Sorokin}}, \bibinfo {author} {\bibfnamefont {M.}~\bibnamefont {Clausen}},
  \bibinfo {author} {\bibfnamefont {Y.}~\bibnamefont {Bozhko}}, \bibinfo
  {author} {\bibfnamefont {B.}~\bibnamefont {Stephenson}}, \bibinfo {author}
  {\bibfnamefont {H.}~\bibnamefont {Graafsma}}, \bibinfo {author}
  {\bibfnamefont {A.}~\bibnamefont {Maquet}}, \bibinfo {author} {\bibfnamefont
  {A.}~\bibnamefont {Zolotov}}, \bibinfo {author} {\bibfnamefont
  {M.}~\bibnamefont {Schloesser}}, \bibinfo {author} {\bibfnamefont
  {J.}~\bibnamefont {Sekutowicz}}, \bibinfo {author} {\bibfnamefont
  {E.}~\bibnamefont {Matyushevskiy}}, \bibinfo {author} {\bibfnamefont
  {M.}~\bibnamefont {Koerfer}}, \bibinfo {author} {\bibfnamefont {M.~R.}\
  \bibnamefont {Howells}}, \bibinfo {author} {\bibfnamefont {M.}~\bibnamefont
  {Dohlus}}, \bibinfo {author} {\bibfnamefont {I.}~\bibnamefont {McNulty}},
  \bibinfo {author} {\bibfnamefont {A.}~\bibnamefont {Oppelt}}, \bibinfo
  {author} {\bibfnamefont {F.}~\bibnamefont {Rosmej}}, \bibinfo {author}
  {\bibfnamefont {S.}~\bibnamefont {Khodyachykh}}, \bibinfo {author}
  {\bibfnamefont {R.}~\bibnamefont {Wichmann}}, \bibinfo {author}
  {\bibfnamefont {G.}~\bibnamefont {Amatuni}}, \bibinfo {author} {\bibfnamefont
  {J.}~\bibnamefont {Wojtkiewicz}}, \bibinfo {author} {\bibfnamefont
  {P.}~\bibnamefont {Meulen}}, \bibinfo {author} {\bibfnamefont
  {A.}~\bibnamefont {Eckhardt}}, \bibinfo {author} {\bibfnamefont
  {T.}~\bibnamefont {Limberg}}, \bibinfo {author} {\bibfnamefont
  {N.}~\bibnamefont {Mildner}}, \bibinfo {author} {\bibfnamefont
  {O.}~\bibnamefont {Krebs}}, \bibinfo {author} {\bibfnamefont
  {R.}~\bibnamefont {Reininger}}, \bibinfo {author} {\bibfnamefont
  {W.}~\bibnamefont {Koehler}}, \bibinfo {author} {\bibfnamefont
  {O.}~\bibnamefont {Hensler}}, \bibinfo {author} {\bibfnamefont
  {L.}~\bibnamefont {Staykov}}, \bibinfo {author} {\bibfnamefont
  {F.}~\bibnamefont {Schotte}}, \bibinfo {author} {\bibfnamefont
  {K.}~\bibnamefont {Jensch}}, \bibinfo {author} {\bibfnamefont
  {V.}~\bibnamefont {Ziemann}}, \bibinfo {author} {\bibfnamefont
  {E.}~\bibnamefont {Springate}}, \bibinfo {author} {\bibfnamefont
  {B.}~\bibnamefont {Keil}}, \bibinfo {author} {\bibfnamefont {P.}~\bibnamefont
  {Audebert}}, \bibinfo {author} {\bibfnamefont {H.~B.}\ \bibnamefont
  {Pedersen}}, \bibinfo {author} {\bibfnamefont {H.}~\bibnamefont {Danared}},
  \bibinfo {author} {\bibfnamefont {C.}~\bibnamefont {Magne}}, \bibinfo
  {author} {\bibfnamefont {A.}~\bibnamefont {Leuschner}}, \bibinfo {author}
  {\bibfnamefont {B.}~\bibnamefont {Beutner}}, \bibinfo {author} {\bibfnamefont
  {R.}~\bibnamefont {Follath}}, \bibinfo {author} {\bibfnamefont
  {K.}~\bibnamefont {Zapfe}}, \bibinfo {author} {\bibfnamefont
  {T.}~\bibnamefont {Hott}}, \bibinfo {author} {\bibfnamefont {K.}~\bibnamefont
  {Ludwig}},\ and\ \bibinfo {author} {\bibfnamefont {M.}~\bibnamefont
  {Richter}},\ }\href {https://bib-pubdb1.desy.de/record/77248#} {\emph
  {\bibinfo {title} {{XFEL}: The {E}uropean X-Ray Free-Electron Laser --
  Technical Design Report}}}\ (\bibinfo  {publisher} {Deutsches
  Elektronen-Synchrotron DESY},\ \bibinfo {address} {Hamburg, Germany},\
  \bibinfo {year} {2006})\BibitemShut {NoStop}%
\bibitem [{\citenamefont {Tschentscher}\ \emph {et~al.}(2017)\citenamefont
  {Tschentscher}, \citenamefont {Bressler}, \citenamefont {Grünert},
  \citenamefont {Madsen}, \citenamefont {Mancuso}, \citenamefont {Meyer},
  \citenamefont {Scherz}, \citenamefont {Sinn},\ and\ \citenamefont
  {Zastrau}}]{Tschentscher:ApplSci}%
  \BibitemOpen
  \bibfield  {author} {\bibinfo {author} {\bibfnamefont {T.}~\bibnamefont
  {Tschentscher}}, \bibinfo {author} {\bibfnamefont {C.}~\bibnamefont
  {Bressler}}, \bibinfo {author} {\bibfnamefont {J.}~\bibnamefont {Grünert}},
  \bibinfo {author} {\bibfnamefont {A.}~\bibnamefont {Madsen}}, \bibinfo
  {author} {\bibfnamefont {A.~P.}\ \bibnamefont {Mancuso}}, \bibinfo {author}
  {\bibfnamefont {M.}~\bibnamefont {Meyer}}, \bibinfo {author} {\bibfnamefont
  {A.}~\bibnamefont {Scherz}}, \bibinfo {author} {\bibfnamefont
  {H.}~\bibnamefont {Sinn}},\ and\ \bibinfo {author} {\bibfnamefont
  {U.}~\bibnamefont {Zastrau}},\ }\bibfield  {title} {\bibinfo {title} {Photon
  beam transport and scientific instruments at the {E}uropean {XFEL}},\ }\href
  {https://doi.org/10.3390/app7060592} {\bibfield  {journal} {\bibinfo
  {journal} {Appl. Sci.}\ }\textbf {\bibinfo {volume} {7}},\ \bibinfo {pages}
  {592} (\bibinfo {year} {2017})}\BibitemShut {NoStop}%
\bibitem [{SLA(2015)}]{SLAC:New-Science-LCLS-II:2016}%
  \BibitemOpen
  \href
  {https://portal.slac.stanford.edu/sites/lcls_public/Documents/LCLS-IIScienceOpportunities_final.pdf}
  {\emph {\bibinfo {title} {New Science Opportunities Enabled By {LCLS}-{II}
  X-Ray Lasers}}},\ \bibinfo {type} {Tech. Rep.}\ \bibinfo {number}
  {SLAC-R-1053}\ (\bibinfo  {institution} {SLAC National Accelerator
  Laboratory},\ \bibinfo {address} {Menlo Park, CA, USA},\ \bibinfo {year}
  {2015})\BibitemShut {NoStop}%
\bibitem [{\citenamefont {Henrich}\ \emph {et~al.}(2011)\citenamefont
  {Henrich}, \citenamefont {Becker}, \citenamefont {Dinapoli}, \citenamefont
  {Goettlicher}, \citenamefont {Graafsma}, \citenamefont {Hirsemann},
  \citenamefont {Klanner}, \citenamefont {Krueger}, \citenamefont {Mazzocco},
  \citenamefont {Mozzanica}, \citenamefont {Perrey}, \citenamefont {Potdevin},
  \citenamefont {Schmitt}, \citenamefont {Shi}, \citenamefont {Srivastava},
  \citenamefont {Trunk},\ and\ \citenamefont {Youngman}}]{Henrich:NIMA633:S11}%
  \BibitemOpen
  \bibfield  {author} {\bibinfo {author} {\bibfnamefont {B.}~\bibnamefont
  {Henrich}}, \bibinfo {author} {\bibfnamefont {J.}~\bibnamefont {Becker}},
  \bibinfo {author} {\bibfnamefont {R.}~\bibnamefont {Dinapoli}}, \bibinfo
  {author} {\bibfnamefont {P.}~\bibnamefont {Goettlicher}}, \bibinfo {author}
  {\bibfnamefont {H.}~\bibnamefont {Graafsma}}, \bibinfo {author}
  {\bibfnamefont {H.}~\bibnamefont {Hirsemann}}, \bibinfo {author}
  {\bibfnamefont {R.}~\bibnamefont {Klanner}}, \bibinfo {author} {\bibfnamefont
  {H.}~\bibnamefont {Krueger}}, \bibinfo {author} {\bibfnamefont
  {R.}~\bibnamefont {Mazzocco}}, \bibinfo {author} {\bibfnamefont
  {A.}~\bibnamefont {Mozzanica}}, \bibinfo {author} {\bibfnamefont
  {H.}~\bibnamefont {Perrey}}, \bibinfo {author} {\bibfnamefont
  {G.}~\bibnamefont {Potdevin}}, \bibinfo {author} {\bibfnamefont
  {B.}~\bibnamefont {Schmitt}}, \bibinfo {author} {\bibfnamefont
  {X.}~\bibnamefont {Shi}}, \bibinfo {author} {\bibfnamefont {A.}~\bibnamefont
  {Srivastava}}, \bibinfo {author} {\bibfnamefont {U.}~\bibnamefont {Trunk}},\
  and\ \bibinfo {author} {\bibfnamefont {C.}~\bibnamefont {Youngman}},\
  }\bibfield  {title} {\bibinfo {title} {The adaptive gain integrating pixel
  detector {AGIPD} a detector for the {E}uropean {XFEL}},\ }\href
  {https://doi.org/10.1016/j.nima.2010.06.107} {\bibfield  {journal} {\bibinfo
  {journal} {Nucl. Instrum. Meth. A}\ }\textbf {\bibinfo {volume} {633}},\
  \bibinfo {pages} {S11} (\bibinfo {year} {2011})}\BibitemShut {NoStop}%
\bibitem [{\citenamefont {Blaj}\ \emph {et~al.}(2016)\citenamefont {Blaj},
  \citenamefont {Caragiulo}, \citenamefont {Carini}, \citenamefont {Dragone},
  \citenamefont {Haller}, \citenamefont {Hart}, \citenamefont {Hasi},
  \citenamefont {Herbst}, \citenamefont {Kenney}, \citenamefont {Markovic},
  \citenamefont {Nishimura}, \citenamefont {Pines}, \citenamefont {Segal},
  \citenamefont {Tamma},\ and\ \citenamefont
  {Tomada}}]{Blaj:AIPConfProc1741:040012}%
  \BibitemOpen
  \bibfield  {author} {\bibinfo {author} {\bibfnamefont {G.}~\bibnamefont
  {Blaj}}, \bibinfo {author} {\bibfnamefont {P.}~\bibnamefont {Caragiulo}},
  \bibinfo {author} {\bibfnamefont {G.}~\bibnamefont {Carini}}, \bibinfo
  {author} {\bibfnamefont {A.}~\bibnamefont {Dragone}}, \bibinfo {author}
  {\bibfnamefont {G.}~\bibnamefont {Haller}}, \bibinfo {author} {\bibfnamefont
  {P.}~\bibnamefont {Hart}}, \bibinfo {author} {\bibfnamefont {J.}~\bibnamefont
  {Hasi}}, \bibinfo {author} {\bibfnamefont {R.}~\bibnamefont {Herbst}},
  \bibinfo {author} {\bibfnamefont {C.}~\bibnamefont {Kenney}}, \bibinfo
  {author} {\bibfnamefont {B.}~\bibnamefont {Markovic}}, \bibinfo {author}
  {\bibfnamefont {K.}~\bibnamefont {Nishimura}}, \bibinfo {author}
  {\bibfnamefont {J.}~\bibnamefont {Pines}}, \bibinfo {author} {\bibfnamefont
  {J.}~\bibnamefont {Segal}}, \bibinfo {author} {\bibfnamefont
  {C.}~\bibnamefont {Tamma}},\ and\ \bibinfo {author} {\bibfnamefont
  {A.}~\bibnamefont {Tomada}},\ }\bibfield  {title} {\bibinfo {title} {Future
  of {ePix} detectors for high repetition rate {FEL}s},\ }\href
  {https://doi.org/10.1063/1.4952884} {\bibfield  {journal} {\bibinfo
  {journal} {AIP Conf. Proc.}\ }\textbf {\bibinfo {volume} {1741}},\ \bibinfo
  {pages} {040012} (\bibinfo {year} {2016})}\BibitemShut {NoStop}%
\bibitem [{\citenamefont {Markovic}\ \emph {et~al.}(2016)\citenamefont
  {Markovic}, \citenamefont {Caragiulo}, \citenamefont {Dragone}, \citenamefont
  {Tamma}, \citenamefont {Osipov}, \citenamefont {Bostedt}, \citenamefont
  {Kwiatkowski}, \citenamefont {Segal}, \citenamefont {Hasi}, \citenamefont
  {Blaj}, \citenamefont {Kenney},\ and\ \citenamefont
  {Haller}}]{Markovic:IEEE}%
  \BibitemOpen
  \bibfield  {author} {\bibinfo {author} {\bibfnamefont {B.}~\bibnamefont
  {Markovic}}, \bibinfo {author} {\bibfnamefont {P.}~\bibnamefont {Caragiulo}},
  \bibinfo {author} {\bibfnamefont {A.}~\bibnamefont {Dragone}}, \bibinfo
  {author} {\bibfnamefont {C.}~\bibnamefont {Tamma}}, \bibinfo {author}
  {\bibfnamefont {T.}~\bibnamefont {Osipov}}, \bibinfo {author} {\bibfnamefont
  {C.}~\bibnamefont {Bostedt}}, \bibinfo {author} {\bibfnamefont
  {M.}~\bibnamefont {Kwiatkowski}}, \bibinfo {author} {\bibfnamefont
  {J.}~\bibnamefont {Segal}}, \bibinfo {author} {\bibfnamefont
  {J.}~\bibnamefont {Hasi}}, \bibinfo {author} {\bibfnamefont {G.}~\bibnamefont
  {Blaj}}, \bibinfo {author} {\bibfnamefont {C.}~\bibnamefont {Kenney}},\ and\
  \bibinfo {author} {\bibfnamefont {G.}~\bibnamefont {Haller}},\ }\bibfield
  {title} {\bibinfo {title} {{Design and characterization of the tPix
  prototype: A spatial and time resolving front-end ASIC for electron and ion
  spectroscopy experiments at LCLS}},\ }in\ \href
  {https://doi.org/10.1109/NSSMIC.2016.8069725} {\emph {\bibinfo {booktitle}
  {2016 IEEE Nuclear Science Symposium, Medical Imaging Conference and
  Room-Temperature Semiconductor Detector Workshop (NSS/MIC/RTSD)}}}\ (\bibinfo
  {year} {2016})\ pp.\ \bibinfo {pages} {1--4}\BibitemShut {NoStop}%
\bibitem [{\citenamefont {Porro}\ \emph {et~al.}(2021)\citenamefont {Porro},
  \citenamefont {Andricek}, \citenamefont {Aschauer}, \citenamefont {Castoldi},
  \citenamefont {Donato}, \citenamefont {Engelke}, \citenamefont {Erdinger},
  \citenamefont {Fiorini}, \citenamefont {Fischer}, \citenamefont {Graafsma},
  \citenamefont {Grande}, \citenamefont {Guazzoni}, \citenamefont {Hansen},
  \citenamefont {Hauf}, \citenamefont {Kalavakuru}, \citenamefont {Klaer},
  \citenamefont {Tangl}, \citenamefont {Kugel}, \citenamefont {Kuster},
  \citenamefont {Lechner}, \citenamefont {Lomidze}, \citenamefont
  {Maffessanti}, \citenamefont {Manghisoni}, \citenamefont {Nidhi},
  \citenamefont {Okrent}, \citenamefont {Re}, \citenamefont {Reckleben},
  \citenamefont {Riceputi}, \citenamefont {Richter}, \citenamefont {Samartsev},
  \citenamefont {Schlee}, \citenamefont {Soldat}, \citenamefont {Strüder},
  \citenamefont {Szymanski}, \citenamefont {Turcato}, \citenamefont
  {Weidenspointner},\ and\ \citenamefont {Wunderer}}]{Porro:NS68:1334}%
  \BibitemOpen
  \bibfield  {author} {\bibinfo {author} {\bibfnamefont {M.}~\bibnamefont
  {Porro}}, \bibinfo {author} {\bibfnamefont {L.}~\bibnamefont {Andricek}},
  \bibinfo {author} {\bibfnamefont {S.}~\bibnamefont {Aschauer}}, \bibinfo
  {author} {\bibfnamefont {A.}~\bibnamefont {Castoldi}}, \bibinfo {author}
  {\bibfnamefont {M.}~\bibnamefont {Donato}}, \bibinfo {author} {\bibfnamefont
  {J.}~\bibnamefont {Engelke}}, \bibinfo {author} {\bibfnamefont
  {F.}~\bibnamefont {Erdinger}}, \bibinfo {author} {\bibfnamefont
  {C.}~\bibnamefont {Fiorini}}, \bibinfo {author} {\bibfnamefont
  {P.}~\bibnamefont {Fischer}}, \bibinfo {author} {\bibfnamefont
  {H.}~\bibnamefont {Graafsma}}, \bibinfo {author} {\bibfnamefont
  {A.}~\bibnamefont {Grande}}, \bibinfo {author} {\bibfnamefont
  {C.}~\bibnamefont {Guazzoni}}, \bibinfo {author} {\bibfnamefont
  {K.}~\bibnamefont {Hansen}}, \bibinfo {author} {\bibfnamefont
  {S.}~\bibnamefont {Hauf}}, \bibinfo {author} {\bibfnamefont {P.}~\bibnamefont
  {Kalavakuru}}, \bibinfo {author} {\bibfnamefont {H.}~\bibnamefont {Klaer}},
  \bibinfo {author} {\bibfnamefont {M.}~\bibnamefont {Tangl}}, \bibinfo
  {author} {\bibfnamefont {A.}~\bibnamefont {Kugel}}, \bibinfo {author}
  {\bibfnamefont {M.}~\bibnamefont {Kuster}}, \bibinfo {author} {\bibfnamefont
  {P.}~\bibnamefont {Lechner}}, \bibinfo {author} {\bibfnamefont
  {D.}~\bibnamefont {Lomidze}}, \bibinfo {author} {\bibfnamefont
  {S.}~\bibnamefont {Maffessanti}}, \bibinfo {author} {\bibfnamefont
  {M.}~\bibnamefont {Manghisoni}}, \bibinfo {author} {\bibfnamefont
  {S.}~\bibnamefont {Nidhi}}, \bibinfo {author} {\bibfnamefont
  {F.}~\bibnamefont {Okrent}}, \bibinfo {author} {\bibfnamefont
  {V.}~\bibnamefont {Re}}, \bibinfo {author} {\bibfnamefont {C.}~\bibnamefont
  {Reckleben}}, \bibinfo {author} {\bibfnamefont {E.}~\bibnamefont {Riceputi}},
  \bibinfo {author} {\bibfnamefont {R.}~\bibnamefont {Richter}}, \bibinfo
  {author} {\bibfnamefont {A.}~\bibnamefont {Samartsev}}, \bibinfo {author}
  {\bibfnamefont {S.}~\bibnamefont {Schlee}}, \bibinfo {author} {\bibfnamefont
  {J.}~\bibnamefont {Soldat}}, \bibinfo {author} {\bibfnamefont
  {L.}~\bibnamefont {Strüder}}, \bibinfo {author} {\bibfnamefont
  {J.}~\bibnamefont {Szymanski}}, \bibinfo {author} {\bibfnamefont
  {M.}~\bibnamefont {Turcato}}, \bibinfo {author} {\bibfnamefont
  {G.}~\bibnamefont {Weidenspointner}},\ and\ \bibinfo {author} {\bibfnamefont
  {C.~B.}\ \bibnamefont {Wunderer}},\ }\bibfield  {title} {\bibinfo {title}
  {{The MiniSDD-Based 1-Mpixel Camera of the DSSC Project for the European
  XFEL}},\ }\href {https://doi.org/10.1109/TNS.2021.3076602} {\bibfield
  {journal} {\bibinfo  {journal} {IEEE\ Trans. Nucl. Sci.}\ }\textbf {\bibinfo
  {volume} {68}},\ \bibinfo {pages} {1334} (\bibinfo {year}
  {2021})}\BibitemShut {NoStop}%
\bibitem [{\citenamefont {Whitaker}(2003)}]{Whitaker:Imaging:2003}%
  \BibitemOpen
  \bibfield  {author} {\bibinfo {author} {\bibfnamefont {B.}~\bibnamefont
  {Whitaker}},\ }\href@noop {} {\emph {\bibinfo {title} {Imaging in Molecular
  Dynamics: Technology and Applications}}}\ (\bibinfo  {publisher} {Cambridge
  University Press},\ \bibinfo {address} {missing},\ \bibinfo {year}
  {2003})\BibitemShut {NoStop}%
\bibitem [{\citenamefont {Chandler}\ and\ \citenamefont
  {Houston}(1987)}]{Chandler:JCP87:1445}%
  \BibitemOpen
  \bibfield  {author} {\bibinfo {author} {\bibfnamefont {D.}~\bibnamefont
  {Chandler}}\ and\ \bibinfo {author} {\bibfnamefont {P.}~\bibnamefont
  {Houston}},\ }\bibfield  {title} {\bibinfo {title} {{Two-dimensional Imaging
  Of State-selected Photodissociation Products Detected By Multiphoton
  Ionization}},\ }\href {https://doi.org/10.1063/1.453276} {\bibfield
  {journal} {\bibinfo  {journal} {J. Chem. Phys.}\ }\textbf {\bibinfo {volume}
  {87}},\ \bibinfo {pages} {1445} (\bibinfo {year} {1987})}\BibitemShut
  {NoStop}%
\bibitem [{\citenamefont {Eppink}\ and\ \citenamefont
  {Parker}(1997)}]{Eppink:RSI68:3477}%
  \BibitemOpen
  \bibfield  {author} {\bibinfo {author} {\bibfnamefont {A.~T. J.~B.}\
  \bibnamefont {Eppink}}\ and\ \bibinfo {author} {\bibfnamefont {D.~H.}\
  \bibnamefont {Parker}},\ }\bibfield  {title} {\bibinfo {title} {Velocity map
  imaging of ions and electrons using electrostatic lenses: Application in
  photoelectron and photofragment ion imaging of molecular oxygen},\ }\href
  {https://doi.org/10.1063/1.1148310} {\bibfield  {journal} {\bibinfo
  {journal} {Rev. Sci. Instrum.}\ }\textbf {\bibinfo {volume} {68}},\ \bibinfo
  {pages} {3477} (\bibinfo {year} {1997})}\BibitemShut {NoStop}%
\bibitem [{\citenamefont {Chandler}\ \emph {et~al.}(2017)\citenamefont
  {Chandler}, \citenamefont {Houston},\ and\ \citenamefont
  {Parker}}]{Chandler:JCP147:013601}%
  \BibitemOpen
  \bibfield  {author} {\bibinfo {author} {\bibfnamefont {D.~W.}\ \bibnamefont
  {Chandler}}, \bibinfo {author} {\bibfnamefont {P.~L.}\ \bibnamefont
  {Houston}},\ and\ \bibinfo {author} {\bibfnamefont {D.~H.}\ \bibnamefont
  {Parker}},\ }\bibfield  {title} {\bibinfo {title} {Perspective: Advanced
  particle imaging},\ }\href {https://doi.org/10.1063/1.4983623} {\bibfield
  {journal} {\bibinfo  {journal} {The Journal of Chemical Physics}\ }\textbf
  {\bibinfo {volume} {147}},\ \bibinfo {pages} {013601} (\bibinfo {year}
  {2017})}\BibitemShut {NoStop}%
\bibitem [{\citenamefont {Suits}\ and\ \citenamefont
  {Continetti}(2000)}]{Suits:ACS770:1}%
  \BibitemOpen
  \bibfield  {author} {\bibinfo {author} {\bibfnamefont {A.~G.}\ \bibnamefont
  {Suits}}\ and\ \bibinfo {author} {\bibfnamefont {R.~E.}\ \bibnamefont
  {Continetti}},\ }\bibinfo {title} {Imaging in chemical dynamics: The state of
  the art},\ in\ \href {https://doi.org/10.1021/bk-2001-0770.ch001} {\emph
  {\bibinfo {booktitle} {Imaging in Chemical Dynamics}}},\ \bibinfo {series}
  {ACS Symposium Series}, Vol.\ \bibinfo {volume} {770}\ (\bibinfo  {publisher}
  {American Chemical Society},\ \bibinfo {year} {2000})\ p.\ \bibinfo {pages}
  {1–18}\BibitemShut {NoStop}%
\bibitem [{\citenamefont {Reid}(2012)}]{Reid:MolPhys3:131}%
  \BibitemOpen
  \bibfield  {author} {\bibinfo {author} {\bibfnamefont {K.~L.}\ \bibnamefont
  {Reid}},\ }\bibfield  {title} {\bibinfo {title} {Photoelectron angular
  distributions: developments in applications to isolated molecular systems},\
  }\href {https://doi.org/10.1080/00268976.2011.640292} {\bibfield  {journal}
  {\bibinfo  {journal} {Mol. Phys.}\ }\textbf {\bibinfo {volume} {3}},\
  \bibinfo {pages} {131} (\bibinfo {year} {2012})}\BibitemShut {NoStop}%
\bibitem [{\citenamefont {Takahashi}\ \emph {et~al.}(2000)\citenamefont
  {Takahashi}, \citenamefont {Cave},\ and\ \citenamefont
  {Eland}}]{Takahashi:RSI71:1337}%
  \BibitemOpen
  \bibfield  {author} {\bibinfo {author} {\bibfnamefont {M.}~\bibnamefont
  {Takahashi}}, \bibinfo {author} {\bibfnamefont {J.~P.}\ \bibnamefont
  {Cave}},\ and\ \bibinfo {author} {\bibfnamefont {J.~H.~D.}\ \bibnamefont
  {Eland}},\ }\bibfield  {title} {\bibinfo {title} {Velocity imaging
  photoionization coincidence apparatus for the study of angular correlations
  between electrons and fragment ions},\ }\href
  {https://doi.org/10.1063/1.1150460} {\bibfield  {journal} {\bibinfo
  {journal} {Rev. Sci. Instrum.}\ }\textbf {\bibinfo {volume} {71}},\ \bibinfo
  {pages} {1337} (\bibinfo {year} {2000})},\ \Eprint
  {https://arxiv.org/abs/https://doi.org/10.1063/1.1150460}
  {https://doi.org/10.1063/1.1150460} \BibitemShut {NoStop}%
\bibitem [{\citenamefont {Larsen}\ \emph {et~al.}(1998)\citenamefont {Larsen},
  \citenamefont {M{\o}rkbak}, \citenamefont {Olesen}, \citenamefont {Bjerre},
  \citenamefont {Machholm}, \citenamefont {Keiding},\ and\ \citenamefont
  {Stapelfeldt}}]{Larsen:JCP109:8857}%
  \BibitemOpen
  \bibfield  {author} {\bibinfo {author} {\bibfnamefont {J.~J.}\ \bibnamefont
  {Larsen}}, \bibinfo {author} {\bibfnamefont {N.~J.}\ \bibnamefont
  {M{\o}rkbak}}, \bibinfo {author} {\bibfnamefont {J.}~\bibnamefont {Olesen}},
  \bibinfo {author} {\bibfnamefont {N.}~\bibnamefont {Bjerre}}, \bibinfo
  {author} {\bibfnamefont {M.}~\bibnamefont {Machholm}}, \bibinfo {author}
  {\bibfnamefont {S.~R.}\ \bibnamefont {Keiding}},\ and\ \bibinfo {author}
  {\bibfnamefont {H.}~\bibnamefont {Stapelfeldt}},\ }\bibfield  {title}
  {\bibinfo {title} {Femtosecond photodissociation dynamics of $\text{I}_2$
  studied by ion imaging},\ }\href {https://doi.org/10.1063/1.477557}
  {\bibfield  {journal} {\bibinfo  {journal} {J. Chem. Phys.}\ }\textbf
  {\bibinfo {volume} {109}},\ \bibinfo {pages} {8857} (\bibinfo {year}
  {1998})},\ \Eprint {https://arxiv.org/abs/https://doi.org/10.1063/1.477557}
  {https://doi.org/10.1063/1.477557} \BibitemShut {NoStop}%
\bibitem [{\citenamefont {Roeterdink}\ and\ \citenamefont
  {Janssen}(2002)}]{Roeterdink:PCCP4:601}%
  \BibitemOpen
  \bibfield  {author} {\bibinfo {author} {\bibfnamefont {W.~G.}\ \bibnamefont
  {Roeterdink}}\ and\ \bibinfo {author} {\bibfnamefont {M.~H.~M.}\ \bibnamefont
  {Janssen}},\ }\bibfield  {title} {\bibinfo {title} {{Femtosecond velocity map
  imaging of dissociative ionization dynamics in CF$_3$I}},\ }\href
  {https://doi.org/10.1039/b107897f} {\bibfield  {journal} {\bibinfo  {journal}
  {Phys. Chem. Chem. Phys.}\ }\textbf {\bibinfo {volume} {4}},\ \bibinfo
  {pages} {601} (\bibinfo {year} {2002})}\BibitemShut {NoStop}%
\bibitem [{\citenamefont {Trippel}\ \emph {et~al.}(2013)\citenamefont
  {Trippel}, \citenamefont {Mullins}, \citenamefont {M{\"u}ller}, \citenamefont
  {Kienitz}, \citenamefont {D{\l}ugo{\l}\k{e}cki},\ and\ \citenamefont
  {K{\"u}pper}}]{Trippel:MP111:1738}%
  \BibitemOpen
  \bibfield  {author} {\bibinfo {author} {\bibfnamefont {S.}~\bibnamefont
  {Trippel}}, \bibinfo {author} {\bibfnamefont {T.~G.}\ \bibnamefont
  {Mullins}}, \bibinfo {author} {\bibfnamefont {N.~L.~M.}\ \bibnamefont
  {M{\"u}ller}}, \bibinfo {author} {\bibfnamefont {J.~S.}\ \bibnamefont
  {Kienitz}}, \bibinfo {author} {\bibfnamefont {K.}~\bibnamefont
  {D{\l}ugo{\l}\k{e}cki}},\ and\ \bibinfo {author} {\bibfnamefont
  {J.}~\bibnamefont {K{\"u}pper}},\ }\bibfield  {title} {\bibinfo {title}
  {Strongly aligned and oriented molecular samples at a {kHz} repetition
  rate},\ }\href {https://doi.org/10.1080/00268976.2013.780334} {\bibfield
  {journal} {\bibinfo  {journal} {Mol. Phys.}\ }\textbf {\bibinfo {volume}
  {111}},\ \bibinfo {pages} {1738} (\bibinfo {year} {2013})},\ \Eprint
  {https://arxiv.org/abs/1301.1826} {arXiv:1301.1826 [physics]} \BibitemShut
  {NoStop}%
\bibitem [{\citenamefont {Aseyev}\ \emph {et~al.}(2003)\citenamefont {Aseyev},
  \citenamefont {Ni}, \citenamefont {Frasinski}, \citenamefont {Muller},\ and\
  \citenamefont {Vrakking}}]{Aseyev:PRL91:223902}%
  \BibitemOpen
  \bibfield  {author} {\bibinfo {author} {\bibfnamefont {S.~A.}\ \bibnamefont
  {Aseyev}}, \bibinfo {author} {\bibfnamefont {Y.}~\bibnamefont {Ni}}, \bibinfo
  {author} {\bibfnamefont {L.~J.}\ \bibnamefont {Frasinski}}, \bibinfo {author}
  {\bibfnamefont {H.~G.}\ \bibnamefont {Muller}},\ and\ \bibinfo {author}
  {\bibfnamefont {M.}~\bibnamefont {Vrakking}},\ }\bibfield  {title} {\bibinfo
  {title} {{Attosecond angle-resolved photoelectron spectroscopy}},\ }\href
  {https://doi.org/10.1103/physrevlett.91.223902} {\bibfield  {journal}
  {\bibinfo  {journal} {Phys. Rev. Lett.}\ }\textbf {\bibinfo {volume} {91}},\
  \bibinfo {pages} {223902} (\bibinfo {year} {2003})}\BibitemShut {NoStop}%
\bibitem [{\citenamefont {Rading}\ \emph {et~al.}(2018)\citenamefont {Rading},
  \citenamefont {Lahl}, \citenamefont {Maclot}, \citenamefont {Campi},
  \citenamefont {Coudert-Alteirac}, \citenamefont {Oostenrijk}, \citenamefont
  {Peschel}, \citenamefont {Wikmark}, \citenamefont {Rudawski}, \citenamefont
  {Gisselbrecht},\ and\ \citenamefont {Johnsson}}]{Rading:AS8:998}%
  \BibitemOpen
  \bibfield  {author} {\bibinfo {author} {\bibfnamefont {L.}~\bibnamefont
  {Rading}}, \bibinfo {author} {\bibfnamefont {J.}~\bibnamefont {Lahl}},
  \bibinfo {author} {\bibfnamefont {S.}~\bibnamefont {Maclot}}, \bibinfo
  {author} {\bibfnamefont {F.}~\bibnamefont {Campi}}, \bibinfo {author}
  {\bibfnamefont {H.}~\bibnamefont {Coudert-Alteirac}}, \bibinfo {author}
  {\bibfnamefont {B.}~\bibnamefont {Oostenrijk}}, \bibinfo {author}
  {\bibfnamefont {J.}~\bibnamefont {Peschel}}, \bibinfo {author} {\bibfnamefont
  {H.}~\bibnamefont {Wikmark}}, \bibinfo {author} {\bibfnamefont
  {P.}~\bibnamefont {Rudawski}}, \bibinfo {author} {\bibfnamefont
  {M.}~\bibnamefont {Gisselbrecht}},\ and\ \bibinfo {author} {\bibfnamefont
  {P.}~\bibnamefont {Johnsson}},\ }\bibfield  {title} {\bibinfo {title} {A
  versatile velocity map ion-electron covariance imaging spectrometer for
  high-intensity {XUV} experiments},\ }\href
  {https://doi.org/10.3390/app8060998} {\bibfield  {journal} {\bibinfo
  {journal} {Appl. Sci.}\ }\textbf {\bibinfo {volume} {8}},\ \bibinfo {pages}
  {998} (\bibinfo {year} {2018})},\ \Eprint {https://arxiv.org/abs/1901.03077}
  {arXiv:1901.03077 [physics]} \BibitemShut {NoStop}%
\bibitem [{\citenamefont {Kling}\ \emph {et~al.}(2014)\citenamefont {Kling},
  \citenamefont {Paul}, \citenamefont {Gura}, \citenamefont {Laurent},
  \citenamefont {De}, \citenamefont {Li}, \citenamefont {Wang}, \citenamefont
  {Ahn}, \citenamefont {Kim}, \citenamefont {Kim}, \citenamefont {Litvinyuk},
  \citenamefont {Cocke}, \citenamefont {Ben-Itzhak}, \citenamefont {Kim},\ and\
  \citenamefont {Kling}}]{Kling:JInst9:P05005}%
  \BibitemOpen
  \bibfield  {author} {\bibinfo {author} {\bibfnamefont {N.~G.}\ \bibnamefont
  {Kling}}, \bibinfo {author} {\bibfnamefont {D.}~\bibnamefont {Paul}},
  \bibinfo {author} {\bibfnamefont {A.}~\bibnamefont {Gura}}, \bibinfo {author}
  {\bibfnamefont {G.}~\bibnamefont {Laurent}}, \bibinfo {author} {\bibfnamefont
  {S.}~\bibnamefont {De}}, \bibinfo {author} {\bibfnamefont {H.}~\bibnamefont
  {Li}}, \bibinfo {author} {\bibfnamefont {Z.}~\bibnamefont {Wang}}, \bibinfo
  {author} {\bibfnamefont {B.}~\bibnamefont {Ahn}}, \bibinfo {author}
  {\bibfnamefont {C.~H.}\ \bibnamefont {Kim}}, \bibinfo {author} {\bibfnamefont
  {T.~K.}\ \bibnamefont {Kim}}, \bibinfo {author} {\bibfnamefont {I.~V.}\
  \bibnamefont {Litvinyuk}}, \bibinfo {author} {\bibfnamefont {C.~L.}\
  \bibnamefont {Cocke}}, \bibinfo {author} {\bibfnamefont {I.}~\bibnamefont
  {Ben-Itzhak}}, \bibinfo {author} {\bibfnamefont {D.}~\bibnamefont {Kim}},\
  and\ \bibinfo {author} {\bibfnamefont {M.~F.}\ \bibnamefont {Kling}},\
  }\bibfield  {title} {\bibinfo {title} {{Thick-lens velocity-map imaging
  spectrometer with high resolution for high-energy charged particles}},\
  }\href {https://doi.org/10.1088/1748-0221/9/05/p05005} {\bibfield  {journal}
  {\bibinfo  {journal} {J. Instrum.}\ }\textbf {\bibinfo {volume} {9}}\bibinfo
  {number} { (05)},\ \bibinfo {pages} {P05005}}\BibitemShut {NoStop}%
\bibitem [{\citenamefont {Garcia}\ \emph {et~al.}(2005)\citenamefont {Garcia},
  \citenamefont {Nahon}, \citenamefont {Harding}, \citenamefont {Mikajlo},\
  and\ \citenamefont {Powis}}]{Garcia:RSI76:053302}%
  \BibitemOpen
\bibfield  {number} {  }\bibfield  {author} {\bibinfo {author} {\bibfnamefont
  {G.~A.}\ \bibnamefont {Garcia}}, \bibinfo {author} {\bibfnamefont
  {L.}~\bibnamefont {Nahon}}, \bibinfo {author} {\bibfnamefont {C.~J.}\
  \bibnamefont {Harding}}, \bibinfo {author} {\bibfnamefont {E.~A.}\
  \bibnamefont {Mikajlo}},\ and\ \bibinfo {author} {\bibfnamefont
  {I.}~\bibnamefont {Powis}},\ }\bibfield  {title} {\bibinfo {title} {A
  refocusing modified velocity map imaging electron/ion spectrometer adapted to
  synchrotron radiation studies},\ }\href {https://doi.org/10.1063/1.1900646}
  {\bibfield  {journal} {\bibinfo  {journal} {Rev. Sci. Instrum.}\ }\textbf
  {\bibinfo {volume} {76}},\ \bibinfo {pages} {053302} (\bibinfo {year}
  {2005})}\BibitemShut {NoStop}%
\bibitem [{\citenamefont {Rolles}\ \emph {et~al.}(2007)\citenamefont {Rolles},
  \citenamefont {Pešić}, \citenamefont {Perri}, \citenamefont {Bilodeau},
  \citenamefont {Ackerman}, \citenamefont {Rude}, \citenamefont {Kilcoyne},
  \citenamefont {Bozek},\ and\ \citenamefont {Berrah}}]{Rolles:NIMB261:170}%
  \BibitemOpen
  \bibfield  {author} {\bibinfo {author} {\bibfnamefont {D.}~\bibnamefont
  {Rolles}}, \bibinfo {author} {\bibfnamefont {Z.}~\bibnamefont {Pešić}},
  \bibinfo {author} {\bibfnamefont {M.}~\bibnamefont {Perri}}, \bibinfo
  {author} {\bibfnamefont {R.}~\bibnamefont {Bilodeau}}, \bibinfo {author}
  {\bibfnamefont {G.}~\bibnamefont {Ackerman}}, \bibinfo {author}
  {\bibfnamefont {B.}~\bibnamefont {Rude}}, \bibinfo {author} {\bibfnamefont
  {A.}~\bibnamefont {Kilcoyne}}, \bibinfo {author} {\bibfnamefont
  {J.}~\bibnamefont {Bozek}},\ and\ \bibinfo {author} {\bibfnamefont
  {N.}~\bibnamefont {Berrah}},\ }\bibfield  {title} {\bibinfo {title} {A
  velocity map imaging spectrometer for electron–ion and ion–ion
  coincidence experiments with synchrotron radiation},\ }\href
  {https://doi.org/10.1016/j.nimb.2007.04.186} {\bibfield  {journal} {\bibinfo
  {journal} {Nucl. Instrum. Meth. B}\ }\textbf {\bibinfo {volume} {261}},\
  \bibinfo {pages} {170} (\bibinfo {year} {2007})},\ \bibinfo {note} {the
  Application of Accelerators in Research and Industry}\BibitemShut {NoStop}%
\bibitem [{\citenamefont {Johnsson}\ \emph {et~al.}(2008)\citenamefont
  {Johnsson}, \citenamefont {Siu}, \citenamefont {Gijsbertsen}, \citenamefont
  {Verhoeven}, \citenamefont {Meijer}, \citenamefont {van~der Zande},\ and\
  \citenamefont {Vrakking}}]{Johnsson:JModOpt55:2693}%
  \BibitemOpen
  \bibfield  {author} {\bibinfo {author} {\bibfnamefont {P.}~\bibnamefont
  {Johnsson}}, \bibinfo {author} {\bibfnamefont {W.}~\bibnamefont {Siu}},
  \bibinfo {author} {\bibfnamefont {A.}~\bibnamefont {Gijsbertsen}}, \bibinfo
  {author} {\bibfnamefont {J.}~\bibnamefont {Verhoeven}}, \bibinfo {author}
  {\bibfnamefont {A.~S.}\ \bibnamefont {Meijer}}, \bibinfo {author}
  {\bibfnamefont {W.}~\bibnamefont {van~der Zande}},\ and\ \bibinfo {author}
  {\bibfnamefont {M.~J.~J.}\ \bibnamefont {Vrakking}},\ }\bibfield  {title}
  {\bibinfo {title} {Velocity map imaging of atomic and molecular processes at
  the free electron laser in {H}amburg {(FLASH)}},\ }\href
  {https://doi.org/10.1080/09500340802393062} {\bibfield  {journal} {\bibinfo
  {journal} {J Mod Opt}\ }\textbf {\bibinfo {volume} {55}},\ \bibinfo {pages}
  {2693} (\bibinfo {year} {2008})}\BibitemShut {NoStop}%
\bibitem [{\citenamefont {Burt}\ \emph {et~al.}(2017)\citenamefont {Burt},
  \citenamefont {Boll}, \citenamefont {Lee}, \citenamefont {Amini},
  \citenamefont {K{\"o}ckert}, \citenamefont {Vallance}, \citenamefont
  {Gentleman}, \citenamefont {Mackenzie}, \citenamefont {Bari}, \citenamefont
  {Bomme}, \citenamefont {D{\"u}sterer}, \citenamefont {Erk}, \citenamefont
  {Manschwetus}, \citenamefont {M{\"u}ller}, \citenamefont {Rompotis},
  \citenamefont {Savelyev}, \citenamefont {Schirmel}, \citenamefont {Techert},
  \citenamefont {Treusch}, \citenamefont {K{\"u}pper}, \citenamefont {Trippel},
  \citenamefont {Wiese}, \citenamefont {Stapelfeldt}, \citenamefont {Cunha~de
  Miranda}, \citenamefont {Guillemin}, \citenamefont {Ismail}, \citenamefont
  {Journel}, \citenamefont {Marchenko}, \citenamefont {Palaudoux},
  \citenamefont {Penent}, \citenamefont {Piancastelli}, \citenamefont {Simon},
  \citenamefont {Travnikova}, \citenamefont {Brau{\ss}e}, \citenamefont
  {Goldsztejn}, \citenamefont {Rouz{\'e}e}, \citenamefont {G{\'e}l{\'e}oc},
  \citenamefont {Geneaux}, \citenamefont {Ruchon}, \citenamefont {Underwood},
  \citenamefont {Holland}, \citenamefont {Mereshchenko}, \citenamefont
  {Olshin}, \citenamefont {Johnsson}, \citenamefont {Maclot}, \citenamefont
  {Lahl}, \citenamefont {Rudenko}, \citenamefont {Ziaee}, \citenamefont
  {Brouard},\ and\ \citenamefont {Rolles}}]{Burt:PRA96:043415}%
  \BibitemOpen
  \bibfield  {author} {\bibinfo {author} {\bibfnamefont {M.}~\bibnamefont
  {Burt}}, \bibinfo {author} {\bibfnamefont {R.}~\bibnamefont {Boll}}, \bibinfo
  {author} {\bibfnamefont {J.~W.~L.}\ \bibnamefont {Lee}}, \bibinfo {author}
  {\bibfnamefont {K.}~\bibnamefont {Amini}}, \bibinfo {author} {\bibfnamefont
  {H.}~\bibnamefont {K{\"o}ckert}}, \bibinfo {author} {\bibfnamefont
  {C.}~\bibnamefont {Vallance}}, \bibinfo {author} {\bibfnamefont {A.~S.}\
  \bibnamefont {Gentleman}}, \bibinfo {author} {\bibfnamefont {S.~R.}\
  \bibnamefont {Mackenzie}}, \bibinfo {author} {\bibfnamefont {S.}~\bibnamefont
  {Bari}}, \bibinfo {author} {\bibfnamefont {C.}~\bibnamefont {Bomme}},
  \bibinfo {author} {\bibfnamefont {S.}~\bibnamefont {D{\"u}sterer}}, \bibinfo
  {author} {\bibfnamefont {B.}~\bibnamefont {Erk}}, \bibinfo {author}
  {\bibfnamefont {B.}~\bibnamefont {Manschwetus}}, \bibinfo {author}
  {\bibfnamefont {E.}~\bibnamefont {M{\"u}ller}}, \bibinfo {author}
  {\bibfnamefont {D.}~\bibnamefont {Rompotis}}, \bibinfo {author}
  {\bibfnamefont {E.}~\bibnamefont {Savelyev}}, \bibinfo {author}
  {\bibfnamefont {N.}~\bibnamefont {Schirmel}}, \bibinfo {author}
  {\bibfnamefont {S.}~\bibnamefont {Techert}}, \bibinfo {author} {\bibfnamefont
  {R.}~\bibnamefont {Treusch}}, \bibinfo {author} {\bibfnamefont
  {J.}~\bibnamefont {K{\"u}pper}}, \bibinfo {author} {\bibfnamefont
  {S.}~\bibnamefont {Trippel}}, \bibinfo {author} {\bibfnamefont
  {J.}~\bibnamefont {Wiese}}, \bibinfo {author} {\bibfnamefont
  {H.}~\bibnamefont {Stapelfeldt}}, \bibinfo {author} {\bibfnamefont
  {B.}~\bibnamefont {Cunha~de Miranda}}, \bibinfo {author} {\bibfnamefont
  {R.}~\bibnamefont {Guillemin}}, \bibinfo {author} {\bibfnamefont
  {I.}~\bibnamefont {Ismail}}, \bibinfo {author} {\bibfnamefont
  {L.}~\bibnamefont {Journel}}, \bibinfo {author} {\bibfnamefont
  {T.}~\bibnamefont {Marchenko}}, \bibinfo {author} {\bibfnamefont
  {J.}~\bibnamefont {Palaudoux}}, \bibinfo {author} {\bibfnamefont
  {F.}~\bibnamefont {Penent}}, \bibinfo {author} {\bibfnamefont {M.~N.}\
  \bibnamefont {Piancastelli}}, \bibinfo {author} {\bibfnamefont
  {M.}~\bibnamefont {Simon}}, \bibinfo {author} {\bibfnamefont
  {O.}~\bibnamefont {Travnikova}}, \bibinfo {author} {\bibfnamefont
  {F.}~\bibnamefont {Brau{\ss}e}}, \bibinfo {author} {\bibfnamefont
  {G.}~\bibnamefont {Goldsztejn}}, \bibinfo {author} {\bibfnamefont
  {A.}~\bibnamefont {Rouz{\'e}e}}, \bibinfo {author} {\bibfnamefont
  {M.}~\bibnamefont {G{\'e}l{\'e}oc}}, \bibinfo {author} {\bibfnamefont
  {R.}~\bibnamefont {Geneaux}}, \bibinfo {author} {\bibfnamefont
  {T.}~\bibnamefont {Ruchon}}, \bibinfo {author} {\bibfnamefont
  {J.}~\bibnamefont {Underwood}}, \bibinfo {author} {\bibfnamefont {D.~M.~P.}\
  \bibnamefont {Holland}}, \bibinfo {author} {\bibfnamefont {A.~S.}\
  \bibnamefont {Mereshchenko}}, \bibinfo {author} {\bibfnamefont {P.~K.}\
  \bibnamefont {Olshin}}, \bibinfo {author} {\bibfnamefont {P.}~\bibnamefont
  {Johnsson}}, \bibinfo {author} {\bibfnamefont {S.}~\bibnamefont {Maclot}},
  \bibinfo {author} {\bibfnamefont {J.}~\bibnamefont {Lahl}}, \bibinfo {author}
  {\bibfnamefont {A.}~\bibnamefont {Rudenko}}, \bibinfo {author} {\bibfnamefont
  {F.}~\bibnamefont {Ziaee}}, \bibinfo {author} {\bibfnamefont
  {M.}~\bibnamefont {Brouard}},\ and\ \bibinfo {author} {\bibfnamefont
  {D.}~\bibnamefont {Rolles}},\ }\bibfield  {title} {\bibinfo {title}
  {Coulomb-explosion imaging of concurrent {CH$_2$BrI} photodissociation
  dynamics},\ }\href {https://doi.org/10.1103/PhysRevA.96.043415} {\bibfield
  {journal} {\bibinfo  {journal} {Phys. Rev. A}\ }\textbf {\bibinfo {volume}
  {96}},\ \bibinfo {pages} {043415} (\bibinfo {year} {2017})}\BibitemShut
  {NoStop}%
\bibitem [{\citenamefont {Erk}\ \emph {et~al.}(2018)\citenamefont {Erk},
  \citenamefont {M{\"{u}}ller}, \citenamefont {Bomme}, \citenamefont {Boll},
  \citenamefont {Brenner}, \citenamefont {Chapman}, \citenamefont {Correa},
  \citenamefont {D{\"{u}}sterer}, \citenamefont {Dziarzhytski}, \citenamefont
  {Eisebitt}, \citenamefont {Graafsma}, \citenamefont {Grunewald},
  \citenamefont {Gumprecht}, \citenamefont {Hartmann}, \citenamefont {Hauser},
  \citenamefont {Keitel}, \citenamefont {von Korff~Schmising}, \citenamefont
  {Kuhlmann}, \citenamefont {Manschwetus}, \citenamefont {Mercadier},
  \citenamefont {M{\"{u}}ller}, \citenamefont {Passow}, \citenamefont
  {Pl{\"{o}}njes}, \citenamefont {Ramm}, \citenamefont {Rompotis},
  \citenamefont {Rudenko}, \citenamefont {Rupp}, \citenamefont {Sauppe},
  \citenamefont {Siewert}, \citenamefont {Schlosser}, \citenamefont
  {Str{\"{u}}der}, \citenamefont {Swiderski}, \citenamefont {Techert},
  \citenamefont {Tiedtke}, \citenamefont {Tilp}, \citenamefont {Treusch},
  \citenamefont {Schlichting}, \citenamefont {Ullrich}, \citenamefont
  {Moshammer}, \citenamefont {M{\"{o}}ller},\ and\ \citenamefont
  {Rolles}}]{Erk:JSR25:1529}%
  \BibitemOpen
  \bibfield  {author} {\bibinfo {author} {\bibfnamefont {B.}~\bibnamefont
  {Erk}}, \bibinfo {author} {\bibfnamefont {J.~P.}\ \bibnamefont
  {M{\"{u}}ller}}, \bibinfo {author} {\bibfnamefont {C.}~\bibnamefont {Bomme}},
  \bibinfo {author} {\bibfnamefont {R.}~\bibnamefont {Boll}}, \bibinfo {author}
  {\bibfnamefont {G.}~\bibnamefont {Brenner}}, \bibinfo {author} {\bibfnamefont
  {H.~N.}\ \bibnamefont {Chapman}}, \bibinfo {author} {\bibfnamefont
  {J.}~\bibnamefont {Correa}}, \bibinfo {author} {\bibfnamefont
  {S.}~\bibnamefont {D{\"{u}}sterer}}, \bibinfo {author} {\bibfnamefont
  {S.}~\bibnamefont {Dziarzhytski}}, \bibinfo {author} {\bibfnamefont
  {S.}~\bibnamefont {Eisebitt}}, \bibinfo {author} {\bibfnamefont
  {H.}~\bibnamefont {Graafsma}}, \bibinfo {author} {\bibfnamefont
  {S.}~\bibnamefont {Grunewald}}, \bibinfo {author} {\bibfnamefont
  {L.}~\bibnamefont {Gumprecht}}, \bibinfo {author} {\bibfnamefont
  {R.}~\bibnamefont {Hartmann}}, \bibinfo {author} {\bibfnamefont
  {G.}~\bibnamefont {Hauser}}, \bibinfo {author} {\bibfnamefont
  {B.}~\bibnamefont {Keitel}}, \bibinfo {author} {\bibfnamefont
  {C.}~\bibnamefont {von Korff~Schmising}}, \bibinfo {author} {\bibfnamefont
  {M.}~\bibnamefont {Kuhlmann}}, \bibinfo {author} {\bibfnamefont
  {B.}~\bibnamefont {Manschwetus}}, \bibinfo {author} {\bibfnamefont
  {L.}~\bibnamefont {Mercadier}}, \bibinfo {author} {\bibfnamefont
  {E.}~\bibnamefont {M{\"{u}}ller}}, \bibinfo {author} {\bibfnamefont
  {C.}~\bibnamefont {Passow}}, \bibinfo {author} {\bibfnamefont
  {E.}~\bibnamefont {Pl{\"{o}}njes}}, \bibinfo {author} {\bibfnamefont
  {D.}~\bibnamefont {Ramm}}, \bibinfo {author} {\bibfnamefont {D.}~\bibnamefont
  {Rompotis}}, \bibinfo {author} {\bibfnamefont {A.}~\bibnamefont {Rudenko}},
  \bibinfo {author} {\bibfnamefont {D.}~\bibnamefont {Rupp}}, \bibinfo {author}
  {\bibfnamefont {M.}~\bibnamefont {Sauppe}}, \bibinfo {author} {\bibfnamefont
  {F.}~\bibnamefont {Siewert}}, \bibinfo {author} {\bibfnamefont
  {D.}~\bibnamefont {Schlosser}}, \bibinfo {author} {\bibfnamefont
  {L.}~\bibnamefont {Str{\"{u}}der}}, \bibinfo {author} {\bibfnamefont
  {A.}~\bibnamefont {Swiderski}}, \bibinfo {author} {\bibfnamefont
  {S.}~\bibnamefont {Techert}}, \bibinfo {author} {\bibfnamefont
  {K.}~\bibnamefont {Tiedtke}}, \bibinfo {author} {\bibfnamefont
  {T.}~\bibnamefont {Tilp}}, \bibinfo {author} {\bibfnamefont {R.}~\bibnamefont
  {Treusch}}, \bibinfo {author} {\bibfnamefont {I.}~\bibnamefont
  {Schlichting}}, \bibinfo {author} {\bibfnamefont {J.}~\bibnamefont
  {Ullrich}}, \bibinfo {author} {\bibfnamefont {R.}~\bibnamefont {Moshammer}},
  \bibinfo {author} {\bibfnamefont {T.}~\bibnamefont {M{\"{o}}ller}},\ and\
  \bibinfo {author} {\bibfnamefont {D.}~\bibnamefont {Rolles}},\ }\bibfield
  {title} {\bibinfo {title} {{CAMP@FLASH}: an end-station for imaging,
  electron- and ion-spectroscopy, and pump{--}probe experiments at the {FLASH}
  free-electron laser},\ }\href {https://doi.org/10.1107/S1600577518008585}
  {\bibfield  {journal} {\bibinfo  {journal} {J. Synchrotron\ Rad.}\ }\textbf
  {\bibinfo {volume} {25}},\ \bibinfo {pages} {1529} (\bibinfo {year}
  {2018})}\BibitemShut {NoStop}%
\bibitem [{\citenamefont {Osipov}\ \emph {et~al.}(2018)\citenamefont {Osipov},
  \citenamefont {Bostedt}, \citenamefont {Castagna}, \citenamefont {Ferguson},
  \citenamefont {Bucher}, \citenamefont {Montero}, \citenamefont {Swiggers},
  \citenamefont {Obaid}, \citenamefont {Rolles}, \citenamefont {Rudenko},
  \citenamefont {Bozek},\ and\ \citenamefont {Berrah}}]{Osipov:RSI89:035112}%
  \BibitemOpen
  \bibfield  {author} {\bibinfo {author} {\bibfnamefont {T.}~\bibnamefont
  {Osipov}}, \bibinfo {author} {\bibfnamefont {C.}~\bibnamefont {Bostedt}},
  \bibinfo {author} {\bibfnamefont {J.-C.}\ \bibnamefont {Castagna}}, \bibinfo
  {author} {\bibfnamefont {K.~R.}\ \bibnamefont {Ferguson}}, \bibinfo {author}
  {\bibfnamefont {M.}~\bibnamefont {Bucher}}, \bibinfo {author} {\bibfnamefont
  {S.~C.}\ \bibnamefont {Montero}}, \bibinfo {author} {\bibfnamefont {M.~L.}\
  \bibnamefont {Swiggers}}, \bibinfo {author} {\bibfnamefont {R.}~\bibnamefont
  {Obaid}}, \bibinfo {author} {\bibfnamefont {D.}~\bibnamefont {Rolles}},
  \bibinfo {author} {\bibfnamefont {A.}~\bibnamefont {Rudenko}}, \bibinfo
  {author} {\bibfnamefont {J.~D.}\ \bibnamefont {Bozek}},\ and\ \bibinfo
  {author} {\bibfnamefont {N.}~\bibnamefont {Berrah}},\ }\bibfield  {title}
  {\bibinfo {title} {{The LAMP instrument at the Linac Coherent Light Source
  free-electron laser}},\ }\href {https://doi.org/10.1063/1.5017727} {\bibfield
   {journal} {\bibinfo  {journal} {Rev. Sci. Instrum.}\ }\textbf {\bibinfo
  {volume} {89}},\ \bibinfo {pages} {035112} (\bibinfo {year}
  {2018})}\BibitemShut {NoStop}%
\bibitem [{\citenamefont {Mikosch}\ \emph {et~al.}(2008)\citenamefont
  {Mikosch}, \citenamefont {Trippel}, \citenamefont {Eichhorn}, \citenamefont
  {Otto}, \citenamefont {Lourderaj}, \citenamefont {Zhang}, \citenamefont
  {Hase}, \citenamefont {Weidem{\"u}ller},\ and\ \citenamefont
  {Wester}}]{Mikosch:Science319:183}%
  \BibitemOpen
  \bibfield  {author} {\bibinfo {author} {\bibfnamefont {J.}~\bibnamefont
  {Mikosch}}, \bibinfo {author} {\bibfnamefont {S.}~\bibnamefont {Trippel}},
  \bibinfo {author} {\bibfnamefont {C.}~\bibnamefont {Eichhorn}}, \bibinfo
  {author} {\bibfnamefont {R.}~\bibnamefont {Otto}}, \bibinfo {author}
  {\bibfnamefont {U.}~\bibnamefont {Lourderaj}}, \bibinfo {author}
  {\bibfnamefont {J.~X.}\ \bibnamefont {Zhang}}, \bibinfo {author}
  {\bibfnamefont {W.~L.}\ \bibnamefont {Hase}}, \bibinfo {author}
  {\bibfnamefont {M.}~\bibnamefont {Weidem{\"u}ller}},\ and\ \bibinfo {author}
  {\bibfnamefont {R.}~\bibnamefont {Wester}},\ }\bibfield  {title} {\bibinfo
  {title} {Imaging nucleophilic substitution dynamics},\ }\href
  {https://doi.org/10.1126/science.1150238} {\bibfield  {journal} {\bibinfo
  {journal} {Science}\ }\textbf {\bibinfo {volume} {319}},\ \bibinfo {pages}
  {183} (\bibinfo {year} {2008})}\BibitemShut {NoStop}%
\bibitem [{\citenamefont {Chichinin}\ \emph {et~al.}(2009)\citenamefont
  {Chichinin}, \citenamefont {Gericke}, \citenamefont {Kauczok},\ and\
  \citenamefont {Maul}}]{Chichinin:IRPC28:607}%
  \BibitemOpen
  \bibfield  {author} {\bibinfo {author} {\bibfnamefont {A.~I.}\ \bibnamefont
  {Chichinin}}, \bibinfo {author} {\bibfnamefont {K.~H.}\ \bibnamefont
  {Gericke}}, \bibinfo {author} {\bibfnamefont {S.}~\bibnamefont {Kauczok}},\
  and\ \bibinfo {author} {\bibfnamefont {C.}~\bibnamefont {Maul}},\ }\bibfield
  {title} {\bibinfo {title} {Imaging chemical reactions --- 3{D} velocity
  mapping},\ }\href {https://doi.org/10.1080/01442350903235045} {\bibfield
  {journal} {\bibinfo  {journal} {Int. Rev. Phys. Chem.}\ }\textbf {\bibinfo
  {volume} {28}},\ \bibinfo {pages} {607} (\bibinfo {year} {2009})}\BibitemShut
  {NoStop}%
\bibitem [{\citenamefont {Debrah}\ \emph {et~al.}(2020)\citenamefont {Debrah},
  \citenamefont {Stewart}, \citenamefont {Basnayake}, \citenamefont
  {Nomerotski}, \citenamefont {Svihra}, \citenamefont {Lee},\ and\
  \citenamefont {Li}}]{Debrah:RSI91:023316}%
  \BibitemOpen
  \bibfield  {author} {\bibinfo {author} {\bibfnamefont {D.~A.}\ \bibnamefont
  {Debrah}}, \bibinfo {author} {\bibfnamefont {G.~A.}\ \bibnamefont {Stewart}},
  \bibinfo {author} {\bibfnamefont {G.}~\bibnamefont {Basnayake}}, \bibinfo
  {author} {\bibfnamefont {A.}~\bibnamefont {Nomerotski}}, \bibinfo {author}
  {\bibfnamefont {P.}~\bibnamefont {Svihra}}, \bibinfo {author} {\bibfnamefont
  {S.~K.}\ \bibnamefont {Lee}},\ and\ \bibinfo {author} {\bibfnamefont
  {W.}~\bibnamefont {Li}},\ }\bibfield  {title} {\bibinfo {title} {Developing a
  camera-based 3{D} momentum imaging system capable of 1~{M}hits/s},\ }\href
  {https://doi.org/10.1063/1.5138731} {\bibfield  {journal} {\bibinfo
  {journal} {Rev. Sci. Instrum.}\ }\textbf {\bibinfo {volume} {91}},\ \bibinfo
  {pages} {023316} (\bibinfo {year} {2020})}\BibitemShut {NoStop}%
\bibitem [{\citenamefont {Cheng}\ \emph {et~al.}(2020)\citenamefont {Cheng},
  \citenamefont {Forbes}, \citenamefont {Howard}, \citenamefont {Spanner},
  \citenamefont {Bucksbaum},\ and\ \citenamefont
  {Weinacht}}]{Cheng:PRA102:052813}%
  \BibitemOpen
  \bibfield  {author} {\bibinfo {author} {\bibfnamefont {C.}~\bibnamefont
  {Cheng}}, \bibinfo {author} {\bibfnamefont {R.}~\bibnamefont {Forbes}},
  \bibinfo {author} {\bibfnamefont {A.~J.}\ \bibnamefont {Howard}}, \bibinfo
  {author} {\bibfnamefont {M.}~\bibnamefont {Spanner}}, \bibinfo {author}
  {\bibfnamefont {P.~H.}\ \bibnamefont {Bucksbaum}},\ and\ \bibinfo {author}
  {\bibfnamefont {T.}~\bibnamefont {Weinacht}},\ }\bibfield  {title} {\bibinfo
  {title} {Momentum-resolved above-threshold ionization of deuterated water},\
  }\href {https://doi.org/10.1103/PhysRevA.102.052813} {\bibfield  {journal}
  {\bibinfo  {journal} {Phys. Rev. A}\ }\textbf {\bibinfo {volume} {102}},\
  \bibinfo {pages} {052813} (\bibinfo {year} {2020})},\ \bibinfo {note}
  {publisher: American Physical Society}\BibitemShut {NoStop}%
\bibitem [{\citenamefont {Oelsner}\ \emph {et~al.}(2001)\citenamefont
  {Oelsner}, \citenamefont {Schmidt}, \citenamefont {Schicketanz},
  \citenamefont {Klais}, \citenamefont {Sch\"onhense}, \citenamefont {Mergel},
  \citenamefont {Jagutzki},\ and\ \citenamefont
  {Schmidt-\"ocking}}]{Oelsner:RSI72:3968}%
  \BibitemOpen
  \bibfield  {author} {\bibinfo {author} {\bibfnamefont {A.}~\bibnamefont
  {Oelsner}}, \bibinfo {author} {\bibfnamefont {O.}~\bibnamefont {Schmidt}},
  \bibinfo {author} {\bibfnamefont {M.}~\bibnamefont {Schicketanz}}, \bibinfo
  {author} {\bibfnamefont {M.}~\bibnamefont {Klais}}, \bibinfo {author}
  {\bibfnamefont {G.}~\bibnamefont {Sch\"onhense}}, \bibinfo {author}
  {\bibfnamefont {V.}~\bibnamefont {Mergel}}, \bibinfo {author} {\bibfnamefont
  {O.}~\bibnamefont {Jagutzki}},\ and\ \bibinfo {author} {\bibfnamefont
  {H.}~\bibnamefont {Schmidt-\"ocking}},\ }\bibfield  {title} {\bibinfo {title}
  {{Microspectroscopy and imaging using a delay line detector in time-of-flight
  photoemission microscopy}},\ }\href {https://doi.org/10.1063/1.1405781}
  {\bibfield  {journal} {\bibinfo  {journal} {Rev. Sci. Instrum.}\ }\textbf
  {\bibinfo {volume} {72}},\ \bibinfo {pages} {3968} (\bibinfo {year}
  {2001})}\BibitemShut {NoStop}%
\bibitem [{\citenamefont {Chichinin}\ \emph {et~al.}(2002)\citenamefont
  {Chichinin}, \citenamefont {Einfeld}, \citenamefont {Maul},\ and\
  \citenamefont {Gericke}}]{Chichinin:RSI73:1856}%
  \BibitemOpen
  \bibfield  {author} {\bibinfo {author} {\bibfnamefont {A.~I.}\ \bibnamefont
  {Chichinin}}, \bibinfo {author} {\bibfnamefont {T.}~\bibnamefont {Einfeld}},
  \bibinfo {author} {\bibfnamefont {C.}~\bibnamefont {Maul}},\ and\ \bibinfo
  {author} {\bibfnamefont {K.-H.}\ \bibnamefont {Gericke}},\ }\bibfield
  {title} {\bibinfo {title} {Three-dimensional imaging technique for direct
  observation of the complete velocity distribution of state-selected
  photodissociation products},\ }\href {https://doi.org/10.1063/1.1453505}
  {\bibfield  {journal} {\bibinfo  {journal} {Rev. Sci. Instrum.}\ }\textbf
  {\bibinfo {volume} {73}},\ \bibinfo {pages} {1856} (\bibinfo {year}
  {2002})},\ \Eprint {https://arxiv.org/abs/https://doi.org/10.1063/1.1453505}
  {https://doi.org/10.1063/1.1453505} \BibitemShut {NoStop}%
\bibitem [{\citenamefont {Montgomery~Smith}\ \emph {et~al.}(1988)\citenamefont
  {Montgomery~Smith}, \citenamefont {Keefer},\ and\ \citenamefont
  {Sudharsanan}}]{Montgomery:JQSRT39:367}%
  \BibitemOpen
  \bibfield  {author} {\bibinfo {author} {\bibfnamefont {L.}~\bibnamefont
  {Montgomery~Smith}}, \bibinfo {author} {\bibfnamefont {D.~R.}\ \bibnamefont
  {Keefer}},\ and\ \bibinfo {author} {\bibfnamefont {S.~I.}\ \bibnamefont
  {Sudharsanan}},\ }\bibfield  {title} {\bibinfo {title} {Abel inversion using
  transform techniques},\ }\href {https://doi.org/10.1016/0022-4073(88)90101-X}
  {\bibfield  {journal} {\bibinfo  {journal} {J. Quant. Spectrosc. Radiat.
  Transf.}\ }\textbf {\bibinfo {volume} {39}},\ \bibinfo {pages} {367}
  (\bibinfo {year} {1988})}\BibitemShut {NoStop}%
\bibitem [{\citenamefont {Vrakking}(2001)}]{Vrakking:RSI72:4084}%
  \BibitemOpen
  \bibfield  {author} {\bibinfo {author} {\bibfnamefont {M.~J.~J.}\
  \bibnamefont {Vrakking}},\ }\bibfield  {title} {\bibinfo {title} {An
  iterative procedure for the inversion of two-dimensional ion/photoelectron
  imaging experiments},\ }\href {https://doi.org/10.1063/1.1406923} {\bibfield
  {journal} {\bibinfo  {journal} {Rev. Sci. Instrum.}\ }\textbf {\bibinfo
  {volume} {72}},\ \bibinfo {pages} {4084} (\bibinfo {year}
  {2001})}\BibitemShut {NoStop}%
\bibitem [{\citenamefont {Garc{\'\i}a}\ \emph {et~al.}(2004)\citenamefont
  {Garc{\'\i}a}, \citenamefont {Nahon},\ and\ \citenamefont
  {Powis}}]{Garcia:RSI75:4989}%
  \BibitemOpen
  \bibfield  {author} {\bibinfo {author} {\bibfnamefont {G.}~\bibnamefont
  {Garc{\'\i}a}}, \bibinfo {author} {\bibfnamefont {L.}~\bibnamefont {Nahon}},\
  and\ \bibinfo {author} {\bibfnamefont {I.}~\bibnamefont {Powis}},\ }\bibfield
   {title} {\bibinfo {title} {Two-dimensional charged particle image inversion
  using a polar basis function expansion},\ }\href
  {https://doi.org/10.1063/1.1807578} {\bibfield  {journal} {\bibinfo
  {journal} {Rev. Sci. Instrum.}\ }\textbf {\bibinfo {volume} {75}},\ \bibinfo
  {pages} {4989} (\bibinfo {year} {2004})}\BibitemShut {NoStop}%
\bibitem [{\citenamefont {Gebhardt}\ \emph {et~al.}(2001)\citenamefont
  {Gebhardt}, \citenamefont {Rakitzis}, \citenamefont {Samartzis},
  \citenamefont {Ladopoulos},\ and\ \citenamefont
  {Kitsopoulos}}]{Gebhardt:RSI72:3848}%
  \BibitemOpen
  \bibfield  {author} {\bibinfo {author} {\bibfnamefont {C.~R.}\ \bibnamefont
  {Gebhardt}}, \bibinfo {author} {\bibfnamefont {T.~P.}\ \bibnamefont
  {Rakitzis}}, \bibinfo {author} {\bibfnamefont {P.~C.}\ \bibnamefont
  {Samartzis}}, \bibinfo {author} {\bibfnamefont {V.}~\bibnamefont
  {Ladopoulos}},\ and\ \bibinfo {author} {\bibfnamefont {T.~N.}\ \bibnamefont
  {Kitsopoulos}},\ }\bibfield  {title} {\bibinfo {title} {Slice imaging: A new
  approach to ion imaging and velocity mapping},\ }\href
  {https://doi.org/10.1063/1.1403010} {\bibfield  {journal} {\bibinfo
  {journal} {Rev. Sci. Instrum.}\ }\textbf {\bibinfo {volume} {72}},\ \bibinfo
  {pages} {3848} (\bibinfo {year} {2001})}\BibitemShut {NoStop}%
\bibitem [{\citenamefont {Amini}\ \emph {et~al.}(2015)\citenamefont {Amini},
  \citenamefont {Blake}, \citenamefont {Brouard}, \citenamefont {Burt},
  \citenamefont {Halford}, \citenamefont {Lauer}, \citenamefont {Slater},
  \citenamefont {Lee},\ and\ \citenamefont {Vallance}}]{Amini:RSI86:103113}%
  \BibitemOpen
  \bibfield  {author} {\bibinfo {author} {\bibfnamefont {K.}~\bibnamefont
  {Amini}}, \bibinfo {author} {\bibfnamefont {S.}~\bibnamefont {Blake}},
  \bibinfo {author} {\bibfnamefont {M.}~\bibnamefont {Brouard}}, \bibinfo
  {author} {\bibfnamefont {M.~B.}\ \bibnamefont {Burt}}, \bibinfo {author}
  {\bibfnamefont {E.}~\bibnamefont {Halford}}, \bibinfo {author} {\bibfnamefont
  {A.}~\bibnamefont {Lauer}}, \bibinfo {author} {\bibfnamefont {C.~S.}\
  \bibnamefont {Slater}}, \bibinfo {author} {\bibfnamefont {J.~W.~L.}\
  \bibnamefont {Lee}},\ and\ \bibinfo {author} {\bibfnamefont {C.}~\bibnamefont
  {Vallance}},\ }\bibfield  {title} {\bibinfo {title} {Three-dimensional
  imaging of carbonyl sulfide and ethyl iodide photodissociation using the
  pixel imaging mass spectrometry camera},\ }\href
  {https://doi.org/10.1063/1.4934544} {\bibfield  {journal} {\bibinfo
  {journal} {Review of Scientific Instruments}\ }\textbf {\bibinfo {volume}
  {86}},\ \bibinfo {pages} {103113} (\bibinfo {year} {2015})},\ \Eprint
  {https://arxiv.org/abs/https://doi.org/10.1063/1.4934544}
  {https://doi.org/10.1063/1.4934544} \BibitemShut {NoStop}%
\bibitem [{\citenamefont {Dörner}\ \emph {et~al.}(2000)\citenamefont
  {Dörner}, \citenamefont {Mergel}, \citenamefont {Jagutzki}, \citenamefont
  {Spielberger}, \citenamefont {Ullrich}, \citenamefont {Moshammer},\ and\
  \citenamefont {Schmidt-Böcking}}]{Doerner:PhysRep330:95}%
  \BibitemOpen
  \bibfield  {author} {\bibinfo {author} {\bibfnamefont {R.}~\bibnamefont
  {Dörner}}, \bibinfo {author} {\bibfnamefont {V.}~\bibnamefont {Mergel}},
  \bibinfo {author} {\bibfnamefont {O.}~\bibnamefont {Jagutzki}}, \bibinfo
  {author} {\bibfnamefont {L.}~\bibnamefont {Spielberger}}, \bibinfo {author}
  {\bibfnamefont {J.}~\bibnamefont {Ullrich}}, \bibinfo {author} {\bibfnamefont
  {R.}~\bibnamefont {Moshammer}},\ and\ \bibinfo {author} {\bibfnamefont
  {H.}~\bibnamefont {Schmidt-Böcking}},\ }\bibfield  {title} {\bibinfo {title}
  {Cold target recoil ion momentum spectroscopy: a ‘momentum microscope’ to
  view atomic collision dynamics},\ }\href
  {https://doi.org/10.1016/S0370-1573(99)00109-X} {\bibfield  {journal}
  {\bibinfo  {journal} {Phys. Rep.}\ }\textbf {\bibinfo {volume} {330}},\
  \bibinfo {pages} {95} (\bibinfo {year} {2000})}\BibitemShut {NoStop}%
\bibitem [{\citenamefont {Ullrich}\ \emph {et~al.}(2003)\citenamefont
  {Ullrich}, \citenamefont {Moshammer}, \citenamefont {Dorn}, \citenamefont
  {D{\"o}rner}, \citenamefont {Schmidt},\ and\ \citenamefont
  {Schmidt-B{\"o}cking}}]{Ullrich:RPP66:1463}%
  \BibitemOpen
  \bibfield  {author} {\bibinfo {author} {\bibfnamefont {J.}~\bibnamefont
  {Ullrich}}, \bibinfo {author} {\bibfnamefont {R.}~\bibnamefont {Moshammer}},
  \bibinfo {author} {\bibfnamefont {A.}~\bibnamefont {Dorn}}, \bibinfo {author}
  {\bibfnamefont {R.}~\bibnamefont {D{\"o}rner}}, \bibinfo {author}
  {\bibfnamefont {L.~P.~H.}\ \bibnamefont {Schmidt}},\ and\ \bibinfo {author}
  {\bibfnamefont {H.}~\bibnamefont {Schmidt-B{\"o}cking}},\ }\bibfield  {title}
  {\bibinfo {title} {Recoil-ion and electron momentum spectroscopy:
  {R}eaction-microscopes},\ }\href {https://doi.org/10.1088/0034-4885/66/9/203}
  {\bibfield  {journal} {\bibinfo  {journal} {Rep. Prog. Phys.}\ }\textbf
  {\bibinfo {volume} {66}},\ \bibinfo {pages} {1463} (\bibinfo {year}
  {2003})}\BibitemShut {NoStop}%
\bibitem [{\citenamefont {Ullrich}\ \emph {et~al.}(1997)\citenamefont
  {Ullrich}, \citenamefont {Moshammer}, \citenamefont {D{\"o}rner},
  \citenamefont {Jagutzki}, \citenamefont {Mergel}, \citenamefont
  {Schmidt-B{\"o}cking},\ and\ \citenamefont
  {Spielberger}}]{Ullrich:JPB30:2917}%
  \BibitemOpen
  \bibfield  {author} {\bibinfo {author} {\bibfnamefont {J.}~\bibnamefont
  {Ullrich}}, \bibinfo {author} {\bibfnamefont {R.}~\bibnamefont {Moshammer}},
  \bibinfo {author} {\bibfnamefont {R.}~\bibnamefont {D{\"o}rner}}, \bibinfo
  {author} {\bibfnamefont {O.}~\bibnamefont {Jagutzki}}, \bibinfo {author}
  {\bibfnamefont {V.}~\bibnamefont {Mergel}}, \bibinfo {author} {\bibfnamefont
  {H.}~\bibnamefont {Schmidt-B{\"o}cking}},\ and\ \bibinfo {author}
  {\bibfnamefont {L.}~\bibnamefont {Spielberger}},\ }\bibfield  {title}
  {\bibinfo {title} {Recoil-ion momentum spectroscopy},\ }\href
  {https://doi.org/10.1088/0953-4075/30/13/006} {\bibfield  {journal} {\bibinfo
   {journal} {J. Phys. B}\ }\textbf {\bibinfo {volume} {30}},\ \bibinfo {pages}
  {2917} (\bibinfo {year} {1997})}\BibitemShut {NoStop}%
\bibitem [{\citenamefont {Davies}\ \emph {et~al.}(1999)\citenamefont {Davies},
  \citenamefont {LeClaire}, \citenamefont {Continetti},\ and\ \citenamefont
  {Hayden}}]{Davies:JCP111:1}%
  \BibitemOpen
  \bibfield  {author} {\bibinfo {author} {\bibfnamefont {J.~A.}\ \bibnamefont
  {Davies}}, \bibinfo {author} {\bibfnamefont {J.~E.}\ \bibnamefont
  {LeClaire}}, \bibinfo {author} {\bibfnamefont {R.~E.}\ \bibnamefont
  {Continetti}},\ and\ \bibinfo {author} {\bibfnamefont {C.~C.}\ \bibnamefont
  {Hayden}},\ }\bibfield  {title} {\bibinfo {title} {Femtosecond time-resolved
  photoelectron–photoion coincidence imaging studies of dissociation
  dynamics},\ }\href {https://doi.org/10.1063/1.479248} {\bibfield  {journal}
  {\bibinfo  {journal} {J. Chem. Phys.}\ }\textbf {\bibinfo {volume} {111}},\
  \bibinfo {pages} {1} (\bibinfo {year} {1999})},\ \Eprint
  {https://arxiv.org/abs/https://doi.org/10.1063/1.479248}
  {https://doi.org/10.1063/1.479248} \BibitemShut {NoStop}%
\bibitem [{\citenamefont {Lafosse}\ \emph {et~al.}(2000)\citenamefont
  {Lafosse}, \citenamefont {Lebech}, \citenamefont {Brenot}, \citenamefont
  {Guyon}, \citenamefont {Jagutzki}, \citenamefont {Spielberger}, \citenamefont
  {Vervloet}, \citenamefont {Houver},\ and\ \citenamefont
  {Dowek}}]{Lafosse:PRL84:5987}%
  \BibitemOpen
  \bibfield  {author} {\bibinfo {author} {\bibfnamefont {A.}~\bibnamefont
  {Lafosse}}, \bibinfo {author} {\bibfnamefont {M.}~\bibnamefont {Lebech}},
  \bibinfo {author} {\bibfnamefont {J.~C.}\ \bibnamefont {Brenot}}, \bibinfo
  {author} {\bibfnamefont {P.~M.}\ \bibnamefont {Guyon}}, \bibinfo {author}
  {\bibfnamefont {O.}~\bibnamefont {Jagutzki}}, \bibinfo {author}
  {\bibfnamefont {L.}~\bibnamefont {Spielberger}}, \bibinfo {author}
  {\bibfnamefont {M.}~\bibnamefont {Vervloet}}, \bibinfo {author}
  {\bibfnamefont {J.~C.}\ \bibnamefont {Houver}},\ and\ \bibinfo {author}
  {\bibfnamefont {D.}~\bibnamefont {Dowek}},\ }\bibfield  {title} {\bibinfo
  {title} {Vector correlations in dissociative photoionization of diatomic
  molecules in the {VUV} range: Strong anisotropies in electron emission from
  spatially oriented {NO} molecules},\ }\href
  {https://doi.org/10.1103/PhysRevLett.84.5987} {\bibfield  {journal} {\bibinfo
   {journal} {Phys. Rev. Lett.}\ }\textbf {\bibinfo {volume} {84}},\ \bibinfo
  {pages} {5987} (\bibinfo {year} {2000})}\BibitemShut {NoStop}%
\bibitem [{\citenamefont {Dowek}\ \emph {et~al.}(2002)\citenamefont {Dowek},
  \citenamefont {Brenot}, \citenamefont {Guyon}, \citenamefont {Houver},
  \citenamefont {Lafosse}, \citenamefont {Lebech}, \citenamefont {Jagutzki},\
  and\ \citenamefont {Spielberger}}]{Dowek:NIMA477:323}%
  \BibitemOpen
  \bibfield  {author} {\bibinfo {author} {\bibfnamefont {D.}~\bibnamefont
  {Dowek}}, \bibinfo {author} {\bibfnamefont {J.}~\bibnamefont {Brenot}},
  \bibinfo {author} {\bibfnamefont {P.}~\bibnamefont {Guyon}}, \bibinfo
  {author} {\bibfnamefont {J.}~\bibnamefont {Houver}}, \bibinfo {author}
  {\bibfnamefont {A.}~\bibnamefont {Lafosse}}, \bibinfo {author} {\bibfnamefont
  {M.}~\bibnamefont {Lebech}}, \bibinfo {author} {\bibfnamefont
  {O.}~\bibnamefont {Jagutzki}},\ and\ \bibinfo {author} {\bibfnamefont
  {L.}~\bibnamefont {Spielberger}},\ }\bibfield  {title} {\bibinfo {title}
  {Imaging a chemical bond in the molecule frame using microchannel plate based
  position sensitive detectors and coincidence techniques},\ }\href
  {https://doi.org/10.1016/S0168-9002(01)01847-2} {\bibfield  {journal}
  {\bibinfo  {journal} {Nucl. Instrum. Meth. A}\ }\textbf {\bibinfo {volume}
  {477}},\ \bibinfo {pages} {323} (\bibinfo {year} {2002})},\ \bibinfo {note}
  {5th Int. Conf. on Position-Sensitive Detectors}\BibitemShut {NoStop}%
\bibitem [{\citenamefont {Ablikim}\ \emph {et~al.}(2019)\citenamefont
  {Ablikim}, \citenamefont {Bomme}, \citenamefont {Osipov}, \citenamefont
  {Xiong}, \citenamefont {Obaid}, \citenamefont {Bilodeau}, \citenamefont
  {Kling}, \citenamefont {Dumitriu}, \citenamefont {Augustin}, \citenamefont
  {Pathak}, \citenamefont {Schnorr}, \citenamefont {Kilcoyne}, \citenamefont
  {Berrah},\ and\ \citenamefont {Rolles}}]{Ablikim:RSI90:055103}%
  \BibitemOpen
  \bibfield  {author} {\bibinfo {author} {\bibfnamefont {U.}~\bibnamefont
  {Ablikim}}, \bibinfo {author} {\bibfnamefont {C.}~\bibnamefont {Bomme}},
  \bibinfo {author} {\bibfnamefont {T.}~\bibnamefont {Osipov}}, \bibinfo
  {author} {\bibfnamefont {H.}~\bibnamefont {Xiong}}, \bibinfo {author}
  {\bibfnamefont {R.}~\bibnamefont {Obaid}}, \bibinfo {author} {\bibfnamefont
  {R.~C.}\ \bibnamefont {Bilodeau}}, \bibinfo {author} {\bibfnamefont {N.~G.}\
  \bibnamefont {Kling}}, \bibinfo {author} {\bibfnamefont {I.}~\bibnamefont
  {Dumitriu}}, \bibinfo {author} {\bibfnamefont {S.}~\bibnamefont {Augustin}},
  \bibinfo {author} {\bibfnamefont {S.}~\bibnamefont {Pathak}}, \bibinfo
  {author} {\bibfnamefont {K.}~\bibnamefont {Schnorr}}, \bibinfo {author}
  {\bibfnamefont {D.}~\bibnamefont {Kilcoyne}}, \bibinfo {author}
  {\bibfnamefont {N.}~\bibnamefont {Berrah}},\ and\ \bibinfo {author}
  {\bibfnamefont {D.}~\bibnamefont {Rolles}},\ }\bibfield  {title} {\bibinfo
  {title} {A coincidence velocity map imaging spectrometer for ions and
  high-energy electrons to study inner-shell photoionization of gas-phase
  molecules},\ }\href {https://doi.org/10.1063/1.5093420} {\bibfield  {journal}
  {\bibinfo  {journal} {Rev. Sci. Instrum.}\ }\textbf {\bibinfo {volume}
  {90}},\ \bibinfo {pages} {055103} (\bibinfo {year} {2019})}\BibitemShut
  {NoStop}%
\bibitem [{\citenamefont {Pešić}\ \emph {et~al.}(2007)\citenamefont
  {Pešić}, \citenamefont {Rolles}, \citenamefont {Perri}, \citenamefont
  {Bilodeau}, \citenamefont {Ackerman}, \citenamefont {Rude}, \citenamefont
  {Kilcoyne}, \citenamefont {Bozek},\ and\ \citenamefont
  {Berrah}}]{Pesic:JESRP155:155}%
  \BibitemOpen
  \bibfield  {author} {\bibinfo {author} {\bibfnamefont {Z.}~\bibnamefont
  {Pešić}}, \bibinfo {author} {\bibfnamefont {D.}~\bibnamefont {Rolles}},
  \bibinfo {author} {\bibfnamefont {M.}~\bibnamefont {Perri}}, \bibinfo
  {author} {\bibfnamefont {R.}~\bibnamefont {Bilodeau}}, \bibinfo {author}
  {\bibfnamefont {G.}~\bibnamefont {Ackerman}}, \bibinfo {author}
  {\bibfnamefont {B.}~\bibnamefont {Rude}}, \bibinfo {author} {\bibfnamefont
  {A.}~\bibnamefont {Kilcoyne}}, \bibinfo {author} {\bibfnamefont
  {J.}~\bibnamefont {Bozek}},\ and\ \bibinfo {author} {\bibfnamefont
  {N.}~\bibnamefont {Berrah}},\ }\bibfield  {title} {\bibinfo {title} {Velocity
  map ion imaging applied to studies of molecular fragmentation with
  synchrotron radiation},\ }\href
  {https://doi.org/10.1016/j.elspec.2006.11.046} {\bibfield  {journal}
  {\bibinfo  {journal} {J. Electron. Spectrosc. Relat. Phenom.}\ }\textbf
  {\bibinfo {volume} {155}},\ \bibinfo {pages} {155} (\bibinfo {year}
  {2007})}\BibitemShut {NoStop}%
\bibitem [{\citenamefont {Bodi}\ \emph {et~al.}(2009)\citenamefont {Bodi},
  \citenamefont {Johnson}, \citenamefont {Gerber}, \citenamefont {Gengeliczki},
  \citenamefont {Sztáray},\ and\ \citenamefont {Baer}}]{Bodi:RSI80:034101}%
  \BibitemOpen
  \bibfield  {author} {\bibinfo {author} {\bibfnamefont {A.}~\bibnamefont
  {Bodi}}, \bibinfo {author} {\bibfnamefont {M.}~\bibnamefont {Johnson}},
  \bibinfo {author} {\bibfnamefont {T.}~\bibnamefont {Gerber}}, \bibinfo
  {author} {\bibfnamefont {Z.}~\bibnamefont {Gengeliczki}}, \bibinfo {author}
  {\bibfnamefont {B.}~\bibnamefont {Sztáray}},\ and\ \bibinfo {author}
  {\bibfnamefont {T.}~\bibnamefont {Baer}},\ }\bibfield  {title} {\bibinfo
  {title} {Imaging photoelectron photoion coincidence spectroscopy with
  velocity focusing electron optics},\ }\href
  {https://doi.org/10.1063/1.3082016} {\bibfield  {journal} {\bibinfo
  {journal} {Rev. Sci. Instrum.}\ }\textbf {\bibinfo {volume} {80}},\ \bibinfo
  {pages} {034101} (\bibinfo {year} {2009})},\ \Eprint
  {https://arxiv.org/abs/https://doi.org/10.1063/1.3082016}
  {https://doi.org/10.1063/1.3082016} \BibitemShut {NoStop}%
\bibitem [{\citenamefont {O’Keeffe}\ \emph {et~al.}(2011)\citenamefont
  {O’Keeffe}, \citenamefont {Bolognesi}, \citenamefont {Coreno},
  \citenamefont {Moise}, \citenamefont {Richter}, \citenamefont {Cautero},
  \citenamefont {Stebel}, \citenamefont {Sergo}, \citenamefont {Pravica},
  \citenamefont {Ovcharenko},\ and\ \citenamefont
  {Avaldi}}]{Keeffe:RSI82:033109}%
  \BibitemOpen
  \bibfield  {author} {\bibinfo {author} {\bibfnamefont {P.}~\bibnamefont
  {O’Keeffe}}, \bibinfo {author} {\bibfnamefont {P.}~\bibnamefont
  {Bolognesi}}, \bibinfo {author} {\bibfnamefont {M.}~\bibnamefont {Coreno}},
  \bibinfo {author} {\bibfnamefont {A.}~\bibnamefont {Moise}}, \bibinfo
  {author} {\bibfnamefont {R.}~\bibnamefont {Richter}}, \bibinfo {author}
  {\bibfnamefont {G.}~\bibnamefont {Cautero}}, \bibinfo {author} {\bibfnamefont
  {L.}~\bibnamefont {Stebel}}, \bibinfo {author} {\bibfnamefont
  {R.}~\bibnamefont {Sergo}}, \bibinfo {author} {\bibfnamefont
  {L.}~\bibnamefont {Pravica}}, \bibinfo {author} {\bibfnamefont
  {Y.}~\bibnamefont {Ovcharenko}},\ and\ \bibinfo {author} {\bibfnamefont
  {L.}~\bibnamefont {Avaldi}},\ }\bibfield  {title} {\bibinfo {title} {A
  photoelectron velocity map imaging spectrometer for experiments combining
  synchrotron and laser radiations},\ }\href
  {https://doi.org/10.1063/1.3563723} {\bibfield  {journal} {\bibinfo
  {journal} {Rev. Sci. Instrum.}\ }\textbf {\bibinfo {volume} {82}},\ \bibinfo
  {pages} {033109} (\bibinfo {year} {2011})},\ \Eprint
  {https://arxiv.org/abs/https://doi.org/10.1063/1.3563723}
  {https://doi.org/10.1063/1.3563723} \BibitemShut {NoStop}%
\bibitem [{\citenamefont {Str{\"u}der}\ \emph {et~al.}(2010)\citenamefont
  {Str{\"u}der}, \citenamefont {Epp}, \citenamefont {Rolles}, \citenamefont
  {Hartmann}, \citenamefont {Holl}, \citenamefont {Lutz}, \citenamefont
  {Soltau}, \citenamefont {Eckart}, \citenamefont {Reich}, \citenamefont
  {Heinzinger}, \citenamefont {Thamm}, \citenamefont {Rudenko}, \citenamefont
  {Krasniqi}, \citenamefont {K{\"u}hnel}, \citenamefont {Bauer}, \citenamefont
  {Schroeter}, \citenamefont {Moshammer}, \citenamefont {Techert},
  \citenamefont {Miessner}, \citenamefont {Porro}, \citenamefont {Haelker},
  \citenamefont {Meidinger}, \citenamefont {Kimmel}, \citenamefont
  {Andritschke}, \citenamefont {Schopper}, \citenamefont {Weidenspointner},
  \citenamefont {Ziegler}, \citenamefont {Pietschner}, \citenamefont
  {Herrmann}, \citenamefont {Pietsch}, \citenamefont {Walenta}, \citenamefont
  {Leitenberger}, \citenamefont {Bostedt}, \citenamefont {Moeller},
  \citenamefont {Rupp}, \citenamefont {Adolph}, \citenamefont {Graafsma},
  \citenamefont {Hirsemann}, \citenamefont {Gaertner}, \citenamefont {Richter},
  \citenamefont {Foucar}, \citenamefont {Shoeman}, \citenamefont
  {Schlichting},\ and\ \citenamefont {Ullrich}}]{Strueder:NIMA614:483}%
  \BibitemOpen
  \bibfield  {author} {\bibinfo {author} {\bibfnamefont {L.}~\bibnamefont
  {Str{\"u}der}}, \bibinfo {author} {\bibfnamefont {S.}~\bibnamefont {Epp}},
  \bibinfo {author} {\bibfnamefont {D.}~\bibnamefont {Rolles}}, \bibinfo
  {author} {\bibfnamefont {R.}~\bibnamefont {Hartmann}}, \bibinfo {author}
  {\bibfnamefont {P.}~\bibnamefont {Holl}}, \bibinfo {author} {\bibfnamefont
  {G.}~\bibnamefont {Lutz}}, \bibinfo {author} {\bibfnamefont {H.}~\bibnamefont
  {Soltau}}, \bibinfo {author} {\bibfnamefont {R.}~\bibnamefont {Eckart}},
  \bibinfo {author} {\bibfnamefont {C.}~\bibnamefont {Reich}}, \bibinfo
  {author} {\bibfnamefont {K.}~\bibnamefont {Heinzinger}}, \bibinfo {author}
  {\bibfnamefont {C.}~\bibnamefont {Thamm}}, \bibinfo {author} {\bibfnamefont
  {A.}~\bibnamefont {Rudenko}}, \bibinfo {author} {\bibfnamefont
  {F.}~\bibnamefont {Krasniqi}}, \bibinfo {author} {\bibfnamefont
  {K.}~\bibnamefont {K{\"u}hnel}}, \bibinfo {author} {\bibfnamefont
  {C.}~\bibnamefont {Bauer}}, \bibinfo {author} {\bibfnamefont {C.-D.}\
  \bibnamefont {Schroeter}}, \bibinfo {author} {\bibfnamefont {R.}~\bibnamefont
  {Moshammer}}, \bibinfo {author} {\bibfnamefont {S.}~\bibnamefont {Techert}},
  \bibinfo {author} {\bibfnamefont {D.}~\bibnamefont {Miessner}}, \bibinfo
  {author} {\bibfnamefont {M.}~\bibnamefont {Porro}}, \bibinfo {author}
  {\bibfnamefont {O.}~\bibnamefont {Haelker}}, \bibinfo {author} {\bibfnamefont
  {N.}~\bibnamefont {Meidinger}}, \bibinfo {author} {\bibfnamefont
  {N.}~\bibnamefont {Kimmel}}, \bibinfo {author} {\bibfnamefont
  {R.}~\bibnamefont {Andritschke}}, \bibinfo {author} {\bibfnamefont
  {F.}~\bibnamefont {Schopper}}, \bibinfo {author} {\bibfnamefont
  {G.}~\bibnamefont {Weidenspointner}}, \bibinfo {author} {\bibfnamefont
  {A.}~\bibnamefont {Ziegler}}, \bibinfo {author} {\bibfnamefont
  {D.}~\bibnamefont {Pietschner}}, \bibinfo {author} {\bibfnamefont
  {S.}~\bibnamefont {Herrmann}}, \bibinfo {author} {\bibfnamefont
  {U.}~\bibnamefont {Pietsch}}, \bibinfo {author} {\bibfnamefont
  {A.}~\bibnamefont {Walenta}}, \bibinfo {author} {\bibfnamefont
  {W.}~\bibnamefont {Leitenberger}}, \bibinfo {author} {\bibfnamefont
  {C.}~\bibnamefont {Bostedt}}, \bibinfo {author} {\bibfnamefont
  {T.}~\bibnamefont {Moeller}}, \bibinfo {author} {\bibfnamefont
  {D.}~\bibnamefont {Rupp}}, \bibinfo {author} {\bibfnamefont {M.}~\bibnamefont
  {Adolph}}, \bibinfo {author} {\bibfnamefont {H.}~\bibnamefont {Graafsma}},
  \bibinfo {author} {\bibfnamefont {H.}~\bibnamefont {Hirsemann}}, \bibinfo
  {author} {\bibfnamefont {K.}~\bibnamefont {Gaertner}}, \bibinfo {author}
  {\bibfnamefont {R.}~\bibnamefont {Richter}}, \bibinfo {author} {\bibfnamefont
  {L.}~\bibnamefont {Foucar}}, \bibinfo {author} {\bibfnamefont {R.~L.}\
  \bibnamefont {Shoeman}}, \bibinfo {author} {\bibfnamefont {I.}~\bibnamefont
  {Schlichting}},\ and\ \bibinfo {author} {\bibfnamefont {J.}~\bibnamefont
  {Ullrich}},\ }\bibfield  {title} {\bibinfo {title} {Large-format, high-speed,
  {X}-ray {pnCCDs} combined with electron and ion imaging spectrometers in a
  multipurpose chamber for experiments at 4th generation light sources},\
  }\href {https://doi.org/10.1016/j.nima.2009.12.053} {\bibfield  {journal}
  {\bibinfo  {journal} {Nucl. Instrum. Meth. A}\ }\textbf {\bibinfo {volume}
  {614}},\ \bibinfo {pages} {483} (\bibinfo {year} {2010})}\BibitemShut
  {NoStop}%
\bibitem [{\citenamefont {Rolles}\ \emph {et~al.}(2014)\citenamefont {Rolles},
  \citenamefont {Boll}, \citenamefont {Adolph}, \citenamefont {Aquila},
  \citenamefont {Bostedt}, \citenamefont {Bozek}, \citenamefont {Chapman},
  \citenamefont {Coffee}, \citenamefont {Coppola}, \citenamefont {Decleva},
  \citenamefont {Delmas}, \citenamefont {Epp}, \citenamefont {Erk},
  \citenamefont {Filsinger}, \citenamefont {Foucar}, \citenamefont {Gumprecht},
  \citenamefont {H{\"o}mke}, \citenamefont {Gorkhover}, \citenamefont
  {Holmegaard}, \citenamefont {Johnsson}, \citenamefont {Kaiser}, \citenamefont
  {Krasniqi}, \citenamefont {K{\"u}hnel}, \citenamefont {Maurer}, \citenamefont
  {Messerschmidt}, \citenamefont {Moshammer}, \citenamefont {Quevedo},
  \citenamefont {Rajkovic}, \citenamefont {Rouz{\'e}e}, \citenamefont {Rudek},
  \citenamefont {Schlichting}, \citenamefont {Schmidt}, \citenamefont {Schorb},
  \citenamefont {Schr{\"o}ter}, \citenamefont {Schulz}, \citenamefont
  {Stapelfeldt}, \citenamefont {Stener}, \citenamefont {Stern}, \citenamefont
  {Techert}, \citenamefont {Th{\o}gersen}, \citenamefont {Vrakking},
  \citenamefont {Rudenko}, \citenamefont {K{\"u}pper},\ and\ \citenamefont
  {Ullrich}}]{Rolles:JPB47:124035}%
  \BibitemOpen
  \bibfield  {author} {\bibinfo {author} {\bibfnamefont {D.}~\bibnamefont
  {Rolles}}, \bibinfo {author} {\bibfnamefont {R.}~\bibnamefont {Boll}},
  \bibinfo {author} {\bibfnamefont {M.}~\bibnamefont {Adolph}}, \bibinfo
  {author} {\bibfnamefont {A.}~\bibnamefont {Aquila}}, \bibinfo {author}
  {\bibfnamefont {C.}~\bibnamefont {Bostedt}}, \bibinfo {author} {\bibfnamefont
  {J.}~\bibnamefont {Bozek}}, \bibinfo {author} {\bibfnamefont
  {H.}~\bibnamefont {Chapman}}, \bibinfo {author} {\bibfnamefont
  {R.}~\bibnamefont {Coffee}}, \bibinfo {author} {\bibfnamefont
  {N.}~\bibnamefont {Coppola}}, \bibinfo {author} {\bibfnamefont
  {P.}~\bibnamefont {Decleva}}, \bibinfo {author} {\bibfnamefont
  {T.}~\bibnamefont {Delmas}}, \bibinfo {author} {\bibfnamefont
  {S.}~\bibnamefont {Epp}}, \bibinfo {author} {\bibfnamefont {B.}~\bibnamefont
  {Erk}}, \bibinfo {author} {\bibfnamefont {F.}~\bibnamefont {Filsinger}},
  \bibinfo {author} {\bibfnamefont {L.}~\bibnamefont {Foucar}}, \bibinfo
  {author} {\bibfnamefont {L.}~\bibnamefont {Gumprecht}}, \bibinfo {author}
  {\bibfnamefont {A.}~\bibnamefont {H{\"o}mke}}, \bibinfo {author}
  {\bibfnamefont {T.}~\bibnamefont {Gorkhover}}, \bibinfo {author}
  {\bibfnamefont {L.}~\bibnamefont {Holmegaard}}, \bibinfo {author}
  {\bibfnamefont {P.}~\bibnamefont {Johnsson}}, \bibinfo {author}
  {\bibfnamefont {C.}~\bibnamefont {Kaiser}}, \bibinfo {author} {\bibfnamefont
  {F.}~\bibnamefont {Krasniqi}}, \bibinfo {author} {\bibfnamefont {K.-U.}\
  \bibnamefont {K{\"u}hnel}}, \bibinfo {author} {\bibfnamefont
  {J.}~\bibnamefont {Maurer}}, \bibinfo {author} {\bibfnamefont
  {M.}~\bibnamefont {Messerschmidt}}, \bibinfo {author} {\bibfnamefont
  {R.}~\bibnamefont {Moshammer}}, \bibinfo {author} {\bibfnamefont
  {W.}~\bibnamefont {Quevedo}}, \bibinfo {author} {\bibfnamefont
  {I.}~\bibnamefont {Rajkovic}}, \bibinfo {author} {\bibfnamefont
  {A.}~\bibnamefont {Rouz{\'e}e}}, \bibinfo {author} {\bibfnamefont
  {B.}~\bibnamefont {Rudek}}, \bibinfo {author} {\bibfnamefont
  {I.}~\bibnamefont {Schlichting}}, \bibinfo {author} {\bibfnamefont
  {C.}~\bibnamefont {Schmidt}}, \bibinfo {author} {\bibfnamefont
  {S.}~\bibnamefont {Schorb}}, \bibinfo {author} {\bibfnamefont {C.~D.}\
  \bibnamefont {Schr{\"o}ter}}, \bibinfo {author} {\bibfnamefont
  {J.}~\bibnamefont {Schulz}}, \bibinfo {author} {\bibfnamefont
  {H.}~\bibnamefont {Stapelfeldt}}, \bibinfo {author} {\bibfnamefont
  {M.}~\bibnamefont {Stener}}, \bibinfo {author} {\bibfnamefont
  {S.}~\bibnamefont {Stern}}, \bibinfo {author} {\bibfnamefont
  {S.}~\bibnamefont {Techert}}, \bibinfo {author} {\bibfnamefont
  {J.}~\bibnamefont {Th{\o}gersen}}, \bibinfo {author} {\bibfnamefont
  {M.~J.~J.}\ \bibnamefont {Vrakking}}, \bibinfo {author} {\bibfnamefont
  {A.}~\bibnamefont {Rudenko}}, \bibinfo {author} {\bibfnamefont
  {J.}~\bibnamefont {K{\"u}pper}},\ and\ \bibinfo {author} {\bibfnamefont
  {J.}~\bibnamefont {Ullrich}},\ }\bibfield  {title} {\bibinfo {title}
  {Femtosecond x-ray photoelectron diffraction on gas-phase dibromobenzene
  molecules},\ }\href {https://doi.org/10.1088/0953-4075/47/12/124035}
  {\bibfield  {journal} {\bibinfo  {journal} {J. Phys. B}\ }\textbf {\bibinfo
  {volume} {47}},\ \bibinfo {pages} {124035} (\bibinfo {year}
  {2014})}\BibitemShut {NoStop}%
\bibitem [{\citenamefont {Vredenborg}\ \emph {et~al.}(2008)\citenamefont
  {Vredenborg}, \citenamefont {Roeterdink},\ and\ \citenamefont
  {Janssen}}]{Vredenborg:RSI79:0034}%
  \BibitemOpen
  \bibfield  {author} {\bibinfo {author} {\bibfnamefont {A.}~\bibnamefont
  {Vredenborg}}, \bibinfo {author} {\bibfnamefont {W.~G.}\ \bibnamefont
  {Roeterdink}},\ and\ \bibinfo {author} {\bibfnamefont {M.~H.~M.}\
  \bibnamefont {Janssen}},\ }\bibfield  {title} {\bibinfo {title} {A
  photoelectron-photoion coincidence imaging apparatus for femtosecond
  time-resolved molecular dynamics with electron time-of-flight resolution of
  $\sigma$=18~ps and energy resolution $\delta e/e$=3.5~\%},\ }\href
  {https://doi.org/10.1063/1.2949142} {\bibfield  {journal} {\bibinfo
  {journal} {Rev. Sci. Instrum.}\ }\textbf {\bibinfo {volume} {79}},\ \bibinfo
  {pages} {063108} (\bibinfo {year} {2008})},\ \bibinfo {note} {publisher:
  American Institute of Physics}\BibitemShut {NoStop}%
\bibitem [{\citenamefont {Garcia}\ \emph {et~al.}(2013)\citenamefont {Garcia},
  \citenamefont {Cunha~de Miranda}, \citenamefont {Tia}, \citenamefont {Daly},\
  and\ \citenamefont {Nahon}}]{Garcia:RSI84:053112}%
  \BibitemOpen
  \bibfield  {author} {\bibinfo {author} {\bibfnamefont {G.~A.}\ \bibnamefont
  {Garcia}}, \bibinfo {author} {\bibfnamefont {B.~K.}\ \bibnamefont {Cunha~de
  Miranda}}, \bibinfo {author} {\bibfnamefont {M.}~\bibnamefont {Tia}},
  \bibinfo {author} {\bibfnamefont {S.}~\bibnamefont {Daly}},\ and\ \bibinfo
  {author} {\bibfnamefont {L.}~\bibnamefont {Nahon}},\ }\bibfield  {title}
  {\bibinfo {title} {Delicious iii: A multipurpose double imaging particle
  coincidence spectrometer for gas phase vacuum ultraviolet photodynamics
  studies},\ }\href {https://doi.org/10.1063/1.4807751} {\bibfield  {journal}
  {\bibinfo  {journal} {Rev. Sci. Instrum.}\ }\textbf {\bibinfo {volume}
  {84}},\ \bibinfo {pages} {053112} (\bibinfo {year} {2013})},\ \Eprint
  {https://arxiv.org/abs/https://doi.org/10.1063/1.4807751}
  {https://doi.org/10.1063/1.4807751} \BibitemShut {NoStop}%
\bibitem [{\citenamefont {Bomme}\ \emph {et~al.}(2013)\citenamefont {Bomme},
  \citenamefont {Guillemin}, \citenamefont {Marin}, \citenamefont {Journel},
  \citenamefont {Marchenko}, \citenamefont {Dowek}, \citenamefont {Trcera},
  \citenamefont {Pilette}, \citenamefont {Avila}, \citenamefont {Ringuenet},
  \citenamefont {Kushawaha},\ and\ \citenamefont {Simon}}]{Bomme:RSI84:103104}%
  \BibitemOpen
  \bibfield  {author} {\bibinfo {author} {\bibfnamefont {C.}~\bibnamefont
  {Bomme}}, \bibinfo {author} {\bibfnamefont {R.}~\bibnamefont {Guillemin}},
  \bibinfo {author} {\bibfnamefont {T.}~\bibnamefont {Marin}}, \bibinfo
  {author} {\bibfnamefont {L.}~\bibnamefont {Journel}}, \bibinfo {author}
  {\bibfnamefont {T.}~\bibnamefont {Marchenko}}, \bibinfo {author}
  {\bibfnamefont {D.}~\bibnamefont {Dowek}}, \bibinfo {author} {\bibfnamefont
  {N.}~\bibnamefont {Trcera}}, \bibinfo {author} {\bibfnamefont
  {B.}~\bibnamefont {Pilette}}, \bibinfo {author} {\bibfnamefont
  {A.}~\bibnamefont {Avila}}, \bibinfo {author} {\bibfnamefont
  {H.}~\bibnamefont {Ringuenet}}, \bibinfo {author} {\bibfnamefont {R.~K.}\
  \bibnamefont {Kushawaha}},\ and\ \bibinfo {author} {\bibfnamefont
  {M.}~\bibnamefont {Simon}},\ }\bibfield  {title} {\bibinfo {title} {Double
  momentum spectrometer for ion-electron vector correlations in dissociative
  photoionization.},\ }\href {https://doi.org/10.1063/1.4824194} {\bibfield
  {journal} {\bibinfo  {journal} {Rev. Sci. Instrum.}\ }\textbf {\bibinfo
  {volume} {84}},\ \bibinfo {pages} {103104} (\bibinfo {year}
  {2013})}\BibitemShut {NoStop}%
\bibitem [{\citenamefont {Sztáray}\ \emph {et~al.}(2017)\citenamefont
  {Sztáray}, \citenamefont {Voronova}, \citenamefont {Torma}, \citenamefont
  {Covert}, \citenamefont {Bodi}, \citenamefont {Hemberger}, \citenamefont
  {Gerber},\ and\ \citenamefont {Osborn}}]{Sztaray:JCP147:013944}%
  \BibitemOpen
  \bibfield  {author} {\bibinfo {author} {\bibfnamefont {B.}~\bibnamefont
  {Sztáray}}, \bibinfo {author} {\bibfnamefont {K.}~\bibnamefont {Voronova}},
  \bibinfo {author} {\bibfnamefont {K.~G.}\ \bibnamefont {Torma}}, \bibinfo
  {author} {\bibfnamefont {K.~J.}\ \bibnamefont {Covert}}, \bibinfo {author}
  {\bibfnamefont {A.}~\bibnamefont {Bodi}}, \bibinfo {author} {\bibfnamefont
  {P.}~\bibnamefont {Hemberger}}, \bibinfo {author} {\bibfnamefont
  {T.}~\bibnamefont {Gerber}},\ and\ \bibinfo {author} {\bibfnamefont {D.~L.}\
  \bibnamefont {Osborn}},\ }\bibfield  {title} {\bibinfo {title} {{CRF-PEPICO:
  Double velocity map imaging photoelectron photoion coincidence spectroscopy
  for reaction kinetics studies}},\ }\href {https://doi.org/10.1063/1.4984304}
  {\bibfield  {journal} {\bibinfo  {journal} {J. Chem. Phys.}\ }\textbf
  {\bibinfo {volume} {147}},\ \bibinfo {pages} {013944} (\bibinfo {year}
  {2017})},\ \Eprint {https://arxiv.org/abs/https://doi.org/10.1063/1.4984304}
  {https://doi.org/10.1063/1.4984304} \BibitemShut {NoStop}%
\bibitem [{\citenamefont {Hosaka}\ \emph {et~al.}(2006)\citenamefont {Hosaka},
  \citenamefont {Adachi}, \citenamefont {Golovin}, \citenamefont {Takahashi},
  \citenamefont {Watanabe},\ and\ \citenamefont
  {Yagishita}}]{Hosaka:JJAP45:1841}%
  \BibitemOpen
  \bibfield  {author} {\bibinfo {author} {\bibfnamefont {K.}~\bibnamefont
  {Hosaka}}, \bibinfo {author} {\bibfnamefont {J.}~\bibnamefont {Adachi}},
  \bibinfo {author} {\bibfnamefont {A.~V.}\ \bibnamefont {Golovin}}, \bibinfo
  {author} {\bibfnamefont {M.}~\bibnamefont {Takahashi}}, \bibinfo {author}
  {\bibfnamefont {N.}~\bibnamefont {Watanabe}},\ and\ \bibinfo {author}
  {\bibfnamefont {A.}~\bibnamefont {Yagishita}},\ }\bibfield  {title} {\bibinfo
  {title} {Coincidence velocity imaging apparatus for study of angular
  correlations between photoelectrons and photofragments},\ }\href
  {https://doi.org/10.1143/jjap.45.1841} {\bibfield  {journal} {\bibinfo
  {journal} {Jpn. J. Appl. Phys.}\ }\textbf {\bibinfo {volume} {45}},\ \bibinfo
  {pages} {1841} (\bibinfo {year} {2006})}\BibitemShut {NoStop}%
\bibitem [{\citenamefont {Tang}\ \emph {et~al.}(2009)\citenamefont {Tang},
  \citenamefont {Zhou}, \citenamefont {Niu}, \citenamefont {Liu}, \citenamefont
  {Sun}, \citenamefont {Shan}, \citenamefont {Liu},\ and\ \citenamefont
  {Sheng}}]{Tang:RSI80:1131101}%
  \BibitemOpen
  \bibfield  {author} {\bibinfo {author} {\bibfnamefont {X.}~\bibnamefont
  {Tang}}, \bibinfo {author} {\bibfnamefont {X.}~\bibnamefont {Zhou}}, \bibinfo
  {author} {\bibfnamefont {M.}~\bibnamefont {Niu}}, \bibinfo {author}
  {\bibfnamefont {S.}~\bibnamefont {Liu}}, \bibinfo {author} {\bibfnamefont
  {J.}~\bibnamefont {Sun}}, \bibinfo {author} {\bibfnamefont {X.}~\bibnamefont
  {Shan}}, \bibinfo {author} {\bibfnamefont {F.}~\bibnamefont {Liu}},\ and\
  \bibinfo {author} {\bibfnamefont {L.}~\bibnamefont {Sheng}},\ }\bibfield
  {title} {\bibinfo {title} {A threshold photoelectron-photoion coincidence
  spectrometer with double velocity imaging using synchrotron radiation},\
  }\href {https://doi.org/10.1063/1.3250872} {\bibfield  {journal} {\bibinfo
  {journal} {Rev. Sci. Instrum.}\ }\textbf {\bibinfo {volume} {80}},\ \bibinfo
  {pages} {113101} (\bibinfo {year} {2009})},\ \Eprint
  {https://arxiv.org/abs/https://doi.org/10.1063/1.3250872}
  {https://doi.org/10.1063/1.3250872} \BibitemShut {NoStop}%
\bibitem [{\citenamefont {Bodi}\ \emph {et~al.}(2012)\citenamefont {Bodi},
  \citenamefont {Hemberger}, \citenamefont {Gerber},\ and\ \citenamefont
  {Sztáray}}]{Bodi:RSI83:083105}%
  \BibitemOpen
  \bibfield  {author} {\bibinfo {author} {\bibfnamefont {A.}~\bibnamefont
  {Bodi}}, \bibinfo {author} {\bibfnamefont {P.}~\bibnamefont {Hemberger}},
  \bibinfo {author} {\bibfnamefont {T.}~\bibnamefont {Gerber}},\ and\ \bibinfo
  {author} {\bibfnamefont {B.}~\bibnamefont {Sztáray}},\ }\bibfield  {title}
  {\bibinfo {title} {A new double imaging velocity focusing coincidence
  experiment: {i2PEPICO}},\ }\href {https://doi.org/10.1063/1.4742769}
  {\bibfield  {journal} {\bibinfo  {journal} {Rev. Sci. Instrum.}\ }\textbf
  {\bibinfo {volume} {83}},\ \bibinfo {pages} {083105} (\bibinfo {year}
  {2012})},\ \Eprint {https://arxiv.org/abs/https://doi.org/10.1063/1.4742769}
  {https://doi.org/10.1063/1.4742769} \BibitemShut {NoStop}%
\bibitem [{\citenamefont {Clark}\ \emph {et~al.}(2012)\citenamefont {Clark},
  \citenamefont {Crooks}, \citenamefont {Sedgwick}, \citenamefont {Turchetta},
  \citenamefont {Lee}, \citenamefont {John}, \citenamefont {Wilman},
  \citenamefont {Hill}, \citenamefont {Halford}, \citenamefont {Slater},
  \citenamefont {Winter}, \citenamefont {Yuen}, \citenamefont {Gardiner},
  \citenamefont {Lipciuc}, \citenamefont {Brouard}, \citenamefont
  {Nomerotski},\ and\ \citenamefont {Vallance}}]{Clark:JPCA116:10897}%
  \BibitemOpen
  \bibfield  {author} {\bibinfo {author} {\bibfnamefont {A.~T.}\ \bibnamefont
  {Clark}}, \bibinfo {author} {\bibfnamefont {J.~P.}\ \bibnamefont {Crooks}},
  \bibinfo {author} {\bibfnamefont {I.}~\bibnamefont {Sedgwick}}, \bibinfo
  {author} {\bibfnamefont {R.}~\bibnamefont {Turchetta}}, \bibinfo {author}
  {\bibfnamefont {J.~W.~L.}\ \bibnamefont {Lee}}, \bibinfo {author}
  {\bibfnamefont {J.~J.}\ \bibnamefont {John}}, \bibinfo {author}
  {\bibfnamefont {E.~S.}\ \bibnamefont {Wilman}}, \bibinfo {author}
  {\bibfnamefont {L.}~\bibnamefont {Hill}}, \bibinfo {author} {\bibfnamefont
  {E.}~\bibnamefont {Halford}}, \bibinfo {author} {\bibfnamefont {C.~S.}\
  \bibnamefont {Slater}}, \bibinfo {author} {\bibfnamefont {B.}~\bibnamefont
  {Winter}}, \bibinfo {author} {\bibfnamefont {W.~H.}\ \bibnamefont {Yuen}},
  \bibinfo {author} {\bibfnamefont {S.~H.}\ \bibnamefont {Gardiner}}, \bibinfo
  {author} {\bibfnamefont {M.~L.}\ \bibnamefont {Lipciuc}}, \bibinfo {author}
  {\bibfnamefont {M.}~\bibnamefont {Brouard}}, \bibinfo {author} {\bibfnamefont
  {A.}~\bibnamefont {Nomerotski}},\ and\ \bibinfo {author} {\bibfnamefont
  {C.}~\bibnamefont {Vallance}},\ }\bibfield  {title} {\bibinfo {title}
  {{Multimass Velocity-Map Imaging with the Pixel Imaging Mass Spectrometry
  (PImMS) Sensor: An Ultra-Fast Event-Triggered Camera for Particle Imaging}},\
  }\href {https://doi.org/10.1021/jp309860t} {\bibfield  {journal} {\bibinfo
  {journal} {J. Phys. Chem. A}\ }\textbf {\bibinfo {volume} {116}},\ \bibinfo
  {pages} {10897} (\bibinfo {year} {2012})},\ \Eprint
  {https://arxiv.org/abs/https://doi.org/10.1021/jp309860t}
  {https://doi.org/10.1021/jp309860t} \BibitemShut {NoStop}%
\bibitem [{\citenamefont {John}\ \emph {et~al.}(2012)\citenamefont {John},
  \citenamefont {Brouard}, \citenamefont {Clark}, \citenamefont {Crooks},
  \citenamefont {Halford}, \citenamefont {Hill}, \citenamefont {Lee},
  \citenamefont {Nomerotski}, \citenamefont {Pisarczyk}, \citenamefont
  {Sedgwick}, \citenamefont {Slater}, \citenamefont {Turchetta}, \citenamefont
  {Vallance}, \citenamefont {Wilman}, \citenamefont {Winter},\ and\
  \citenamefont {Yuen}}]{John:JInst7:C08001}%
  \BibitemOpen
  \bibfield  {author} {\bibinfo {author} {\bibfnamefont {J.~J.}\ \bibnamefont
  {John}}, \bibinfo {author} {\bibfnamefont {M.}~\bibnamefont {Brouard}},
  \bibinfo {author} {\bibfnamefont {A.}~\bibnamefont {Clark}}, \bibinfo
  {author} {\bibfnamefont {J.}~\bibnamefont {Crooks}}, \bibinfo {author}
  {\bibfnamefont {E.}~\bibnamefont {Halford}}, \bibinfo {author} {\bibfnamefont
  {L.}~\bibnamefont {Hill}}, \bibinfo {author} {\bibfnamefont {J.~W.~L.}\
  \bibnamefont {Lee}}, \bibinfo {author} {\bibfnamefont {A.}~\bibnamefont
  {Nomerotski}}, \bibinfo {author} {\bibfnamefont {R.}~\bibnamefont
  {Pisarczyk}}, \bibinfo {author} {\bibfnamefont {I.}~\bibnamefont {Sedgwick}},
  \bibinfo {author} {\bibfnamefont {C.~S.}\ \bibnamefont {Slater}}, \bibinfo
  {author} {\bibfnamefont {R.}~\bibnamefont {Turchetta}}, \bibinfo {author}
  {\bibfnamefont {C.}~\bibnamefont {Vallance}}, \bibinfo {author}
  {\bibfnamefont {E.}~\bibnamefont {Wilman}}, \bibinfo {author} {\bibfnamefont
  {B.}~\bibnamefont {Winter}},\ and\ \bibinfo {author} {\bibfnamefont {W.~H.}\
  \bibnamefont {Yuen}},\ }\bibfield  {title} {\bibinfo {title} {{PImMS}, a fast
  event-triggered monolithic pixel detector with storage of multiple
  timestamps},\ }\href {https://doi.org/10.1088/1748-0221/7/08/C08001}
  {\bibfield  {journal} {\bibinfo  {journal} {J. Instrum.}\ }\textbf {\bibinfo
  {volume} {7}}\bibinfo  {number} { (8)},\ \bibinfo {pages}
  {C08001}}\BibitemShut {NoStop}%
\bibitem [{\citenamefont {Slater}\ \emph {et~al.}(2014)\citenamefont {Slater},
  \citenamefont {Blake}, \citenamefont {Brouard}, \citenamefont {Lauer},
  \citenamefont {Vallance}, \citenamefont {John}, \citenamefont {Turchetta},
  \citenamefont {Nomerotski}, \citenamefont {Christensen}, \citenamefont
  {Nielsen}, \citenamefont {Johansson},\ and\ \citenamefont
  {Stapelfeldt}}]{Slater:PRA89:011401}%
  \BibitemOpen
\bibfield  {number} {  }\bibfield  {author} {\bibinfo {author} {\bibfnamefont
  {C.~S.}\ \bibnamefont {Slater}}, \bibinfo {author} {\bibfnamefont
  {S.}~\bibnamefont {Blake}}, \bibinfo {author} {\bibfnamefont
  {M.}~\bibnamefont {Brouard}}, \bibinfo {author} {\bibfnamefont
  {A.}~\bibnamefont {Lauer}}, \bibinfo {author} {\bibfnamefont
  {C.}~\bibnamefont {Vallance}}, \bibinfo {author} {\bibfnamefont {J.~J.}\
  \bibnamefont {John}}, \bibinfo {author} {\bibfnamefont {R.}~\bibnamefont
  {Turchetta}}, \bibinfo {author} {\bibfnamefont {A.}~\bibnamefont
  {Nomerotski}}, \bibinfo {author} {\bibfnamefont {L.}~\bibnamefont
  {Christensen}}, \bibinfo {author} {\bibfnamefont {J.~H.}\ \bibnamefont
  {Nielsen}}, \bibinfo {author} {\bibfnamefont {M.~P.}\ \bibnamefont
  {Johansson}},\ and\ \bibinfo {author} {\bibfnamefont {H.}~\bibnamefont
  {Stapelfeldt}},\ }\bibfield  {title} {\bibinfo {title} {Covariance imaging
  experiments using a pixel-imaging mass-spectrometry camera},\ }\href
  {https://doi.org/10.1103/PhysRevA.89.011401} {\bibfield  {journal} {\bibinfo
  {journal} {Phys. Rev. A}\ }\textbf {\bibinfo {volume} {89}},\ \bibinfo
  {pages} {011401} (\bibinfo {year} {2014})}\BibitemShut {NoStop}%
\bibitem [{\citenamefont {Poikela}\ \emph {et~al.}(2014)\citenamefont
  {Poikela}, \citenamefont {Plosila}, \citenamefont {Westerlund}, \citenamefont
  {Campbell}, \citenamefont {De~Gaspari}, \citenamefont {Llopart},
  \citenamefont {Gromov}, \citenamefont {Kluit}, \citenamefont {van Beuzekom},
  \citenamefont {Zappon}, \citenamefont {Zivkovicd}, \citenamefont {Brezinae},
  \citenamefont {Desche}, \citenamefont {Fue},\ and\ \citenamefont
  {Kruth}}]{Poikela:JInst9:C05013}%
  \BibitemOpen
  \bibfield  {author} {\bibinfo {author} {\bibfnamefont {T.}~\bibnamefont
  {Poikela}}, \bibinfo {author} {\bibfnamefont {J.}~\bibnamefont {Plosila}},
  \bibinfo {author} {\bibfnamefont {T.}~\bibnamefont {Westerlund}}, \bibinfo
  {author} {\bibfnamefont {M.}~\bibnamefont {Campbell}}, \bibinfo {author}
  {\bibfnamefont {M.}~\bibnamefont {De~Gaspari}}, \bibinfo {author}
  {\bibfnamefont {X.}~\bibnamefont {Llopart}}, \bibinfo {author} {\bibfnamefont
  {V.}~\bibnamefont {Gromov}}, \bibinfo {author} {\bibfnamefont
  {R.}~\bibnamefont {Kluit}}, \bibinfo {author} {\bibfnamefont
  {M.}~\bibnamefont {van Beuzekom}}, \bibinfo {author} {\bibfnamefont
  {F.}~\bibnamefont {Zappon}}, \bibinfo {author} {\bibfnamefont
  {V.}~\bibnamefont {Zivkovicd}}, \bibinfo {author} {\bibfnamefont
  {C.}~\bibnamefont {Brezinae}}, \bibinfo {author} {\bibfnamefont
  {K.}~\bibnamefont {Desche}}, \bibinfo {author} {\bibfnamefont
  {Y.}~\bibnamefont {Fue}},\ and\ \bibinfo {author} {\bibfnamefont
  {A.}~\bibnamefont {Kruth}},\ }\bibfield  {title} {\bibinfo {title}
  {{Timepix3}: a {65K} channel hybrid pixel readout chip with simultaneous
  {ToA}/{ToT} and sparse readout},\ }\href
  {https://doi.org/10.1088/1748-0221/9/05/C05013} {\bibfield  {journal}
  {\bibinfo  {journal} {J. Instrum.}\ }\textbf {\bibinfo {volume} {9}}\bibinfo
  {number} { (5)},\ \bibinfo {pages} {C05013}}\BibitemShut {NoStop}%
\bibitem [{\citenamefont {Zhao}\ \emph {et~al.}(2017)\citenamefont {Zhao},
  \citenamefont {van Beuzekom}, \citenamefont {Bouwens}, \citenamefont
  {Byelov}, \citenamefont {Chakaberia}, \citenamefont {Cheng}, \citenamefont
  {Maddox}, \citenamefont {Nomerotski}, \citenamefont {Svihra}, \citenamefont
  {Visser}, \citenamefont {Vrba},\ and\ \citenamefont
  {Weinacht}}]{Zhao:RSI88:113104}%
  \BibitemOpen
\bibfield  {number} {  }\bibfield  {author} {\bibinfo {author} {\bibfnamefont
  {A.}~\bibnamefont {Zhao}}, \bibinfo {author} {\bibfnamefont {M.}~\bibnamefont
  {van Beuzekom}}, \bibinfo {author} {\bibfnamefont {B.}~\bibnamefont
  {Bouwens}}, \bibinfo {author} {\bibfnamefont {D.}~\bibnamefont {Byelov}},
  \bibinfo {author} {\bibfnamefont {I.}~\bibnamefont {Chakaberia}}, \bibinfo
  {author} {\bibfnamefont {C.}~\bibnamefont {Cheng}}, \bibinfo {author}
  {\bibfnamefont {E.}~\bibnamefont {Maddox}}, \bibinfo {author} {\bibfnamefont
  {A.}~\bibnamefont {Nomerotski}}, \bibinfo {author} {\bibfnamefont
  {P.}~\bibnamefont {Svihra}}, \bibinfo {author} {\bibfnamefont
  {J.}~\bibnamefont {Visser}}, \bibinfo {author} {\bibfnamefont
  {V.}~\bibnamefont {Vrba}},\ and\ \bibinfo {author} {\bibfnamefont
  {T.}~\bibnamefont {Weinacht}},\ }\bibfield  {title} {\bibinfo {title}
  {Coincidence velocity map imaging using {Tpx3Cam}, a time stamping optical
  camera with 1.5~ns timing resolution},\ }\href
  {https://doi.org/10.1063/1.4996888} {\bibfield  {journal} {\bibinfo
  {journal} {Rev. Sci. Instrum.}\ }\textbf {\bibinfo {volume} {88}},\ \bibinfo
  {pages} {113104} (\bibinfo {year} {2017})},\ \Eprint
  {https://arxiv.org/abs/1707.06253} {arXiv:1707.06253 [physics]} \BibitemShut
  {NoStop}%
\bibitem [{\citenamefont {Fisher-Levine}\ and\ \citenamefont
  {Nomerotski}()}]{FisherLevine:JInst11:C03016}%
  \BibitemOpen
  \bibfield  {author} {\bibinfo {author} {\bibfnamefont {M.}~\bibnamefont
  {Fisher-Levine}}\ and\ \bibinfo {author} {\bibfnamefont {A.}~\bibnamefont
  {Nomerotski}},\ }\bibfield  {title} {\bibinfo {title} {{TimepixCam}: a fast
  optical imager with time-stamping},\ }\href
  {https://doi.org/10.1088/1748-0221/11/03/C03016} {\bibfield  {journal}
  {\bibinfo  {journal} {J. Instrum.}\ }\textbf {\bibinfo {volume} {11}}\bibinfo
   {number} { (3)},\ \bibinfo {pages} {C03016}}\BibitemShut {NoStop}%
\bibitem [{\citenamefont {Nomerotski}(2019)}]{Nomerotski:NIMA937:26}%
  \BibitemOpen
\bibfield  {number} {  }\bibfield  {author} {\bibinfo {author} {\bibfnamefont
  {A.}~\bibnamefont {Nomerotski}},\ }\bibfield  {title} {\bibinfo {title}
  {Imaging and time stamping of photons with nanosecond resolution in timepix
  based optical cameras},\ }\href {https://doi.org/10.1016/j.nima.2019.05.034}
  {\bibfield  {journal} {\bibinfo  {journal} {Nucl. Instrum. Meth. A}\ }\textbf
  {\bibinfo {volume} {937}},\ \bibinfo {pages} {26} (\bibinfo {year}
  {2019})}\BibitemShut {NoStop}%
\bibitem [{\citenamefont {Roberts}\ \emph {et~al.}(2019)\citenamefont
  {Roberts}, \citenamefont {Svihra}, \citenamefont {Al-Refaie}, \citenamefont
  {Graafsma}, \citenamefont {Küpper}, \citenamefont {Majumdar}, \citenamefont
  {Mavrokoridis}, \citenamefont {Nomerotski}, \citenamefont {Pennicard},
  \citenamefont {Philippou}, \citenamefont {Trippel}, \citenamefont
  {Touramanis},\ and\ \citenamefont {Vann}}]{Roberts:JInst14:P06001}%
  \BibitemOpen
  \bibfield  {author} {\bibinfo {author} {\bibfnamefont {A.}~\bibnamefont
  {Roberts}}, \bibinfo {author} {\bibfnamefont {P.}~\bibnamefont {Svihra}},
  \bibinfo {author} {\bibfnamefont {A.}~\bibnamefont {Al-Refaie}}, \bibinfo
  {author} {\bibfnamefont {H.}~\bibnamefont {Graafsma}}, \bibinfo {author}
  {\bibfnamefont {J.}~\bibnamefont {Küpper}}, \bibinfo {author} {\bibfnamefont
  {K.}~\bibnamefont {Majumdar}}, \bibinfo {author} {\bibfnamefont
  {K.}~\bibnamefont {Mavrokoridis}}, \bibinfo {author} {\bibfnamefont
  {A.}~\bibnamefont {Nomerotski}}, \bibinfo {author} {\bibfnamefont
  {D.}~\bibnamefont {Pennicard}}, \bibinfo {author} {\bibfnamefont
  {B.}~\bibnamefont {Philippou}}, \bibinfo {author} {\bibfnamefont
  {S.}~\bibnamefont {Trippel}}, \bibinfo {author} {\bibfnamefont
  {C.}~\bibnamefont {Touramanis}},\ and\ \bibinfo {author} {\bibfnamefont
  {J.}~\bibnamefont {Vann}},\ }\bibfield  {title} {\bibinfo {title} {First
  demonstration of {3D} optical readout of a {TPC} using a single photon
  sensitive {T}imepix3 based camera},\ }\href
  {https://doi.org/10.1088/1748-0221/14/06/p06001} {\bibfield  {journal}
  {\bibinfo  {journal} {J. Instrum.}\ }\textbf {\bibinfo {volume}
  {14}}\bibfield  {number} {\bibinfo  {number} { (06)},\ \bibinfo {pages}
  {P06001}},\ }\Eprint {https://arxiv.org/abs/1810.09955} {arXiv:1810.09955
  [physics]} \BibitemShut {NoStop}%
\bibitem [{\citenamefont {Vallance}\ \emph {et~al.}(2014)\citenamefont
  {Vallance}, \citenamefont {Brouard}, \citenamefont {Lauer}, \citenamefont
  {Slater}, \citenamefont {Halford}, \citenamefont {Winter}, \citenamefont
  {King}, \citenamefont {Lee}, \citenamefont {Pooley}, \citenamefont
  {Sedgwick}, \citenamefont {Turchetta}, \citenamefont {Nomerotski},
  \citenamefont {John},\ and\ \citenamefont {Hill}}]{Vallance:PCCP16:383}%
  \BibitemOpen
  \bibfield  {author} {\bibinfo {author} {\bibfnamefont {C.}~\bibnamefont
  {Vallance}}, \bibinfo {author} {\bibfnamefont {M.}~\bibnamefont {Brouard}},
  \bibinfo {author} {\bibfnamefont {A.}~\bibnamefont {Lauer}}, \bibinfo
  {author} {\bibfnamefont {C.~S.}\ \bibnamefont {Slater}}, \bibinfo {author}
  {\bibfnamefont {E.}~\bibnamefont {Halford}}, \bibinfo {author} {\bibfnamefont
  {B.}~\bibnamefont {Winter}}, \bibinfo {author} {\bibfnamefont {S.~J.}\
  \bibnamefont {King}}, \bibinfo {author} {\bibfnamefont {J.~W.~L.}\
  \bibnamefont {Lee}}, \bibinfo {author} {\bibfnamefont {D.~E.}\ \bibnamefont
  {Pooley}}, \bibinfo {author} {\bibfnamefont {I.}~\bibnamefont {Sedgwick}},
  \bibinfo {author} {\bibfnamefont {R.}~\bibnamefont {Turchetta}}, \bibinfo
  {author} {\bibfnamefont {A.}~\bibnamefont {Nomerotski}}, \bibinfo {author}
  {\bibfnamefont {J.~J.}\ \bibnamefont {John}},\ and\ \bibinfo {author}
  {\bibfnamefont {L.}~\bibnamefont {Hill}},\ }\bibfield  {title} {\bibinfo
  {title} {{Fast sensors for time-of-flight imaging applications}},\ }\href
  {https://doi.org/10.1039/c3cp53183j} {\bibfield  {journal} {\bibinfo
  {journal} {Phys. Chem. Chem. Phys.}\ }\textbf {\bibinfo {volume} {16}},\
  \bibinfo {pages} {383 } (\bibinfo {year} {2014})}\BibitemShut {NoStop}%
\bibitem [{\citenamefont {Lee}\ \emph {et~al.}(2020)\citenamefont {Lee},
  \citenamefont {K\"ockert}, \citenamefont {Heathcote}, \citenamefont {Popat},
  \citenamefont {Chapman}, \citenamefont {Karras}, \citenamefont {Majchrzak},
  \citenamefont {Springate},\ and\ \citenamefont
  {Vallance}}]{Lee:CommChem3:72}%
  \BibitemOpen
  \bibfield  {author} {\bibinfo {author} {\bibfnamefont {J.~W.~L.}\
  \bibnamefont {Lee}}, \bibinfo {author} {\bibfnamefont {H.}~\bibnamefont
  {K\"ockert}}, \bibinfo {author} {\bibfnamefont {D.}~\bibnamefont
  {Heathcote}}, \bibinfo {author} {\bibfnamefont {D.}~\bibnamefont {Popat}},
  \bibinfo {author} {\bibfnamefont {R.~T.}\ \bibnamefont {Chapman}}, \bibinfo
  {author} {\bibfnamefont {G.}~\bibnamefont {Karras}}, \bibinfo {author}
  {\bibfnamefont {P.}~\bibnamefont {Majchrzak}}, \bibinfo {author}
  {\bibfnamefont {E.}~\bibnamefont {Springate}},\ and\ \bibinfo {author}
  {\bibfnamefont {C.}~\bibnamefont {Vallance}},\ }\bibfield  {title} {\bibinfo
  {title} {Three-dimensional covariance-map imaging of molecular structure and
  dynamics on the ultrafast timescale},\ }\href
  {https://doi.org/10.1038/s42004-020-0320-3} {\bibfield  {journal} {\bibinfo
  {journal} {Comm. Chem}\ }\textbf {\bibinfo {volume} {3}},\ \bibinfo {pages}
  {72} (\bibinfo {year} {2020})}\BibitemShut {NoStop}%
\bibitem [{\citenamefont {Mecseki}\ \emph {et~al.}(2019)\citenamefont
  {Mecseki}, \citenamefont {Windeler}, \citenamefont {Miahnahri}, \citenamefont
  {Robinson}, \citenamefont {Fraser}, \citenamefont {Fry},\ and\ \citenamefont
  {Tavella}}]{Mecseki:OptLett44:1257}%
  \BibitemOpen
  \bibfield  {author} {\bibinfo {author} {\bibfnamefont {K.}~\bibnamefont
  {Mecseki}}, \bibinfo {author} {\bibfnamefont {M.~K.~R.}\ \bibnamefont
  {Windeler}}, \bibinfo {author} {\bibfnamefont {A.}~\bibnamefont {Miahnahri}},
  \bibinfo {author} {\bibfnamefont {J.~S.}\ \bibnamefont {Robinson}}, \bibinfo
  {author} {\bibfnamefont {J.~M.}\ \bibnamefont {Fraser}}, \bibinfo {author}
  {\bibfnamefont {A.~R.}\ \bibnamefont {Fry}},\ and\ \bibinfo {author}
  {\bibfnamefont {F.}~\bibnamefont {Tavella}},\ }\bibfield  {title} {\bibinfo
  {title} {High average power 88~{W OPCPA} system for high-repetition-rate
  experiments at the {LCLS} x-ray free-electron laser},\ }\href
  {https://doi.org/10.1364/OL.44.001257} {\bibfield  {journal} {\bibinfo
  {journal} {Opt. Lett.}\ }\textbf {\bibinfo {volume} {44}},\ \bibinfo {pages}
  {1257} (\bibinfo {year} {2019})}\BibitemShut {NoStop}%
\bibitem [{\citenamefont {Tiedtke}\ \emph {et~al.}(2009)\citenamefont
  {Tiedtke}, \citenamefont {Azima}, \citenamefont {von Bargen}, \citenamefont
  {Bittner}, \citenamefont {Bonfigt}, \citenamefont {Düsterer}, \citenamefont
  {Faatz}, \citenamefont {Frühling}, \citenamefont {Gensch}, \citenamefont
  {Gerth}, \citenamefont {Guerassimova}, \citenamefont {Hahn}, \citenamefont
  {Hans}, \citenamefont {Hesse}, \citenamefont {Honkavaar}, \citenamefont
  {Jastrow}, \citenamefont {Juranic}, \citenamefont {Kapitzki}, \citenamefont
  {Keitel}, \citenamefont {Kracht}, \citenamefont {Kuhlmann}, \citenamefont
  {Li}, \citenamefont {Martins}, \citenamefont {N{\'{u}}{\~{n}}ez},
  \citenamefont {Plönjes}, \citenamefont {Redlin}, \citenamefont {Saldin},
  \citenamefont {Schneidmiller}, \citenamefont {Schneider}, \citenamefont
  {Schreiber}, \citenamefont {Stojanovic}, \citenamefont {Tavella},
  \citenamefont {Toleikis}, \citenamefont {Treusch}, \citenamefont {Weigelt},
  \citenamefont {Wellhöfer}, \citenamefont {Wabnitz}, \citenamefont {Yurkov},\
  and\ \citenamefont {Feldhaus}}]{Tiedtke:NJP11:023029}%
  \BibitemOpen
  \bibfield  {author} {\bibinfo {author} {\bibfnamefont {K.}~\bibnamefont
  {Tiedtke}}, \bibinfo {author} {\bibfnamefont {A.}~\bibnamefont {Azima}},
  \bibinfo {author} {\bibfnamefont {N.}~\bibnamefont {von Bargen}}, \bibinfo
  {author} {\bibfnamefont {L.}~\bibnamefont {Bittner}}, \bibinfo {author}
  {\bibfnamefont {S.}~\bibnamefont {Bonfigt}}, \bibinfo {author} {\bibfnamefont
  {S.}~\bibnamefont {Düsterer}}, \bibinfo {author} {\bibfnamefont
  {B.}~\bibnamefont {Faatz}}, \bibinfo {author} {\bibfnamefont
  {U.}~\bibnamefont {Frühling}}, \bibinfo {author} {\bibfnamefont
  {M.}~\bibnamefont {Gensch}}, \bibinfo {author} {\bibfnamefont
  {C.}~\bibnamefont {Gerth}}, \bibinfo {author} {\bibfnamefont
  {N.}~\bibnamefont {Guerassimova}}, \bibinfo {author} {\bibfnamefont
  {U.}~\bibnamefont {Hahn}}, \bibinfo {author} {\bibfnamefont {T.}~\bibnamefont
  {Hans}}, \bibinfo {author} {\bibfnamefont {M.}~\bibnamefont {Hesse}},
  \bibinfo {author} {\bibfnamefont {K.}~\bibnamefont {Honkavaar}}, \bibinfo
  {author} {\bibfnamefont {U.}~\bibnamefont {Jastrow}}, \bibinfo {author}
  {\bibfnamefont {P.}~\bibnamefont {Juranic}}, \bibinfo {author} {\bibfnamefont
  {S.}~\bibnamefont {Kapitzki}}, \bibinfo {author} {\bibfnamefont
  {B.}~\bibnamefont {Keitel}}, \bibinfo {author} {\bibfnamefont
  {T.}~\bibnamefont {Kracht}}, \bibinfo {author} {\bibfnamefont
  {M.}~\bibnamefont {Kuhlmann}}, \bibinfo {author} {\bibfnamefont {W.~B.}\
  \bibnamefont {Li}}, \bibinfo {author} {\bibfnamefont {M.}~\bibnamefont
  {Martins}}, \bibinfo {author} {\bibfnamefont {T.}~\bibnamefont
  {N{\'{u}}{\~{n}}ez}}, \bibinfo {author} {\bibfnamefont {E.}~\bibnamefont
  {Plönjes}}, \bibinfo {author} {\bibfnamefont {H.}~\bibnamefont {Redlin}},
  \bibinfo {author} {\bibfnamefont {E.~L.}\ \bibnamefont {Saldin}}, \bibinfo
  {author} {\bibfnamefont {E.~A.}\ \bibnamefont {Schneidmiller}}, \bibinfo
  {author} {\bibfnamefont {J.~R.}\ \bibnamefont {Schneider}}, \bibinfo {author}
  {\bibfnamefont {S.}~\bibnamefont {Schreiber}}, \bibinfo {author}
  {\bibfnamefont {N.}~\bibnamefont {Stojanovic}}, \bibinfo {author}
  {\bibfnamefont {F.}~\bibnamefont {Tavella}}, \bibinfo {author} {\bibfnamefont
  {S.}~\bibnamefont {Toleikis}}, \bibinfo {author} {\bibfnamefont
  {R.}~\bibnamefont {Treusch}}, \bibinfo {author} {\bibfnamefont
  {H.}~\bibnamefont {Weigelt}}, \bibinfo {author} {\bibfnamefont
  {M.}~\bibnamefont {Wellhöfer}}, \bibinfo {author} {\bibfnamefont
  {H.}~\bibnamefont {Wabnitz}}, \bibinfo {author} {\bibfnamefont {M.~V.}\
  \bibnamefont {Yurkov}},\ and\ \bibinfo {author} {\bibfnamefont
  {J.}~\bibnamefont {Feldhaus}},\ }\bibfield  {title} {\bibinfo {title} {The
  soft x-ray free-electron laser {FLASH} at {DESY}: beamlines, diagnostics and
  end-stations},\ }\href {https://doi.org/10.1088/1367-2630/11/2/023029}
  {\bibfield  {journal} {\bibinfo  {journal} {New J. Phys.}\ }\textbf {\bibinfo
  {volume} {11}},\ \bibinfo {pages} {023029} (\bibinfo {year}
  {2009})}\BibitemShut {NoStop}%
\bibitem [{\citenamefont {Fernandes}(2019)}]{Fernandes:gitlab}%
  \BibitemOpen
  \bibfield  {author} {\bibinfo {author} {\bibfnamefont {B.}~\bibnamefont
  {Fernandes}},\ }\href {https://git.xfel.eu/gitlab/fpga/trainidUSB} {}\bibinfo
  {howpublished} {Code repository, URL:
  \url{https://git.xfel.eu/gitlab/fpga/trainidUSB}} (\bibinfo {year}
  {2019})\BibitemShut {NoStop}%
\bibitem [{\citenamefont {Savelyev}\ \emph {et~al.}(2017)\citenamefont
  {Savelyev}, \citenamefont {Boll}, \citenamefont {Bomme}, \citenamefont
  {Schirmel}, \citenamefont {Redlin}, \citenamefont {Erk}, \citenamefont
  {D{\"u}sterer}, \citenamefont {M{\"u}ller}, \citenamefont {H{\"o}ppner},
  \citenamefont {Toleikis}, \citenamefont {M{\"u}ller}, \citenamefont
  {Kristin~Czwalinna}, \citenamefont {Treusch}, \citenamefont {Kierspel},
  \citenamefont {Mullins}, \citenamefont {Trippel}, \citenamefont {Wiese},
  \citenamefont {K{\"u}pper}, \citenamefont {Brauße}, \citenamefont
  {Krecinic}, \citenamefont {Rouz{\'e}e}, \citenamefont {Rudawski},
  \citenamefont {Johnsson}, \citenamefont {Amini}, \citenamefont {Lauer},
  \citenamefont {Burt}, \citenamefont {Brouard}, \citenamefont {Christensen},
  \citenamefont {Th{\o}gersen}, \citenamefont {Stapelfeldt}, \citenamefont
  {Berrah}, \citenamefont {M{\"u}ller}, \citenamefont {Ulmer}, \citenamefont
  {Techert}, \citenamefont {Rudenko},\ and\ \citenamefont
  {Rolles}}]{Savelyev:NJP19:043009}%
  \BibitemOpen
  \bibfield  {author} {\bibinfo {author} {\bibfnamefont {E.}~\bibnamefont
  {Savelyev}}, \bibinfo {author} {\bibfnamefont {R.}~\bibnamefont {Boll}},
  \bibinfo {author} {\bibfnamefont {C.}~\bibnamefont {Bomme}}, \bibinfo
  {author} {\bibfnamefont {N.}~\bibnamefont {Schirmel}}, \bibinfo {author}
  {\bibfnamefont {H.}~\bibnamefont {Redlin}}, \bibinfo {author} {\bibfnamefont
  {B.}~\bibnamefont {Erk}}, \bibinfo {author} {\bibfnamefont {S.}~\bibnamefont
  {D{\"u}sterer}}, \bibinfo {author} {\bibfnamefont {E.}~\bibnamefont
  {M{\"u}ller}}, \bibinfo {author} {\bibfnamefont {H.}~\bibnamefont
  {H{\"o}ppner}}, \bibinfo {author} {\bibfnamefont {S.}~\bibnamefont
  {Toleikis}}, \bibinfo {author} {\bibfnamefont {J.}~\bibnamefont
  {M{\"u}ller}}, \bibinfo {author} {\bibfnamefont {M.}~\bibnamefont
  {Kristin~Czwalinna}}, \bibinfo {author} {\bibfnamefont {R.}~\bibnamefont
  {Treusch}}, \bibinfo {author} {\bibfnamefont {T.}~\bibnamefont {Kierspel}},
  \bibinfo {author} {\bibfnamefont {T.}~\bibnamefont {Mullins}}, \bibinfo
  {author} {\bibfnamefont {S.}~\bibnamefont {Trippel}}, \bibinfo {author}
  {\bibfnamefont {J.}~\bibnamefont {Wiese}}, \bibinfo {author} {\bibfnamefont
  {J.}~\bibnamefont {K{\"u}pper}}, \bibinfo {author} {\bibfnamefont
  {F.}~\bibnamefont {Brauße}}, \bibinfo {author} {\bibfnamefont
  {F.}~\bibnamefont {Krecinic}}, \bibinfo {author} {\bibfnamefont
  {A.}~\bibnamefont {Rouz{\'e}e}}, \bibinfo {author} {\bibfnamefont
  {P.}~\bibnamefont {Rudawski}}, \bibinfo {author} {\bibfnamefont
  {P.}~\bibnamefont {Johnsson}}, \bibinfo {author} {\bibfnamefont
  {K.}~\bibnamefont {Amini}}, \bibinfo {author} {\bibfnamefont
  {A.}~\bibnamefont {Lauer}}, \bibinfo {author} {\bibfnamefont
  {M.}~\bibnamefont {Burt}}, \bibinfo {author} {\bibfnamefont {M.}~\bibnamefont
  {Brouard}}, \bibinfo {author} {\bibfnamefont {L.}~\bibnamefont
  {Christensen}}, \bibinfo {author} {\bibfnamefont {J.}~\bibnamefont
  {Th{\o}gersen}}, \bibinfo {author} {\bibfnamefont {H.}~\bibnamefont
  {Stapelfeldt}}, \bibinfo {author} {\bibfnamefont {N.}~\bibnamefont {Berrah}},
  \bibinfo {author} {\bibfnamefont {M.}~\bibnamefont {M{\"u}ller}}, \bibinfo
  {author} {\bibfnamefont {A.}~\bibnamefont {Ulmer}}, \bibinfo {author}
  {\bibfnamefont {S.}~\bibnamefont {Techert}}, \bibinfo {author} {\bibfnamefont
  {A.}~\bibnamefont {Rudenko}},\ and\ \bibinfo {author} {\bibfnamefont
  {D.}~\bibnamefont {Rolles}},\ }\bibfield  {title} {\bibinfo {title}
  {Jitter-correction for {IR/UV-XUV} pump-probe experiments at the {FLASH}
  free-electron laser},\ }\href {https://doi.org/10.1088/1367-2630/aa652d}
  {\bibfield  {journal} {\bibinfo  {journal} {New J. Phys.}\ }\textbf {\bibinfo
  {volume} {19}},\ \bibinfo {pages} {043009} (\bibinfo {year}
  {2017})}\BibitemShut {NoStop}%
\bibitem [{\citenamefont {Ott}\ \emph {et~al.}(2019)\citenamefont {Ott},
  \citenamefont {Aufleger}, \citenamefont {Ding}, \citenamefont {Rebholz},
  \citenamefont {Magunia}, \citenamefont {Hartmann}, \citenamefont {Stoo\ss{}},
  \citenamefont {Wachs}, \citenamefont {Birk}, \citenamefont {Borisova},
  \citenamefont {Meyer}, \citenamefont {Rupprecht}, \citenamefont
  {da~Costa~Castanheira}, \citenamefont {Moshammer}, \citenamefont {Attar},
  \citenamefont {Gaumnitz}, \citenamefont {Loh}, \citenamefont {D\"usterer},
  \citenamefont {Treusch}, \citenamefont {Ullrich}, \citenamefont {Jiang},
  \citenamefont {Meyer}, \citenamefont {Lambropoulos},\ and\ \citenamefont
  {Pfeifer}}]{Ott:PRL123:163201}%
  \BibitemOpen
  \bibfield  {author} {\bibinfo {author} {\bibfnamefont {C.}~\bibnamefont
  {Ott}}, \bibinfo {author} {\bibfnamefont {L.}~\bibnamefont {Aufleger}},
  \bibinfo {author} {\bibfnamefont {T.}~\bibnamefont {Ding}}, \bibinfo {author}
  {\bibfnamefont {M.}~\bibnamefont {Rebholz}}, \bibinfo {author} {\bibfnamefont
  {A.}~\bibnamefont {Magunia}}, \bibinfo {author} {\bibfnamefont
  {M.}~\bibnamefont {Hartmann}}, \bibinfo {author} {\bibfnamefont
  {V.}~\bibnamefont {Stoo\ss{}}}, \bibinfo {author} {\bibfnamefont
  {D.}~\bibnamefont {Wachs}}, \bibinfo {author} {\bibfnamefont
  {P.}~\bibnamefont {Birk}}, \bibinfo {author} {\bibfnamefont {G.~D.}\
  \bibnamefont {Borisova}}, \bibinfo {author} {\bibfnamefont {K.}~\bibnamefont
  {Meyer}}, \bibinfo {author} {\bibfnamefont {P.}~\bibnamefont {Rupprecht}},
  \bibinfo {author} {\bibfnamefont {C.}~\bibnamefont {da~Costa~Castanheira}},
  \bibinfo {author} {\bibfnamefont {R.}~\bibnamefont {Moshammer}}, \bibinfo
  {author} {\bibfnamefont {A.~R.}\ \bibnamefont {Attar}}, \bibinfo {author}
  {\bibfnamefont {T.}~\bibnamefont {Gaumnitz}}, \bibinfo {author}
  {\bibfnamefont {Z.-H.}\ \bibnamefont {Loh}}, \bibinfo {author} {\bibfnamefont
  {S.}~\bibnamefont {D\"usterer}}, \bibinfo {author} {\bibfnamefont
  {R.}~\bibnamefont {Treusch}}, \bibinfo {author} {\bibfnamefont
  {J.}~\bibnamefont {Ullrich}}, \bibinfo {author} {\bibfnamefont
  {Y.}~\bibnamefont {Jiang}}, \bibinfo {author} {\bibfnamefont
  {M.}~\bibnamefont {Meyer}}, \bibinfo {author} {\bibfnamefont
  {P.}~\bibnamefont {Lambropoulos}},\ and\ \bibinfo {author} {\bibfnamefont
  {T.}~\bibnamefont {Pfeifer}},\ }\bibfield  {title} {\bibinfo {title}
  {Strong-field extreme-ultraviolet dressing of atomic double excitation},\
  }\href {https://doi.org/10.1103/PhysRevLett.123.163201} {\bibfield  {journal}
  {\bibinfo  {journal} {Phys. Rev. Lett.}\ }\textbf {\bibinfo {volume} {123}},\
  \bibinfo {pages} {163201} (\bibinfo {year} {2019})}\BibitemShut {NoStop}%
\bibitem [{\citenamefont {Fisher-Levine}\ \emph {et~al.}(2018)\citenamefont
  {Fisher-Levine}, \citenamefont {Boll}, \citenamefont {Ziaee}, \citenamefont
  {Bomme}, \citenamefont {Erk}, \citenamefont {Rompotis}, \citenamefont
  {Marchenko}, \citenamefont {Nomerotski},\ and\ \citenamefont
  {Rolles}}]{FisherLevine:JSR25:336}%
  \BibitemOpen
  \bibfield  {author} {\bibinfo {author} {\bibfnamefont {M.}~\bibnamefont
  {Fisher-Levine}}, \bibinfo {author} {\bibfnamefont {R.}~\bibnamefont {Boll}},
  \bibinfo {author} {\bibfnamefont {F.}~\bibnamefont {Ziaee}}, \bibinfo
  {author} {\bibfnamefont {C.}~\bibnamefont {Bomme}}, \bibinfo {author}
  {\bibfnamefont {B.}~\bibnamefont {Erk}}, \bibinfo {author} {\bibfnamefont
  {D.}~\bibnamefont {Rompotis}}, \bibinfo {author} {\bibfnamefont
  {T.}~\bibnamefont {Marchenko}}, \bibinfo {author} {\bibfnamefont
  {A.}~\bibnamefont {Nomerotski}},\ and\ \bibinfo {author} {\bibfnamefont
  {D.}~\bibnamefont {Rolles}},\ }\bibfield  {title} {\bibinfo {title}
  {Time-resolved ion imaging at free-electron lasers using {TimepixCam}},\
  }\href {https://doi.org/10.1107/S1600577517018306} {\bibfield  {journal}
  {\bibinfo  {journal} {J. Synchrotron\ Rad.}\ }\textbf {\bibinfo {volume}
  {25}},\ \bibinfo {pages} {336} (\bibinfo {year} {2018})}\BibitemShut
  {NoStop}%
\bibitem [{\citenamefont {Johny}\ \emph {et~al.}(2020)\citenamefont {Johny},
  \citenamefont {Schouder}, \citenamefont {Al-Refaie}, \citenamefont {He},
  \citenamefont {Wiese}, \citenamefont {Stapelfeldt}, \citenamefont {Trippel},\
  and\ \citenamefont {Küpper}}]{Johny:protection:inprep}%
  \BibitemOpen
  \bibfield  {author} {\bibinfo {author} {\bibfnamefont {M.}~\bibnamefont
  {Johny}}, \bibinfo {author} {\bibfnamefont {C.~A.}\ \bibnamefont {Schouder}},
  \bibinfo {author} {\bibfnamefont {A.}~\bibnamefont {Al-Refaie}}, \bibinfo
  {author} {\bibfnamefont {L.}~\bibnamefont {He}}, \bibinfo {author}
  {\bibfnamefont {J.}~\bibnamefont {Wiese}}, \bibinfo {author} {\bibfnamefont
  {H.}~\bibnamefont {Stapelfeldt}}, \bibinfo {author} {\bibfnamefont
  {S.}~\bibnamefont {Trippel}},\ and\ \bibinfo {author} {\bibfnamefont
  {J.}~\bibnamefont {Küpper}},\ }\href {https://arxiv.org/pdf/2010.00453.pdf}
  {\bibinfo {title} {Molecular sunscreen: water protects pyrrole from radiation
  damage}} (\bibinfo {year} {2020}),\ \bibinfo {note} {submitted},\ \Eprint
  {https://arxiv.org/abs/2010.00453} {arXiv:2010.00453 [physics]} \BibitemShut
  {NoStop}%
\bibitem [{\citenamefont {Al-Refaie}\ \emph {et~al.}(2019)\citenamefont
  {Al-Refaie}, \citenamefont {Johny}, \citenamefont {Correa}, \citenamefont
  {Pennicard}, \citenamefont {Svihra}, \citenamefont {Nomerotski},
  \citenamefont {Trippel},\ and\ \citenamefont
  {Küpper}}]{AlRefaie:JInst14:P10003}%
  \BibitemOpen
  \bibfield  {author} {\bibinfo {author} {\bibfnamefont {A.}~\bibnamefont
  {Al-Refaie}}, \bibinfo {author} {\bibfnamefont {M.}~\bibnamefont {Johny}},
  \bibinfo {author} {\bibfnamefont {J.}~\bibnamefont {Correa}}, \bibinfo
  {author} {\bibfnamefont {D.}~\bibnamefont {Pennicard}}, \bibinfo {author}
  {\bibfnamefont {P.}~\bibnamefont {Svihra}}, \bibinfo {author} {\bibfnamefont
  {A.}~\bibnamefont {Nomerotski}}, \bibinfo {author} {\bibfnamefont
  {S.}~\bibnamefont {Trippel}},\ and\ \bibinfo {author} {\bibfnamefont
  {J.}~\bibnamefont {Küpper}},\ }\bibfield  {title} {\bibinfo {title}
  {{PymePix}: A python library for {SPIDR} readout of {T}imepix3},\ }\href
  {https://doi.org/10.1088/1748-0221/14/10/P10003} {\bibfield  {journal}
  {\bibinfo  {journal} {J. Instrum.}\ }\textbf {\bibinfo {volume}
  {14}}\bibfield  {number} {\bibinfo  {number} { (10)},\ \bibinfo {pages}
  {P10003}},\ }\Eprint {https://arxiv.org/abs/1905.07999} {arXiv:1905.07999
  [physics]} \BibitemShut {NoStop}%
\bibitem [{\citenamefont {Harris}\ \emph {et~al.}(2020)\citenamefont {Harris},
  \citenamefont {Millman}, \citenamefont {van~der Walt}, \citenamefont
  {Gommers}, \citenamefont {Virtanen}, \citenamefont {Cournapeau},
  \citenamefont {Wieser}, \citenamefont {Taylor}, \citenamefont {Berg},
  \citenamefont {Smith}, \citenamefont {Kern}, \citenamefont {Picus},
  \citenamefont {Hoyer}, \citenamefont {van Kerkwijk}, \citenamefont {Brett},
  \citenamefont {Haldane}, \citenamefont {del R{\'{i}}o}, \citenamefont
  {Wiebe}, \citenamefont {Peterson}, \citenamefont {G{\'{e}}rard-Marchant},
  \citenamefont {Sheppard}, \citenamefont {Reddy}, \citenamefont {Weckesser},
  \citenamefont {Abbasi}, \citenamefont {Gohlke},\ and\ \citenamefont
  {Oliphant}}]{Harris:Nature585:7825}%
  \BibitemOpen
  \bibfield  {author} {\bibinfo {author} {\bibfnamefont {C.~R.}\ \bibnamefont
  {Harris}}, \bibinfo {author} {\bibfnamefont {K.~J.}\ \bibnamefont {Millman}},
  \bibinfo {author} {\bibfnamefont {S.~J.}\ \bibnamefont {van~der Walt}},
  \bibinfo {author} {\bibfnamefont {R.}~\bibnamefont {Gommers}}, \bibinfo
  {author} {\bibfnamefont {P.}~\bibnamefont {Virtanen}}, \bibinfo {author}
  {\bibfnamefont {D.}~\bibnamefont {Cournapeau}}, \bibinfo {author}
  {\bibfnamefont {E.}~\bibnamefont {Wieser}}, \bibinfo {author} {\bibfnamefont
  {J.}~\bibnamefont {Taylor}}, \bibinfo {author} {\bibfnamefont
  {S.}~\bibnamefont {Berg}}, \bibinfo {author} {\bibfnamefont {N.~J.}\
  \bibnamefont {Smith}}, \bibinfo {author} {\bibfnamefont {R.}~\bibnamefont
  {Kern}}, \bibinfo {author} {\bibfnamefont {M.}~\bibnamefont {Picus}},
  \bibinfo {author} {\bibfnamefont {S.}~\bibnamefont {Hoyer}}, \bibinfo
  {author} {\bibfnamefont {M.~H.}\ \bibnamefont {van Kerkwijk}}, \bibinfo
  {author} {\bibfnamefont {M.}~\bibnamefont {Brett}}, \bibinfo {author}
  {\bibfnamefont {A.}~\bibnamefont {Haldane}}, \bibinfo {author} {\bibfnamefont
  {J.~F.}\ \bibnamefont {del R{\'{i}}o}}, \bibinfo {author} {\bibfnamefont
  {M.}~\bibnamefont {Wiebe}}, \bibinfo {author} {\bibfnamefont
  {P.}~\bibnamefont {Peterson}}, \bibinfo {author} {\bibfnamefont
  {P.}~\bibnamefont {G{\'{e}}rard-Marchant}}, \bibinfo {author} {\bibfnamefont
  {K.}~\bibnamefont {Sheppard}}, \bibinfo {author} {\bibfnamefont
  {T.}~\bibnamefont {Reddy}}, \bibinfo {author} {\bibfnamefont
  {W.}~\bibnamefont {Weckesser}}, \bibinfo {author} {\bibfnamefont
  {H.}~\bibnamefont {Abbasi}}, \bibinfo {author} {\bibfnamefont
  {C.}~\bibnamefont {Gohlke}},\ and\ \bibinfo {author} {\bibfnamefont {T.~E.}\
  \bibnamefont {Oliphant}},\ }\bibfield  {title} {\bibinfo {title} {Array
  programming with {NumPy}},\ }\href
  {https://doi.org/10.1038/s41586-020-2649-2} {\bibfield  {journal} {\bibinfo
  {journal} {Nature}\ }\textbf {\bibinfo {volume} {585}},\ \bibinfo {pages}
  {357} (\bibinfo {year} {2020})}\BibitemShut {NoStop}%
\bibitem [{\citenamefont {Pedregosa}\ \emph {et~al.}(2011)\citenamefont
  {Pedregosa}, \citenamefont {Varoquaux}, \citenamefont {Gramfort},
  \citenamefont {Michel}, \citenamefont {Thirion}, \citenamefont {Grisel},
  \citenamefont {Blondel}, \citenamefont {Prettenhofer}, \citenamefont {Weiss},
  \citenamefont {Dubourg}, \citenamefont {Vanderplas}, \citenamefont {Passos},
  \citenamefont {Cournapeau}, \citenamefont {Brucher}, \citenamefont {Perrot},\
  and\ \citenamefont {Duchesnay}}]{Pedregosa:JMLR12:2825}%
  \BibitemOpen
  \bibfield  {author} {\bibinfo {author} {\bibfnamefont {F.}~\bibnamefont
  {Pedregosa}}, \bibinfo {author} {\bibfnamefont {G.}~\bibnamefont
  {Varoquaux}}, \bibinfo {author} {\bibfnamefont {A.}~\bibnamefont {Gramfort}},
  \bibinfo {author} {\bibfnamefont {V.}~\bibnamefont {Michel}}, \bibinfo
  {author} {\bibfnamefont {B.}~\bibnamefont {Thirion}}, \bibinfo {author}
  {\bibfnamefont {O.}~\bibnamefont {Grisel}}, \bibinfo {author} {\bibfnamefont
  {M.}~\bibnamefont {Blondel}}, \bibinfo {author} {\bibfnamefont
  {P.}~\bibnamefont {Prettenhofer}}, \bibinfo {author} {\bibfnamefont
  {R.}~\bibnamefont {Weiss}}, \bibinfo {author} {\bibfnamefont
  {V.}~\bibnamefont {Dubourg}}, \bibinfo {author} {\bibfnamefont
  {J.}~\bibnamefont {Vanderplas}}, \bibinfo {author} {\bibfnamefont
  {A.}~\bibnamefont {Passos}}, \bibinfo {author} {\bibfnamefont
  {D.}~\bibnamefont {Cournapeau}}, \bibinfo {author} {\bibfnamefont
  {M.}~\bibnamefont {Brucher}}, \bibinfo {author} {\bibfnamefont
  {M.}~\bibnamefont {Perrot}},\ and\ \bibinfo {author} {\bibfnamefont
  {E.}~\bibnamefont {Duchesnay}},\ }\bibfield  {title} {\bibinfo {title}
  {Scikit-learn: Machine learning in {P}ython},\ }\href
  {https://dl.acm.org/citation.cfm?id=2078195} {\bibfield  {journal} {\bibinfo
  {journal} {J. Mach. Learn. Res.}\ }\textbf {\bibinfo {volume} {12}},\
  \bibinfo {pages} {2825} (\bibinfo {year} {2011})}\BibitemShut {NoStop}%
\bibitem [{\citenamefont {Virtanen}\ \emph {et~al.}(2020)\citenamefont
  {Virtanen}, \citenamefont {Gommers}, \citenamefont {Oliphant}, \citenamefont
  {Haberland}, \citenamefont {Reddy}, \citenamefont {Cournapeau}, \citenamefont
  {Burovski}, \citenamefont {Peterson}, \citenamefont {Weckesser},
  \citenamefont {Bright}, \citenamefont {{van der Walt}}, \citenamefont
  {Brett}, \citenamefont {Wilson}, \citenamefont {Millman}, \citenamefont
  {Mayorov}, \citenamefont {Nelson}, \citenamefont {Jones}, \citenamefont
  {Kern}, \citenamefont {Larson}, \citenamefont {Carey}, \citenamefont {Polat},
  \citenamefont {Feng}, \citenamefont {Moore}, \citenamefont {{VanderPlas}},
  \citenamefont {Laxalde}, \citenamefont {Perktold}, \citenamefont {Cimrman},
  \citenamefont {Henriksen}, \citenamefont {Quintero}, \citenamefont {Harris},
  \citenamefont {Archibald}, \citenamefont {Ribeiro}, \citenamefont
  {Pedregosa}, \citenamefont {{van Mulbregt}},\ and\ \citenamefont {{SciPy 1.0
  Contributors}}}]{Pauli:NatMeth17:261}%
  \BibitemOpen
  \bibfield  {author} {\bibinfo {author} {\bibfnamefont {P.}~\bibnamefont
  {Virtanen}}, \bibinfo {author} {\bibfnamefont {R.}~\bibnamefont {Gommers}},
  \bibinfo {author} {\bibfnamefont {T.~E.}\ \bibnamefont {Oliphant}}, \bibinfo
  {author} {\bibfnamefont {M.}~\bibnamefont {Haberland}}, \bibinfo {author}
  {\bibfnamefont {T.}~\bibnamefont {Reddy}}, \bibinfo {author} {\bibfnamefont
  {D.}~\bibnamefont {Cournapeau}}, \bibinfo {author} {\bibfnamefont
  {E.}~\bibnamefont {Burovski}}, \bibinfo {author} {\bibfnamefont
  {P.}~\bibnamefont {Peterson}}, \bibinfo {author} {\bibfnamefont
  {W.}~\bibnamefont {Weckesser}}, \bibinfo {author} {\bibfnamefont
  {J.}~\bibnamefont {Bright}}, \bibinfo {author} {\bibfnamefont {S.~J.}\
  \bibnamefont {{van der Walt}}}, \bibinfo {author} {\bibfnamefont
  {M.}~\bibnamefont {Brett}}, \bibinfo {author} {\bibfnamefont
  {J.}~\bibnamefont {Wilson}}, \bibinfo {author} {\bibfnamefont {K.~J.}\
  \bibnamefont {Millman}}, \bibinfo {author} {\bibfnamefont {N.}~\bibnamefont
  {Mayorov}}, \bibinfo {author} {\bibfnamefont {A.~R.~J.}\ \bibnamefont
  {Nelson}}, \bibinfo {author} {\bibfnamefont {E.}~\bibnamefont {Jones}},
  \bibinfo {author} {\bibfnamefont {R.}~\bibnamefont {Kern}}, \bibinfo {author}
  {\bibfnamefont {E.}~\bibnamefont {Larson}}, \bibinfo {author} {\bibfnamefont
  {C.~J.}\ \bibnamefont {Carey}}, \bibinfo {author} {\bibfnamefont
  {{\.I}.}~\bibnamefont {Polat}}, \bibinfo {author} {\bibfnamefont
  {Y.}~\bibnamefont {Feng}}, \bibinfo {author} {\bibfnamefont {E.~W.}\
  \bibnamefont {Moore}}, \bibinfo {author} {\bibfnamefont {J.}~\bibnamefont
  {{VanderPlas}}}, \bibinfo {author} {\bibfnamefont {D.}~\bibnamefont
  {Laxalde}}, \bibinfo {author} {\bibfnamefont {J.}~\bibnamefont {Perktold}},
  \bibinfo {author} {\bibfnamefont {R.}~\bibnamefont {Cimrman}}, \bibinfo
  {author} {\bibfnamefont {I.}~\bibnamefont {Henriksen}}, \bibinfo {author}
  {\bibfnamefont {E.~A.}\ \bibnamefont {Quintero}}, \bibinfo {author}
  {\bibfnamefont {C.~R.}\ \bibnamefont {Harris}}, \bibinfo {author}
  {\bibfnamefont {A.~M.}\ \bibnamefont {Archibald}}, \bibinfo {author}
  {\bibfnamefont {A.~H.}\ \bibnamefont {Ribeiro}}, \bibinfo {author}
  {\bibfnamefont {F.}~\bibnamefont {Pedregosa}}, \bibinfo {author}
  {\bibfnamefont {P.}~\bibnamefont {{van Mulbregt}}},\ and\ \bibinfo {author}
  {\bibnamefont {{SciPy 1.0 Contributors}}},\ }\bibfield  {title} {\bibinfo
  {title} {{{SciPy} 1.0: Fundamental Algorithms for Scientific Computing in
  Python}},\ }\href {https://doi.org/10.1038/s41592-019-0686-2} {\bibfield
  {journal} {\bibinfo  {journal} {Nat. Methods}\ }\textbf {\bibinfo {volume}
  {17}},\ \bibinfo {pages} {261} (\bibinfo {year} {2020})}\BibitemShut
  {NoStop}%
\bibitem [{\citenamefont {Hunter}(2007)}]{Hunter:CISE9:90}%
  \BibitemOpen
  \bibfield  {author} {\bibinfo {author} {\bibfnamefont {J.~D.}\ \bibnamefont
  {Hunter}},\ }\bibfield  {title} {\bibinfo {title} {Matplotlib: a {2D}
  graphics environment},\ }\href {https://doi.org/10.1109/MCSE.2007.55}
  {\bibfield  {journal} {\bibinfo  {journal} {Comp. Sci. \& Eng.}\ }\textbf
  {\bibinfo {volume} {9}},\ \bibinfo {pages} {90} (\bibinfo {year}
  {2007})}\BibitemShut {NoStop}%
\bibitem [{\citenamefont {{W}es {M}c{K}inney}(2010)}]{McKinney::56}%
  \BibitemOpen
  \bibfield  {author} {\bibinfo {author} {\bibnamefont {{W}es {M}c{K}inney}},\
  }\bibfield  {title} {\bibinfo {title} {{D}ata {S}tructures for {S}tatistical
  {C}omputing in {P}ython},\ }in\ \href
  {https://doi.org/10.25080/Majora-92bf1922-00a} {\emph {\bibinfo {booktitle}
  {{P}roceedings of the 9th {P}ython in {S}cience {C}onference}}},\ \bibinfo
  {editor} {edited by\ \bibinfo {editor} {\bibnamefont {{S}t\'efan van~der
  {W}alt}}\ and\ \bibinfo {editor} {\bibnamefont {{J}arrod {M}illman}}}\
  (\bibinfo {year} {2010})\ pp.\ \bibinfo {pages} {56 -- 61}\BibitemShut
  {NoStop}%
\bibitem [{\citenamefont {Hickstein}\ \emph {et~al.}(2019)\citenamefont
  {Hickstein}, \citenamefont {Gibson}, \citenamefont {Yurchak}, \citenamefont
  {Das},\ and\ \citenamefont {Ryazanov}}]{Hickstein:RSI90:065115}%
  \BibitemOpen
  \bibfield  {author} {\bibinfo {author} {\bibfnamefont {D.~D.}\ \bibnamefont
  {Hickstein}}, \bibinfo {author} {\bibfnamefont {S.~T.}\ \bibnamefont
  {Gibson}}, \bibinfo {author} {\bibfnamefont {R.}~\bibnamefont {Yurchak}},
  \bibinfo {author} {\bibfnamefont {D.~D.}\ \bibnamefont {Das}},\ and\ \bibinfo
  {author} {\bibfnamefont {M.}~\bibnamefont {Ryazanov}},\ }\bibfield  {title}
  {\bibinfo {title} {A direct comparison of high-speed methods for the
  numerical {Abel} transform},\ }\href {https://doi.org/10.1063/1.5092635}
  {\bibfield  {journal} {\bibinfo  {journal} {Rev. Sci. Instrum.}\ }\textbf
  {\bibinfo {volume} {90}},\ \bibinfo {pages} {065115} (\bibinfo {year}
  {2019})}\BibitemShut {NoStop}%
\bibitem [{\citenamefont {Tsigaridas}\ \emph {et~al.}(2019)\citenamefont
  {Tsigaridas}, \citenamefont {Beuzekom}, \citenamefont {Graaf}, \citenamefont
  {Hartjes}, \citenamefont {Heijhoff}, \citenamefont {Hessey}, \citenamefont
  {{de Jong}},\ and\ \citenamefont {Prodanovic}}]{Tsigaridas:NIMA930:185}%
  \BibitemOpen
  \bibfield  {author} {\bibinfo {author} {\bibfnamefont {S.}~\bibnamefont
  {Tsigaridas}}, \bibinfo {author} {\bibfnamefont {M.}~\bibnamefont
  {Beuzekom}}, \bibinfo {author} {\bibfnamefont {H.}~\bibnamefont {Graaf}},
  \bibinfo {author} {\bibfnamefont {F.}~\bibnamefont {Hartjes}}, \bibinfo
  {author} {\bibfnamefont {K.}~\bibnamefont {Heijhoff}}, \bibinfo {author}
  {\bibfnamefont {N.}~\bibnamefont {Hessey}}, \bibinfo {author} {\bibfnamefont
  {P.}~\bibnamefont {{de Jong}}},\ and\ \bibinfo {author} {\bibfnamefont
  {V.}~\bibnamefont {Prodanovic}},\ }\bibfield  {title} {\bibinfo {title}
  {Timewalk correction for the {Timepix3} chip obtained with real particle
  data},\ }\href {https://doi.org/10.1016/j.nima.2019.03.077} {\bibfield
  {journal} {\bibinfo  {journal} {Nucl. Instrum. Meth. A}\ }\textbf {\bibinfo
  {volume} {930}},\ \bibinfo {pages} {185} (\bibinfo {year}
  {2019})}\BibitemShut {NoStop}%
\bibitem [{\citenamefont {Turecek}\ \emph {et~al.}(2016)\citenamefont
  {Turecek}, \citenamefont {Jakubek},\ and\ \citenamefont
  {Soukup}}]{Turecek:JInst12:C12065}%
  \BibitemOpen
  \bibfield  {author} {\bibinfo {author} {\bibfnamefont {D.}~\bibnamefont
  {Turecek}}, \bibinfo {author} {\bibfnamefont {J.}~\bibnamefont {Jakubek}},\
  and\ \bibinfo {author} {\bibfnamefont {P.}~\bibnamefont {Soukup}},\
  }\bibfield  {title} {\bibinfo {title} {{USB} 3.0 readout and time-walk
  correction method for {T}imepix3 detector},\ }\href
  {https://doi.org/10.1088/1748-0221/11/12/c12065} {\bibfield  {journal}
  {\bibinfo  {journal} {J. Instrum.}\ }\textbf {\bibinfo {volume} {11}}\bibinfo
   {number} { (12)},\ \bibinfo {pages} {C12065}}\BibitemShut {NoStop}%
\bibitem [{\citenamefont {Pitters}\ \emph {et~al.}(2019)\citenamefont
  {Pitters}, \citenamefont {Tehrani}, \citenamefont {Dannheim}, \citenamefont
  {Fiergolski}, \citenamefont {Hynds}, \citenamefont {Klempt}, \citenamefont
  {Llopart}, \citenamefont {Munker}, \citenamefont {Nürnberg}, \citenamefont
  {Spannagel},\ and\ \citenamefont {Williams}}]{Pitters:JInst14:P05022}%
  \BibitemOpen
\bibfield  {number} {  }\bibfield  {author} {\bibinfo {author} {\bibfnamefont
  {F.}~\bibnamefont {Pitters}}, \bibinfo {author} {\bibfnamefont {N.~A.}\
  \bibnamefont {Tehrani}}, \bibinfo {author} {\bibfnamefont {D.}~\bibnamefont
  {Dannheim}}, \bibinfo {author} {\bibfnamefont {A.}~\bibnamefont
  {Fiergolski}}, \bibinfo {author} {\bibfnamefont {D.}~\bibnamefont {Hynds}},
  \bibinfo {author} {\bibfnamefont {W.}~\bibnamefont {Klempt}}, \bibinfo
  {author} {\bibfnamefont {X.}~\bibnamefont {Llopart}}, \bibinfo {author}
  {\bibfnamefont {M.}~\bibnamefont {Munker}}, \bibinfo {author} {\bibfnamefont
  {A.}~\bibnamefont {Nürnberg}}, \bibinfo {author} {\bibfnamefont
  {S.}~\bibnamefont {Spannagel}},\ and\ \bibinfo {author} {\bibfnamefont
  {M.}~\bibnamefont {Williams}},\ }\bibfield  {title} {\bibinfo {title} {Time
  resolution studies of {Timepix3} assemblies with thin silicon pixel
  sensors},\ }\href {https://doi.org/10.1088/1748-0221/14/05/p05022} {\bibfield
   {journal} {\bibinfo  {journal} {J. Instrum.}\ }\textbf {\bibinfo {volume}
  {14}}\bibinfo  {number} { (05)},\ \bibinfo {pages} {P05022}}\BibitemShut
  {NoStop}%
\bibitem [{\citenamefont {{Scientific Instrument Services Inc.,
  USA}}(2011)}]{Simion:8.1}%
  \BibitemOpen
\bibfield  {number} {  }\bibfield  {author} {\bibinfo {author} {\bibnamefont
  {{Scientific Instrument Services Inc., USA}}},\ }\href@noop {} {\bibinfo
  {title} {Simion 8.1}} (\bibinfo {year} {2011}),\ \bibinfo {note}
  {{URL}:~\url{http://simion.com}}\BibitemShut {NoStop}%
\bibitem [{\citenamefont {Li}\ \emph {et~al.}(2005)\citenamefont {Li},
  \citenamefont {Chambreau}, \citenamefont {Lahankar},\ and\ \citenamefont
  {Suits}}]{Li:RSI76:063106}%
  \BibitemOpen
  \bibfield  {author} {\bibinfo {author} {\bibfnamefont {W.}~\bibnamefont
  {Li}}, \bibinfo {author} {\bibfnamefont {S.~D.}\ \bibnamefont {Chambreau}},
  \bibinfo {author} {\bibfnamefont {S.~A.}\ \bibnamefont {Lahankar}},\ and\
  \bibinfo {author} {\bibfnamefont {A.~G.}\ \bibnamefont {Suits}},\ }\bibfield
  {title} {\bibinfo {title} {Megapixel ion imaging with standard video},\
  }\href {https://doi.org/10.1063/1.1921671} {\bibfield  {journal} {\bibinfo
  {journal} {Rev. Sci. Instrum.}\ }\textbf {\bibinfo {volume} {76}},\ \bibinfo
  {pages} {063106} (\bibinfo {year} {2005})}\BibitemShut {NoStop}%
\bibitem [{\citenamefont {Kella}\ \emph {et~al.}(1993)\citenamefont {Kella},
  \citenamefont {Algranati}, \citenamefont {Feldman}, \citenamefont {Heber},
  \citenamefont {Kovner}, \citenamefont {Malkin}, \citenamefont {Miklazky},
  \citenamefont {Naaman}, \citenamefont {Zajfman}, \citenamefont {Zajfman},\
  and\ \citenamefont {Vager}}]{Kella:NIMA329:440}%
  \BibitemOpen
  \bibfield  {author} {\bibinfo {author} {\bibfnamefont {D.}~\bibnamefont
  {Kella}}, \bibinfo {author} {\bibfnamefont {M.}~\bibnamefont {Algranati}},
  \bibinfo {author} {\bibfnamefont {H.}~\bibnamefont {Feldman}}, \bibinfo
  {author} {\bibfnamefont {O.}~\bibnamefont {Heber}}, \bibinfo {author}
  {\bibfnamefont {H.}~\bibnamefont {Kovner}}, \bibinfo {author} {\bibfnamefont
  {E.}~\bibnamefont {Malkin}}, \bibinfo {author} {\bibfnamefont
  {E.}~\bibnamefont {Miklazky}}, \bibinfo {author} {\bibfnamefont
  {R.}~\bibnamefont {Naaman}}, \bibinfo {author} {\bibfnamefont
  {D.}~\bibnamefont {Zajfman}}, \bibinfo {author} {\bibfnamefont
  {J.}~\bibnamefont {Zajfman}},\ and\ \bibinfo {author} {\bibfnamefont
  {Z.}~\bibnamefont {Vager}},\ }\bibfield  {title} {\bibinfo {title} {{A system
  for Coulomb explosion imaging of small molecules at the Weizmann
  Institute}},\ }\href {https://doi.org/10.1016/0168-9002(93)91279-V}
  {\bibfield  {journal} {\bibinfo  {journal} {Nucl. Instrum. Meth. A}\ }\textbf
  {\bibinfo {volume} {329}},\ \bibinfo {pages} {440} (\bibinfo {year}
  {1993})}\BibitemShut {NoStop}%
\bibitem [{\citenamefont {Wester}\ \emph {et~al.}(1998)\citenamefont {Wester},
  \citenamefont {Albrecht}, \citenamefont {Grieser}, \citenamefont {Knoll},
  \citenamefont {Repnow}, \citenamefont {Schwalm}, \citenamefont {Wolf},
  \citenamefont {Baer}, \citenamefont {Levin}, \citenamefont {Vager},\ and\
  \citenamefont {Zajfman}}]{Wester:NIMA413:379}%
  \BibitemOpen
  \bibfield  {author} {\bibinfo {author} {\bibfnamefont {R.}~\bibnamefont
  {Wester}}, \bibinfo {author} {\bibfnamefont {F.}~\bibnamefont {Albrecht}},
  \bibinfo {author} {\bibfnamefont {M.}~\bibnamefont {Grieser}}, \bibinfo
  {author} {\bibfnamefont {L.}~\bibnamefont {Knoll}}, \bibinfo {author}
  {\bibfnamefont {R.}~\bibnamefont {Repnow}}, \bibinfo {author} {\bibfnamefont
  {D.}~\bibnamefont {Schwalm}}, \bibinfo {author} {\bibfnamefont
  {A.}~\bibnamefont {Wolf}}, \bibinfo {author} {\bibfnamefont {A.}~\bibnamefont
  {Baer}}, \bibinfo {author} {\bibfnamefont {J.}~\bibnamefont {Levin}},
  \bibinfo {author} {\bibfnamefont {Z.}~\bibnamefont {Vager}},\ and\ \bibinfo
  {author} {\bibfnamefont {D.}~\bibnamefont {Zajfman}},\ }\bibfield  {title}
  {\bibinfo {title} {{Coulomb explosion imaging at the heavy ion storage ring
  TSR}},\ }\href {https://doi.org/10.1016/S0168-9002(98)00553-1} {\bibfield
  {journal} {\bibinfo  {journal} {Nucl. Instrum. Meth. A}\ }\textbf {\bibinfo
  {volume} {413}},\ \bibinfo {pages} {379} (\bibinfo {year}
  {1998})}\BibitemShut {NoStop}%
\bibitem [{\citenamefont {Chang}\ \emph {et~al.}(1998)\citenamefont {Chang},
  \citenamefont {Hoetzlein}, \citenamefont {Mueller}, \citenamefont {D.},\ and\
  \citenamefont {Houston}}]{Chang:RSI69:1665}%
  \BibitemOpen
  \bibfield  {author} {\bibinfo {author} {\bibfnamefont {B.~Y.}\ \bibnamefont
  {Chang}}, \bibinfo {author} {\bibfnamefont {R.~C.}\ \bibnamefont
  {Hoetzlein}}, \bibinfo {author} {\bibfnamefont {J.~A.}\ \bibnamefont
  {Mueller}}, \bibinfo {author} {\bibfnamefont {G.~J.}\ \bibnamefont {D.}},\
  and\ \bibinfo {author} {\bibfnamefont {P.~L.}\ \bibnamefont {Houston}},\
  }\bibfield  {title} {\bibinfo {title} {Improved two-dimensional product
  imaging: The real-time ion-counting method},\ }\href
  {https://doi.org/http://doi.org/10.1063/1.1148824} {\bibfield  {journal}
  {\bibinfo  {journal} {Rev. Sci. Instrum.}\ }\textbf {\bibinfo {volume}
  {69}},\ \bibinfo {pages} {1665} (\bibinfo {year} {1998})}\BibitemShut
  {NoStop}%
\bibitem [{\citenamefont {Winter}\ \emph {et~al.}(2014)\citenamefont {Winter},
  \citenamefont {King}, \citenamefont {Brouard},\ and\ \citenamefont
  {Vallance}}]{Winter:RSI85:0034}%
  \BibitemOpen
  \bibfield  {author} {\bibinfo {author} {\bibfnamefont {B.}~\bibnamefont
  {Winter}}, \bibinfo {author} {\bibfnamefont {S.~J.}\ \bibnamefont {King}},
  \bibinfo {author} {\bibfnamefont {M.}~\bibnamefont {Brouard}},\ and\ \bibinfo
  {author} {\bibfnamefont {C.}~\bibnamefont {Vallance}},\ }\bibfield  {title}
  {\bibinfo {title} {A fast microchannel plate-scintillator detector for
  velocity map imaging and imaging mass spectrometry},\ }\href
  {https://doi.org/10.1063/1.4866647} {\bibfield  {journal} {\bibinfo
  {journal} {Rev. Sci. Instrum.}\ }\textbf {\bibinfo {volume} {85}},\ \bibinfo
  {pages} {023306} (\bibinfo {year} {2014})}\BibitemShut {NoStop}%
\bibitem [{\citenamefont {Jiang}\ \emph {et~al.}(2009)\citenamefont {Jiang},
  \citenamefont {Rudenko}, \citenamefont {Kurka}, \citenamefont {Kühnel},
  \citenamefont {Ergler}, \citenamefont {Foucar}, \citenamefont {Sch\"offler},
  \citenamefont {Sch\"ossler}, \citenamefont {Havermeier}, \citenamefont
  {Smolarski}, \citenamefont {Cole}, \citenamefont {D\"orner}, \citenamefont
  {D\"usterer}, \citenamefont {Treusch}, \citenamefont {Gensch}, \citenamefont
  {Schr\"oter}, \citenamefont {Moshammer},\ and\ \citenamefont
  {Ullrich}}]{Jiang:PRL102:123002}%
  \BibitemOpen
  \bibfield  {author} {\bibinfo {author} {\bibfnamefont {Y.~H.}\ \bibnamefont
  {Jiang}}, \bibinfo {author} {\bibfnamefont {A.}~\bibnamefont {Rudenko}},
  \bibinfo {author} {\bibfnamefont {M.}~\bibnamefont {Kurka}}, \bibinfo
  {author} {\bibfnamefont {K.~U.}\ \bibnamefont {Kühnel}}, \bibinfo {author}
  {\bibfnamefont {T.}~\bibnamefont {Ergler}}, \bibinfo {author} {\bibfnamefont
  {L.}~\bibnamefont {Foucar}}, \bibinfo {author} {\bibfnamefont
  {M.}~\bibnamefont {Sch\"offler}}, \bibinfo {author} {\bibfnamefont
  {S.}~\bibnamefont {Sch\"ossler}}, \bibinfo {author} {\bibfnamefont
  {T.}~\bibnamefont {Havermeier}}, \bibinfo {author} {\bibfnamefont
  {M.}~\bibnamefont {Smolarski}}, \bibinfo {author} {\bibfnamefont
  {K.}~\bibnamefont {Cole}}, \bibinfo {author} {\bibfnamefont {R.}~\bibnamefont
  {D\"orner}}, \bibinfo {author} {\bibfnamefont {S.}~\bibnamefont
  {D\"usterer}}, \bibinfo {author} {\bibfnamefont {R.}~\bibnamefont {Treusch}},
  \bibinfo {author} {\bibfnamefont {M.}~\bibnamefont {Gensch}}, \bibinfo
  {author} {\bibfnamefont {C.~D.}\ \bibnamefont {Schr\"oter}}, \bibinfo
  {author} {\bibfnamefont {R.}~\bibnamefont {Moshammer}},\ and\ \bibinfo
  {author} {\bibfnamefont {J.}~\bibnamefont {Ullrich}},\ }\bibfield  {title}
  {\bibinfo {title} {{Few-Photon Multiple Ionization of $\mathrm{N_2}$ by
  Extreme Ultraviolet Free-Electron Laser Radiation}},\ }\href
  {https://doi.org/10.1103/PhysRevLett.102.123002} {\bibfield  {journal}
  {\bibinfo  {journal} {Phys. Rev. Lett.}\ }\textbf {\bibinfo {volume} {102}},\
  \bibinfo {pages} {123002} (\bibinfo {year} {2009})}\BibitemShut {NoStop}%
\bibitem [{\citenamefont {Wu}\ \emph {et~al.}(2011)\citenamefont {Wu},
  \citenamefont {Yang}, \citenamefont {Wu}, \citenamefont {Chen}, \citenamefont
  {Dong}, \citenamefont {Liu}, \citenamefont {Deng}, \citenamefont {Liu},
  \citenamefont {Liu},\ and\ \citenamefont {Gong}}]{Wu:PCCP13:18398}%
  \BibitemOpen
  \bibfield  {author} {\bibinfo {author} {\bibfnamefont {C.}~\bibnamefont
  {Wu}}, \bibinfo {author} {\bibfnamefont {Y.}~\bibnamefont {Yang}}, \bibinfo
  {author} {\bibfnamefont {Z.}~\bibnamefont {Wu}}, \bibinfo {author}
  {\bibfnamefont {B.}~\bibnamefont {Chen}}, \bibinfo {author} {\bibfnamefont
  {H.}~\bibnamefont {Dong}}, \bibinfo {author} {\bibfnamefont {X.}~\bibnamefont
  {Liu}}, \bibinfo {author} {\bibfnamefont {Y.}~\bibnamefont {Deng}}, \bibinfo
  {author} {\bibfnamefont {H.}~\bibnamefont {Liu}}, \bibinfo {author}
  {\bibfnamefont {Y.}~\bibnamefont {Liu}},\ and\ \bibinfo {author}
  {\bibfnamefont {Q.}~\bibnamefont {Gong}},\ }\bibfield  {title} {\bibinfo
  {title} {Coulomb explosion of nitrogen and oxygen molecules through
  non-coulombic states},\ }\href {https://doi.org/10.1039/C1CP21345H}
  {\bibfield  {journal} {\bibinfo  {journal} {Phys. Chem. Chem. Phys.}\
  }\textbf {\bibinfo {volume} {13}},\ \bibinfo {pages} {18398} (\bibinfo {year}
  {2011})}\BibitemShut {NoStop}%
\bibitem [{\citenamefont {Eckstein}\ \emph {et~al.}(2015)\citenamefont
  {Eckstein}, \citenamefont {Yang}, \citenamefont {Kubin}, \citenamefont
  {Frassetto}, \citenamefont {Poletto}, \citenamefont {Ritze}, \citenamefont
  {Vrakking},\ and\ \citenamefont {Kornilov}}]{Eckstein:JPCL6:419}%
  \BibitemOpen
  \bibfield  {author} {\bibinfo {author} {\bibfnamefont {M.}~\bibnamefont
  {Eckstein}}, \bibinfo {author} {\bibfnamefont {C.-H.}\ \bibnamefont {Yang}},
  \bibinfo {author} {\bibfnamefont {M.}~\bibnamefont {Kubin}}, \bibinfo
  {author} {\bibfnamefont {F.}~\bibnamefont {Frassetto}}, \bibinfo {author}
  {\bibfnamefont {L.}~\bibnamefont {Poletto}}, \bibinfo {author} {\bibfnamefont
  {H.-H.}\ \bibnamefont {Ritze}}, \bibinfo {author} {\bibfnamefont {M.~J.~J.}\
  \bibnamefont {Vrakking}},\ and\ \bibinfo {author} {\bibfnamefont
  {O.}~\bibnamefont {Kornilov}},\ }\bibfield  {title} {\bibinfo {title}
  {{Dynamics of $\mathrm{N_2}$ Dissociation upon Inner-Valence Ionization by
  Wavelength-Selected XUV Pulses}},\ }\href {https://doi.org/10.1021/jz5025542}
  {\bibfield  {journal} {\bibinfo  {journal} {J. Phys. Chem. Lett.}\ }\textbf
  {\bibinfo {volume} {6}},\ \bibinfo {pages} {419} (\bibinfo {year}
  {2015})}\BibitemShut {NoStop}%
\bibitem [{\citenamefont {Frasinski}\ \emph {et~al.}(1989)\citenamefont
  {Frasinski}, \citenamefont {Codling},\ and\ \citenamefont
  {Hatherly}}]{Frasinski:Science246:1989}%
  \BibitemOpen
  \bibfield  {author} {\bibinfo {author} {\bibfnamefont {L.~J.}\ \bibnamefont
  {Frasinski}}, \bibinfo {author} {\bibfnamefont {K.}~\bibnamefont {Codling}},\
  and\ \bibinfo {author} {\bibfnamefont {P.~A.}\ \bibnamefont {Hatherly}},\
  }\bibfield  {title} {\bibinfo {title} {Covariance mapping: a correlation
  method applied to multiphoton multiple ionization},\ }\href
  {https://doi.org/10.1126/science.246.4933.1029} {\bibfield  {journal}
  {\bibinfo  {journal} {Science}\ }\textbf {\bibinfo {volume} {246}},\ \bibinfo
  {pages} {1029} (\bibinfo {year} {1989})}\BibitemShut {NoStop}%
\bibitem [{\citenamefont {Pickering}\ \emph {et~al.}(2016)\citenamefont
  {Pickering}, \citenamefont {Amini}, \citenamefont {Brouard}, \citenamefont
  {Burt}, \citenamefont {Bush}, \citenamefont {Christensen}, \citenamefont
  {Lauer}, \citenamefont {Nielsen}, \citenamefont {Slater},\ and\ \citenamefont
  {Stapelfeldt}}]{Pickering:JCP144:161105}%
  \BibitemOpen
  \bibfield  {author} {\bibinfo {author} {\bibfnamefont {J.~D.}\ \bibnamefont
  {Pickering}}, \bibinfo {author} {\bibfnamefont {K.}~\bibnamefont {Amini}},
  \bibinfo {author} {\bibfnamefont {M.}~\bibnamefont {Brouard}}, \bibinfo
  {author} {\bibfnamefont {M.}~\bibnamefont {Burt}}, \bibinfo {author}
  {\bibfnamefont {I.~J.}\ \bibnamefont {Bush}}, \bibinfo {author}
  {\bibfnamefont {L.}~\bibnamefont {Christensen}}, \bibinfo {author}
  {\bibfnamefont {A.}~\bibnamefont {Lauer}}, \bibinfo {author} {\bibfnamefont
  {J.~H.}\ \bibnamefont {Nielsen}}, \bibinfo {author} {\bibfnamefont {C.~S.}\
  \bibnamefont {Slater}},\ and\ \bibinfo {author} {\bibfnamefont
  {H.}~\bibnamefont {Stapelfeldt}},\ }\bibfield  {title} {\bibinfo {title}
  {Communication: Three-fold covariance imaging of laser-induced coulomb
  explosions},\ }\href {https://doi.org/10.1063/1.4947551} {\bibfield
  {journal} {\bibinfo  {journal} {J. Chem. Phys.}\ }\textbf {\bibinfo {volume}
  {144}},\ \bibinfo {pages} {161105} (\bibinfo {year} {2016})}\BibitemShut
  {NoStop}%
\bibitem [{\citenamefont {Frasinski}(2016)}]{Frasinski:JPB49:152004}%
  \BibitemOpen
  \bibfield  {author} {\bibinfo {author} {\bibfnamefont {L.~J.}\ \bibnamefont
  {Frasinski}},\ }\bibfield  {title} {\bibinfo {title} {Covariance mapping
  techniques},\ }\href {https://doi.org/10.1088/0953-4075/49/15/152004}
  {\bibfield  {journal} {\bibinfo  {journal} {J. Phys. B}\ }\textbf {\bibinfo
  {volume} {49}},\ \bibinfo {pages} {152004} (\bibinfo {year}
  {2016})}\BibitemShut {NoStop}%
\bibitem [{\citenamefont {Dooley}\ \emph {et~al.}(2003)\citenamefont {Dooley},
  \citenamefont {Litvinyuk}, \citenamefont {Lee}, \citenamefont {Rayner},
  \citenamefont {Spanner}, \citenamefont {Villeneuve},\ and\ \citenamefont
  {Corkum}}]{Dooley:PRA68:023406}%
  \BibitemOpen
  \bibfield  {author} {\bibinfo {author} {\bibfnamefont {P.~W.}\ \bibnamefont
  {Dooley}}, \bibinfo {author} {\bibfnamefont {I.~V.}\ \bibnamefont
  {Litvinyuk}}, \bibinfo {author} {\bibfnamefont {K.~F.}\ \bibnamefont {Lee}},
  \bibinfo {author} {\bibfnamefont {D.~M.}\ \bibnamefont {Rayner}}, \bibinfo
  {author} {\bibfnamefont {M.}~\bibnamefont {Spanner}}, \bibinfo {author}
  {\bibfnamefont {D.~M.}\ \bibnamefont {Villeneuve}},\ and\ \bibinfo {author}
  {\bibfnamefont {P.~B.}\ \bibnamefont {Corkum}},\ }\bibfield  {title}
  {\bibinfo {title} {Direct imaging of rotational wave-packet dynamics of
  diatomic molecules},\ }\href {https://doi.org/10.1103/PhysRevA.68.023406}
  {\bibfield  {journal} {\bibinfo  {journal} {Phys. Rev. A}\ }\textbf {\bibinfo
  {volume} {68}},\ \bibinfo {pages} {023406} (\bibinfo {year}
  {2003})}\BibitemShut {NoStop}%
\bibitem [{\citenamefont {Kornilov}\ \emph {et~al.}(2013)\citenamefont
  {Kornilov}, \citenamefont {Eckstein}, \citenamefont {Rosenblatt},
  \citenamefont {Schulz}, \citenamefont {Motomura}, \citenamefont
  {Rouz{\'{e}}e}, \citenamefont {Klei}, \citenamefont {Foucar}, \citenamefont
  {Siano}, \citenamefont {Lübcke}, \citenamefont {Schapper}, \citenamefont
  {Johnsson}, \citenamefont {Holland}, \citenamefont {Schlathölter},
  \citenamefont {Marchenko}, \citenamefont {Düsterer}, \citenamefont {Ueda},
  \citenamefont {Vrakking},\ and\ \citenamefont
  {Frasinski}}]{Kornilov:JPB46:164028}%
  \BibitemOpen
  \bibfield  {author} {\bibinfo {author} {\bibfnamefont {O.}~\bibnamefont
  {Kornilov}}, \bibinfo {author} {\bibfnamefont {M.}~\bibnamefont {Eckstein}},
  \bibinfo {author} {\bibfnamefont {M.}~\bibnamefont {Rosenblatt}}, \bibinfo
  {author} {\bibfnamefont {C.~P.}\ \bibnamefont {Schulz}}, \bibinfo {author}
  {\bibfnamefont {K.}~\bibnamefont {Motomura}}, \bibinfo {author}
  {\bibfnamefont {A.}~\bibnamefont {Rouz{\'{e}}e}}, \bibinfo {author}
  {\bibfnamefont {J.}~\bibnamefont {Klei}}, \bibinfo {author} {\bibfnamefont
  {L.}~\bibnamefont {Foucar}}, \bibinfo {author} {\bibfnamefont
  {M.}~\bibnamefont {Siano}}, \bibinfo {author} {\bibfnamefont
  {A.}~\bibnamefont {Lübcke}}, \bibinfo {author} {\bibfnamefont
  {F.}~\bibnamefont {Schapper}}, \bibinfo {author} {\bibfnamefont
  {P.}~\bibnamefont {Johnsson}}, \bibinfo {author} {\bibfnamefont {D.~M.~P.}\
  \bibnamefont {Holland}}, \bibinfo {author} {\bibfnamefont {T.}~\bibnamefont
  {Schlathölter}}, \bibinfo {author} {\bibfnamefont {T.}~\bibnamefont
  {Marchenko}}, \bibinfo {author} {\bibfnamefont {S.}~\bibnamefont
  {Düsterer}}, \bibinfo {author} {\bibfnamefont {K.}~\bibnamefont {Ueda}},
  \bibinfo {author} {\bibfnamefont {M.~J.~J.}\ \bibnamefont {Vrakking}},\ and\
  \bibinfo {author} {\bibfnamefont {L.~J.}\ \bibnamefont {Frasinski}},\
  }\bibfield  {title} {\bibinfo {title} {Coulomb explosion of diatomic
  molecules in intense {XUV} fields mapped by partial covariance},\ }\href
  {https://doi.org/10.1088/0953-4075/46/16/164028} {\bibfield  {journal}
  {\bibinfo  {journal} {J. Phys. B}\ }\textbf {\bibinfo {volume} {46}},\
  \bibinfo {pages} {164028} (\bibinfo {year} {2013})}\BibitemShut {NoStop}%
\bibitem [{\citenamefont {Lehmann}\ \emph {et~al.}(2016)\citenamefont
  {Lehmann}, \citenamefont {Pic\'{o}n}, \citenamefont {Bostedt}, \citenamefont
  {Rudenko}, \citenamefont {Marinelli}, \citenamefont {Moonshiram},
  \citenamefont {Osipov}, \citenamefont {Rolles}, \citenamefont {Berrah},
  \citenamefont {Bomme}, \citenamefont {Bucher}, \citenamefont {Doumy},
  \citenamefont {Erk}, \citenamefont {Ferguson}, \citenamefont {Gorkhover},
  \citenamefont {Ho}, \citenamefont {Kanter}, \citenamefont {Kr\"assig},
  \citenamefont {Krzywinski}, \citenamefont {Lutman}, \citenamefont {March},
  \citenamefont {Ray}, \citenamefont {Young}, \citenamefont {Pratt},\ and\
  \citenamefont {Southworth}}]{Lehmann:PRA94:013426}%
  \BibitemOpen
  \bibfield  {author} {\bibinfo {author} {\bibfnamefont {C.~S.}\ \bibnamefont
  {Lehmann}}, \bibinfo {author} {\bibfnamefont {A.}~\bibnamefont {Pic\'{o}n}},
  \bibinfo {author} {\bibfnamefont {C.}~\bibnamefont {Bostedt}}, \bibinfo
  {author} {\bibfnamefont {A.}~\bibnamefont {Rudenko}}, \bibinfo {author}
  {\bibfnamefont {A.}~\bibnamefont {Marinelli}}, \bibinfo {author}
  {\bibfnamefont {D.}~\bibnamefont {Moonshiram}}, \bibinfo {author}
  {\bibfnamefont {T.}~\bibnamefont {Osipov}}, \bibinfo {author} {\bibfnamefont
  {D.}~\bibnamefont {Rolles}}, \bibinfo {author} {\bibfnamefont
  {N.}~\bibnamefont {Berrah}}, \bibinfo {author} {\bibfnamefont
  {C.}~\bibnamefont {Bomme}}, \bibinfo {author} {\bibfnamefont
  {M.}~\bibnamefont {Bucher}}, \bibinfo {author} {\bibfnamefont
  {G.}~\bibnamefont {Doumy}}, \bibinfo {author} {\bibfnamefont
  {B.}~\bibnamefont {Erk}}, \bibinfo {author} {\bibfnamefont {K.~R.}\
  \bibnamefont {Ferguson}}, \bibinfo {author} {\bibfnamefont {T.}~\bibnamefont
  {Gorkhover}}, \bibinfo {author} {\bibfnamefont {P.~J.}\ \bibnamefont {Ho}},
  \bibinfo {author} {\bibfnamefont {E.~P.}\ \bibnamefont {Kanter}}, \bibinfo
  {author} {\bibfnamefont {B.}~\bibnamefont {Kr\"assig}}, \bibinfo {author}
  {\bibfnamefont {J.}~\bibnamefont {Krzywinski}}, \bibinfo {author}
  {\bibfnamefont {A.~A.}\ \bibnamefont {Lutman}}, \bibinfo {author}
  {\bibfnamefont {A.~M.}\ \bibnamefont {March}}, \bibinfo {author}
  {\bibfnamefont {D.}~\bibnamefont {Ray}}, \bibinfo {author} {\bibfnamefont
  {L.}~\bibnamefont {Young}}, \bibinfo {author} {\bibfnamefont {S.~T.}\
  \bibnamefont {Pratt}},\ and\ \bibinfo {author} {\bibfnamefont {S.~H.}\
  \bibnamefont {Southworth}},\ }\bibfield  {title} {\bibinfo {title} {Ultrafast
  x-ray-induced nuclear dynamics in diatomic molecules using femtosecond
  x-ray-pump--x-ray-probe spectroscopy},\ }\href
  {https://doi.org/10.1103/PhysRevA.94.013426} {\bibfield  {journal} {\bibinfo
  {journal} {Phys. Rev. A}\ }\textbf {\bibinfo {volume} {94}},\ \bibinfo
  {pages} {013426} (\bibinfo {year} {2016})}\BibitemShut {NoStop}%
\bibitem [{\citenamefont {Sorokin}\ \emph {et~al.}(2019)\citenamefont
  {Sorokin}, \citenamefont {Bican}, \citenamefont {Bonfigt}, \citenamefont
  {Brachmanski}, \citenamefont {Braune}, \citenamefont {Jastrow}, \citenamefont
  {Gottwald}, \citenamefont {Kaser}, \citenamefont {Richter},\ and\
  \citenamefont {Tiedtke}}]{Sorokin:JSR26:1092}%
  \BibitemOpen
  \bibfield  {author} {\bibinfo {author} {\bibfnamefont {A.~A.}\ \bibnamefont
  {Sorokin}}, \bibinfo {author} {\bibfnamefont {Y.}~\bibnamefont {Bican}},
  \bibinfo {author} {\bibfnamefont {S.}~\bibnamefont {Bonfigt}}, \bibinfo
  {author} {\bibfnamefont {M.}~\bibnamefont {Brachmanski}}, \bibinfo {author}
  {\bibfnamefont {M.}~\bibnamefont {Braune}}, \bibinfo {author} {\bibfnamefont
  {U.~F.}\ \bibnamefont {Jastrow}}, \bibinfo {author} {\bibfnamefont
  {A.}~\bibnamefont {Gottwald}}, \bibinfo {author} {\bibfnamefont
  {H.}~\bibnamefont {Kaser}}, \bibinfo {author} {\bibfnamefont
  {M.}~\bibnamefont {Richter}},\ and\ \bibinfo {author} {\bibfnamefont
  {K.}~\bibnamefont {Tiedtke}},\ }\bibfield  {title} {\bibinfo {title} {{An
  {X-ray} gas monitor for free-electron lasers}},\ }\href
  {https://doi.org/10.1107/S1600577519005174} {\bibfield  {journal} {\bibinfo
  {journal} {J. Synchrotron\ Rad.}\ }\textbf {\bibinfo {volume} {26}},\
  \bibinfo {pages} {1092} (\bibinfo {year} {2019})}\BibitemShut {NoStop}%
\bibitem [{\citenamefont {Garzetti}\ \emph {et~al.}(2020)\citenamefont
  {Garzetti}, \citenamefont {Lusardi}, \citenamefont {Corna}, \citenamefont
  {Salgaro}, \citenamefont {Busola}, \citenamefont {Geraci}, \citenamefont
  {Brajnik}, \citenamefont {Carrato}, \citenamefont {Cautero}, \citenamefont
  {Cautero}, \citenamefont {Sergo},\ and\ \citenamefont
  {Stebel}}]{Garzetti:IEEE:1}%
  \BibitemOpen
  \bibfield  {author} {\bibinfo {author} {\bibfnamefont {F.}~\bibnamefont
  {Garzetti}}, \bibinfo {author} {\bibfnamefont {N.}~\bibnamefont {Lusardi}},
  \bibinfo {author} {\bibfnamefont {N.}~\bibnamefont {Corna}}, \bibinfo
  {author} {\bibfnamefont {S.}~\bibnamefont {Salgaro}}, \bibinfo {author}
  {\bibfnamefont {N.}~\bibnamefont {Busola}}, \bibinfo {author} {\bibfnamefont
  {A.}~\bibnamefont {Geraci}}, \bibinfo {author} {\bibfnamefont
  {G.}~\bibnamefont {Brajnik}}, \bibinfo {author} {\bibfnamefont
  {S.}~\bibnamefont {Carrato}}, \bibinfo {author} {\bibfnamefont
  {G.}~\bibnamefont {Cautero}}, \bibinfo {author} {\bibfnamefont
  {M.}~\bibnamefont {Cautero}}, \bibinfo {author} {\bibfnamefont
  {R.}~\bibnamefont {Sergo}},\ and\ \bibinfo {author} {\bibfnamefont
  {L.}~\bibnamefont {Stebel}},\ }\bibfield  {title} {\bibinfo {title} {{Fully
  FPGA-based 3D (X,Y,t) imaging system with Cross Delay-Lines detectors and
  Eight-Channels High-Performance Time-to-Digital Converter}},\ }in\ \href
  {https://doi.org/10.1109/NSS/MIC42677.2020.9508048} {\emph {\bibinfo
  {booktitle} {2020 IEEE Nuclear Science Symposium and Medical Imaging
  Conference (NSS/MIC)}}}\ (\bibinfo {year} {2020})\ pp.\ \bibinfo {pages}
  {1--4}\BibitemShut {NoStop}%
\bibitem [{\citenamefont {Ballabriga}\ \emph {et~al.}(2020)\citenamefont
  {Ballabriga}, \citenamefont {Campbell},\ and\ \citenamefont
  {Llopart}}]{Ballabriga:RadMeas136:106271}%
  \BibitemOpen
  \bibfield  {author} {\bibinfo {author} {\bibfnamefont {R.}~\bibnamefont
  {Ballabriga}}, \bibinfo {author} {\bibfnamefont {M.}~\bibnamefont
  {Campbell}},\ and\ \bibinfo {author} {\bibfnamefont {X.}~\bibnamefont
  {Llopart}},\ }\bibfield  {title} {\bibinfo {title} {An introduction to the
  {M}edipix family {ASIC}s},\ }\href
  {https://doi.org/10.1016/j.radmeas.2020.106271} {\bibfield  {journal}
  {\bibinfo  {journal} {Radiat. Meas.}\ }\textbf {\bibinfo {volume} {136}},\
  \bibinfo {pages} {106271} (\bibinfo {year} {2020})}\BibitemShut {NoStop}%
\bibitem [{\citenamefont {Llopart}\ \emph {et~al.}(2021)\citenamefont
  {Llopart}, \citenamefont {Alozy}, \citenamefont {Ballabriga}, \citenamefont
  {Campbell}, \citenamefont {Casanova}, \citenamefont {Gromov}, \citenamefont
  {Heijne}, \citenamefont {Poikela}, \citenamefont {Santin}, \citenamefont
  {Sriskaran}, \citenamefont {Tlustos},\ and\ \citenamefont
  {Vitkovskiy}}]{Llopart:JINST:inprep}%
  \BibitemOpen
  \bibfield  {author} {\bibinfo {author} {\bibfnamefont {X.}~\bibnamefont
  {Llopart}}, \bibinfo {author} {\bibfnamefont {J.}~\bibnamefont {Alozy}},
  \bibinfo {author} {\bibfnamefont {R.}~\bibnamefont {Ballabriga}}, \bibinfo
  {author} {\bibfnamefont {M.}~\bibnamefont {Campbell}}, \bibinfo {author}
  {\bibfnamefont {R.}~\bibnamefont {Casanova}}, \bibinfo {author}
  {\bibfnamefont {V.}~\bibnamefont {Gromov}}, \bibinfo {author} {\bibfnamefont
  {E.}~\bibnamefont {Heijne}}, \bibinfo {author} {\bibfnamefont
  {T.}~\bibnamefont {Poikela}}, \bibinfo {author} {\bibfnamefont
  {E.}~\bibnamefont {Santin}}, \bibinfo {author} {\bibfnamefont
  {V.}~\bibnamefont {Sriskaran}}, \bibinfo {author} {\bibfnamefont
  {L.}~\bibnamefont {Tlustos}},\ and\ \bibinfo {author} {\bibfnamefont
  {A.}~\bibnamefont {Vitkovskiy}},\ }\bibfield  {title} {\bibinfo {title}
  {Timepix4, a large area pixel detector readout chip which can be tiled on 4
  sides providing sub-200ps timestamp binning},\ }\href@noop {} {\bibfield
  {journal} {\bibinfo  {journal} {J. Instrum.}\ }}\bibinfo {note}
  {Submitted}\BibitemShut {NoStop}%
\end{thebibliography}%

\end{document}